\documentclass [10pt]{iopart}
\usepackage[dvips]{graphicx}
\newcommand\sign{\mbox{\, \rm sgn}}

\begin{document}
\jl{4}

\title[Cellular automata]{Cellular automata approach to three-phase traffic theory}

\author{Boris S. Kerner$^1$, Sergey L. Klenov$^2$ and Dietrich E. Wolf $^3$} 

\address{
$^1$ DaimlerChrysler AG, RIC/TN, HPC: T729, 70546 Stuttgart, Germany 
}

\address{
$^2$ Moscow Institute of Physics and Technology, Department of Physics,
141700 Dolgoprudny, Moscow Region, Russia
}

\address{
$^3$ Institut f\"ur Physik, Gehard-Mercator-Universit\"at Duisburg,
 D-47048 Duisburg, Germany}

\date{January 28, 2002}


\begin{abstract}
The cellular automata (CA) approach to traffic modeling  is extended to 
allow for spatially homogeneous steady state solutions that cover a 
two dimensional region in the flow-density plane. Hence these models fulfill
a basic postulate of a three-phase traffic theory proposed by Kerner.
This is achieved by a synchronization distance, within which a vehicle always
tries to adjust its speed to the one of the 
vehicle in front.
In the CA models presented,  the modelling of the free and safe
speeds, the slow-to-start rules
as well as some contributions to noise are based on the ideas of
the Nagel-Schreckenberg type modelling.
It is shown that the proposed CA models can be very transparent and still reproduce 
the two main types of congested patterns 
(the general pattern and the synchronized flow pattern)
as well as their dependence on the flows 
near an on-ramp, in qualitative agreement with the recently developed 
continuum version of the three-phase traffic 
theory [B. S. Kerner and S. L. Klenov. 2002. {\it J. Phys. A: Math. Gen.} {\bf
  35}, L31].  These features are qualitatively different 
than in previously considered CA traffic models.  
The probability of the breakdown phenomenon (i.e., of the phase
transition from free flow to synchronized flow) as function of the
flow rate to the on-ramp and of the flow rate on the road upstream of
the on-ramp is investigated.  The capacity drops at the on-ramp which
occur due to the formation of different congested patterns
 are calculated.
\end{abstract}

\maketitle

\section{Introduction}

Traffic on a highway can be either free or congested.
In empirical investigations congested traffic  shows a very complex 
spatial-temporal behaviour
(see the reviews~\cite{Helbing,Ch,Kerner1999A,Kerner2001}).
Based on a recent empirical study~\cite{KR1996A,Kerner1998} 
Kerner found out that in congested traffic 
two different traffic phases should be distinguished: ``synchronized flow''
and ``wide moving jam''.  
Therefore, there are three traffic phases: 1. free flow, 2. synchronized flow,
3. wide moving jam. 

A wide moving jam is a localized structure moving upstream and limited
by two fronts where the vehicle speed changes sharply, i.e. within a
region that is small compared to the distance between the fronts.
A wide moving jam propagates through either free or synchronized flows and 
through any bottlenecks (e.g. at on-ramps) keeping the velocity
of its downstream front~\cite{Kerner2001,Kerner2000A}. 
In this respect it differs from synchronized flow,
the downstream front of which is usually {\it fixed} at the bottleneck, 
where it occurred.
Such {\it empirical spatial-temporal} features of 
``wide moving jams'' and ``synchronized flow''
are the basis for the distinction of these traffic phases
in congested traffic rather than a
behaviour of traffic data in the flow-density
plane~\cite{Kerner2001,Kerner2000A,Kerner2002B}. 

Wide moving jams do not emerge spontaneously in free flow (with the possible 
exception that synchronized flow is somehow prohibited~\cite{Kerner2000B}). 
Instead, there is a sequence of two first order phase transitions
~\cite{Kerner1998}: First the transition from free flow 
to synchronized flow occurs (called the F$\rightarrow$S-transition),
and only later and usually at a different location moving jams emerge 
in the synchronized flow (the S$\rightarrow$J-transition).

Different explanations of these empirical findings have been proposed by 
various groups in the last years, but so far they remain controversial
(see e.g.,~\cite{Helbing1999,Lee1998,Lee,Lee2000A,Lee2000B,
  Helbing2000,Helbing1999A,Helbing2002A,Tomer,Lubashevsky, 
  Lubashevsky2001,Nelson,Knospe2000,Knospe2001,KKl,Wagner2002A,Fukui}
and the excellent review by Helbing~\cite{Helbing}).

\subsection{The fundamental diagram approach and cellular automata models of 
the Nagel-Schreckenberg (NaSch) type}

Empirical observations show that the higher the vehicle density is the lower 
the average vehicle  speed. The average flow rate, which is the  
product of the average speed and the density, is a function of the density 
which has a maximum.
This curve in the flow-density plane is called the fundamental diagram~\cite{Helbing,Ch,
  May}. 

Apparently the empirical fundamental diagram was the reason that already 
the first  traffic flow models~\cite{LW,GH,New,Wh,Prigogine}  
were based upon the postulate that hypothetical
spatially homogeneous and time-independent 
solutions, 
where all vehicles have the same distances to their neighbours
and move with the same constant speed,
(steady state solutions for short) exist that are related to a
fundamental diagram,  i.e. a curve 
in the flow-density plane. (Steady state model solutions are often called ``equilibrium'' 
solutions or ``equilibrium'' states of a traffic flow model. 
In this paper we will use the term ``steady states''.)
A subset of these steady
states would be unstable with respect to noise or external perturbations.
This postulate underlies almost all 
traffic flow modeling approaches  up to now~\cite{Helbing,Ch} in the sense
that the models are constructed such that in the unperturbed, noiseless limit
they have a fundamental diagram of steady states, i.e. the
steady states form a curve in the flow-density plane.

The congested patterns which are calculated from these models for a
homogeneous road (i.e. a road without bottlenecks)
\cite{Helbing,KK1994,Kerner1995,HK} 
are due to the instability of steady states of the fundamental
diagram within some range of vehicle densities. When perturbed, they
decay into a single or a sequence of wide moving jams (``moving
clusters'' \cite{KK1994}), whose outflows accelerate to rather high
speeds comparable to free flow. This is why we
find it helpful to classify these models as belonging to what 
we call the ``fundamental diagram approach''. 
In the next subsection
another class of models will be described, which belong to the ``three-phase
traffic theory'' and lead to qualitatively different congested
patterns, which 
are
in better agreement with real observations. Its
hallmark is a third kind of flow in addition to the wide moving jams
and free flow, in which all vehicles interact strongly, but
nevertheless drive smoothly at a reduced, rather uniform speed
(synchronized flow). In view of the rich phenomenologies of the two model
classes in particular near an on-ramp, the distinction of the two classes is
admittedly a bit simplified at this point, but a more detailed
comparison will be given in Sec. \ref{Dis}.

Apart from the congested patterns on a homogeneous road the 
dynamical behaviour near an on-ramp differs significantly in 
the fundamental diagram approach~\cite{Helbing,Helbing1999,
Lee1998,Lee,Lee2000A,Lee2000B,Kerner1995,Helbing2000}
and in the three-phase traffic theory~\cite{Kerner2002B,KKl}.
It depends on both the flow rate to the on-ramp $q_{\rm
  on}$ and the initial flow rate on the road upstream of the on-ramp,
$q_{\rm in}$. Therefore it can be conveniently characterized by a
so-called diagram of congested patterns. This is a map with
coordinates $q_{\rm on}$ and $q_{\rm in}$ of the regions, where
different congested patterns upstream of the on-ramp occur.

In this paper we focus on cellular automata (CA) traffic flow  modeling 
which was pioneered by Nagel and Schreckenberg in 1991/1992~\cite{NaSch}.
The original Nagel-Schreckenberg model (NaSch model) as well as
all subsequent modifications and refinements of 
it~\cite{ Knospe2000,Knospe2001,NaSch,Nagel1995,Schreckenberg,Barlovic}
belong to the fundamental diagram approach 
(Fig.~\ref{NaSch_Diagram} (a)). 
See also the reviews by Wolf~\cite{Wolf}, Chowdhury {\it et al.}~\cite{Ch}
and Helbing~\cite{Helbing}. 
In this paper CA models belonging to the three-phase traffic theory
will be proposed, and it will be shown that their congested patterns
differ qualitatively from the ones in the CA models of the NaSch-type.

Within the fundamental diagram approach
there is a subset of traffic models which show spatially homogeneous
high density states with low speed. At first sight these states 
look like ``synchronized flow'', but this interpretation is
incompatible with the observed dynamical behaviour near an
on-ramp, as will be discussed now. In the diagram of congested patterns near an
on-ramp obtained by Helbing {\it et al.}~\cite{Helbing,Helbing1999} 
such high density states with low speed occur upstream of the on-ramp, if
the flow rate $q_{\rm on}$ to the on-ramp is high enough. Helbing {\it et al.}
called these states ``homogeneous congested traffic''
(HCT)~\cite{Helbing1999} and proposed to identify it with synchronized flow~\cite{Helbing}. 
In HCT {\it no} moving jams occur {\it spontaneously}. 

By contrast, in empirical observations of synchronized flow at   
low vehicle speed and  high enough vehicle density moving jams
do emerge spontaneously.  These moving jams are particularly likely to
occur, when due to the high flow rate to the on-ramp 
a strong compression of synchronized flow appears 
(``the pinch effect'')~\cite{Kerner1998,Kerner2002B}. 
In fact, moving jams are {\it only} observed to emerge spontaneously 
in synchronized flow upstream of an on-ramp on which the flow rate is
high enough~\cite{Kerner2002B}. At lower flow rates to the on-ramp 
synchronized flow of higher vehicle speed can occur in which
the nucleation of moving jams was not observed.

\subsection{The three-phase traffic theory}

To explain the above and other empirical results, Kerner introduced 
a three-phase traffic theory which postulates that the steady states 
of synchronized flow cover {\it a 2D region} in the flow-density plane, i.e. 
there is {\it no} fundamental diagram of traffic flow in this 
theory~\cite{Kerner1999A,Kerner1998,Kerner1998A,Kerner1999B,Kerner1999C}.

This is not excluded by the empirical fact mentioned above, that a given
vehicle density determines the {\it average} vehicle speed. 
Even if there is a continuous interval of different 
vehicle speeds at the same distance between vehicles (at the same
density), as is indicated by empirical observations of synchronized flow,
obviously their averaging leads to one value at the given density.
A 2D region of steady states in the flow-density plane is also not
ruled out by car following experiments, where a driver has
the task to follow a specific leading car and not loose contact 
with it (e.g.,~\cite{Koshi}). In such
a situation, the gap between the cars will be biased towards the security gap 
depending on the speed of the leading car. In synchronized flow the situation
is different: The gap between cars can be much larger than the security gap.

As it has recently been postulated on very general grounds
~\cite{Kerner2002B,Kerner2002A} and demonstrated for a microscopic traffic 
model~\cite{KKl},
the fact whether the steady states of a mathematical description of traffic
belong to a curve  or to a 2D-region in the flow-density plane 
{\it qualitatively changes} the basic non-linear 
spatial-temporal features of the congested patterns
which the model allows. 

In particular the diagram of the congested patterns at on-ramps 
is qualitatively different in three-phase traffic theory~\cite{Kerner2002B,Kerner2002A,KKl}   
from the diagram obtained 
in the fundamental diagram approach
~\cite{Helbing,Helbing1999,Lee,Lee2000A,Lee2000B,Helbing2000,Helbing2002A}.
Specifically, at a high flow rate to the on-ramp,
instead of HCT without moving jams, we find that
moving jams always spontaneously emerge in  synchronized flow
of  low vehicle speed and high density, 
whereas synchronized flow of higher 
vehicle speed  can exist for a long time without
an occurrence of moving jams~\cite{Kerner2002B,KKl,Kerner2002A}. This agrees with the 
empirical observations~\cite{Kerner2002B}.

The microscopic model proposed by Kerner and Klenov
in order to derive the diagram of congested patterns
within the three-phase traffic theory is relatively complex~\cite{KKl}.
The main aim of this article is a derivation
of cellular automata models within the 
three-phase traffic theory  (where
steady states of the models cover 2D-regions in 
the flow-density plane) which on the one hand
are the most simple ones and on the other hand are able to reproduce
the diagram of congested patterns found
in~\cite{Kerner2002B,KKl,Kerner2002A}.
The article is organized  as follows.
First, in the next section  several cellular automata models with 
qualitatively different 2D-regions in 
the flow-density plane for steady states
will be introduced. Second, the congested patterns at an on-ramp and
their diagrams 
will be numerically derived for these models and compared with one another
and with the results in~\cite{KKl}.
Third, the probability of the breakdown phenomenon
(the F$\rightarrow$S transition)
at an on-ramp is studied  and the capacity drop is calculated.
Finally, the results of CA-models in the
three-phase traffic theory and in the fundamental 
diagram approach are compared.

\section{Cellular automata models within three-phase traffic theory}

\subsection{Equations of motion}

The starting point of CA-modeling of three-phase traffic theory is
the basic set of rules from~\cite{KKl} which provides a 2D-region of
steady states. Denoting the speed and space coordinate of a vehicle at
discrete time 
$t=n \tau$, 
$n=0,1,2,..$ by $v_{n}$ and $x_{n}$,
respectively, the basic rules are rewritten for CA-models in the form: 
\begin{equation}
v_{n+1}=\max(0, \min(v_{\rm free},  v_{{\rm s},n}, v_{{\rm c},n})),
\quad  x_{n+1}= x_{n}+v_{n+1}\tau.
\label{next}
\end{equation}
$v_{\rm free}$ is the maximum speed of the vehicles. It is
assumed to be the same for all vehicles in this paper.
$v_{{\rm s},n}$ is the save speed which must not be exceeded in order
to avoid collisions. In general it depends on the space gap between vehicles,
$g_{n}=x_{\ell, n}-x_{n}-d$ and the speed $v_{\ell,n}$
\cite{Kr,Wolf}, where the lower index $\ell$ marks 
functions (or values) related to the vehicle in front of the one at
$x_{n}$, the ``leading vehicle'', and $d$ is the vehicle length (assumed to
be the same for all vehicles in this paper). For the
sake of comparability we neglect the $v_{\ell,n}$-dependence 
and choose the same expression as in the standard NaSch-model~\cite{NaSch}:
\begin{equation}
v_{{\rm s},n}= g_{n}/\tau.
\label{safe}
\end{equation}

The crucial difference compared to previous CA-models is that the
acceleration behaviour given by $v_{{\rm c},n}$ (the rule of ``speed
change'') depends on, whether the leading car is within a
``synchronization distance'' $D_{n}=D(v_{n})$ or further away
\cite{KKl}.  
At sufficiently large distances from the leading vehicle, one simply
accelerates with an acceleration $a$, which is assumed to be the same
for all vehicles and independent of time in this paper.
However within the synchronization distance the vehicle tends to
adjust its speed to the one of the leading vehicle, i.e. it
decelerates with deceleration $b$ if it is faster, and
accelerates with $a$, if it becomes slower than the leading vehicle. 
The deceleration $b$ should not be confused with braking for
safety purposes (i.e. in order not to exceed $v_{{\rm s}, n}$). In
practice the speed adjustment within the synchronization distance 
can often be achieved without braking at all simply as a result of
rolling friction of the wheels with the road. Therefore the
deceleration $b$ is typically smaller than the braking capability
of a vehicle.
For simplicity we set $b=a$ in this paper, so that the speed
change per time step within the synchronization distance is given by
$\Delta v_{n} = a\tau \sign(v_{\ell,n}-v_{n})$ ,
where $\sign(x)$ is 1 for $x>0$, 0 for $x=0$ and $-1$ for $x<0$.
In summary,
\begin{equation}
v_{{\rm c},n}=\left\{
\begin{array}{ll}
v_{n}+a\tau &  \textrm{for $x_{\ell,n}-x_{n} > D_{n}$},\\
v_{n}+a\tau \sign(v_{\ell,n}-v_{n}) &  \textrm{for $x_{\ell,n}-x_{n}
  \leq D_{n}$}. \\ 
\end{array} \right. 
\label{next1}
\end{equation}
We want to emphasise that this rule decouples speed and gap between 
vehicles for dense traffic. This can be seen by assuming that
vehicles drive behind each other with the same speed $v$. According to 
(\ref{next1}) neither the speed nor the gaps will change, provided,
all the distances are anywhere between the safe distance $d + v\tau$
and the synchronization distance $D(v) \geq d + v\tau$. There is neither a 
speed-dependent distance, which individual drivers prefer, nor is there 
a distance-dependent optimal speed. This is the principal conceptual
difference between three-phase traffic theory and the fundamental
diagram approach, and it is the reason, why in three phase traffic theory
the steady states fill a two dimensional region in the flow-density-plane,
while in the fundamental diagram approach they lie on a curve. 

Let us contrast (\ref{next1}) with two models belonging to the fundamental 
diagram approach, the NaSch-model with ``comfortable driving'' recently
put forward by Knospe {\em et al.} \cite{Knospe2000,Knospe2001,Knospe2002}
on the one hand, and Wiedemann's modelling approach \cite{Wiedemann1974} 
on the other. Knospe {\em et al.}  \cite{Knospe2000,Knospe2001,Knospe2002} 
put forward an extension of the NaSch CA-model, in which a driver starts 
to brake within some interaction horizon, as soon as he sees the brake 
lights of the vehicle in front being  switched on. If nobody puts on the 
brakes, vehicles would close up to the safe distance, which is a function 
of the speed. In this sense, the term ``comfortable driving'' is a bit
misleading: The behaviour modeled is more accurately described as 
``comfortable braking''. In contrast, the speed synchronisation 
(\ref{next1}) in our model happens always, whether someone brakes 
or not. It reflects what we believe to be the typical way, in which
drivers take into account the vehicle in front of them. 

In Wiedemann's modelling approach \cite{Wiedemann1974} each vehicle has a
prefered following distance, but convergence to it is hindered due to
imperfect perception, so that the actual distance has an oscillatory
behaviour. This model belongs to the fundamental diagram approach,
too, as the steady state solutions have a unique relationship between
speed and distance between vehicles.   

Of course, in models with a fundamental diagram of steady states,
fluctuations and external perturbations let the system evolve in time
through a 2D region in the flow-density plane as well. However, the
dynamics is governed locally by steady state properties, the unstable
steady states acting as ``repellors'' and the stable ones as
``attractors''. If the steady states form a 2D region, part of which
is stable and part of which is metastable, as is the case in
three-phase traffic theory, the dynamics is fundamentally different. 
This leads also to qualitative differences between the patterns of congested
traffic obtained in three-phase traffic theory or in the fundamental
diagram approach, respectively, as will be shown in detail below.

\subsection{Synchronization distance}

The conditions (\ref{next}), (\ref{next1}) are the basis of the
cellular automata models under consideration. It will be shown that
this allows different formulations for fluctuations, acceleration,
deceleration and for the synchronization distance $D_{n}$ which all
lead to qualitatively the same features of congested patterns and the
same diagram of these patterns as postulated in three-phase
traffic theory~\cite{Kerner2002B,Kerner2002A} and in agreement with
the continuum model of Kerner and Klenov in~\cite{KKl}.

In particular, let us consider two different formulations for the dependence
of the   synchronization distance $D_{n}$ on the vehicle speed.
In the first formulation,  
the synchronization
distance $D_{n}$ in (\ref{next1}) is a linear function of the vehicle speed:
\begin{equation}
D(v_{n})=d_{1} + kv_{n}\tau.
\label{D}
\end{equation}
In the second formulation, the synchronization distance $D_{n}$ in
(\ref{next1}) is a non-linear function of the vehicle speed: 
\begin{equation}
D(v_{n})= d + v_{n}\tau  + \beta v^{2}_{n}/2a.
\label{D2}
\end{equation}
In (\ref{D}) and (\ref{D2}) $d_{1}$, $k$ and $\beta$ are positive constants.  
Both formulations lead to 2D-regions of steady states in the
flow-density plane.

\subsection{Steady states}
\label{Sec:SteadyStates}

Whereas in models belonging to the fundamental diagram approach
(e.g.,~\cite{Helbing1999,Lee,Helbing1999A,Tomer,Knospe2000, Knospe2001,NaSch,
  Barlovic,Kr,Nagel,Knospe2002}  and the  reviews~\cite{Ch,Helbing}) a vehicle
would close up to the leading one adjusting its speed and gap as
required by secure driving, in models with the basic structure
(\ref{next}),  (\ref{next1}), 
a driver within the synchronization distance $D_{n}$
adapts his speed to the one of the 
vehicle in front 
without caring,
what the precise gap is, as long as it is safe. This explains why
there is no unique flow-density relationship for steady states in the
present CA models. 

In steady states all accelerations must be zero. Then the time-index
$n$ can be dropped in the above formulas. According to (\ref{next}) --
(\ref{next1}),
there are two possibilities: Either the synchronization distance $D(v) <
g + d$ and the speed is $v = v_{\rm free}$, or 
\begin{equation}
D(v) \geq g + d  \quad {\rm and} \quad v = v_{\ell} \leq \min(v_{\rm
  free}, v_{s}(g,v)). 
\label{2D}
\end{equation}
Thus, in  steady states all speeds are equal $v$. The conditions
$v$ has to fulfill are only equal, if also the gaps $g$ are all
equal. Therefore we defined steady states above as time-independent
{\it and} homogeneous. 

\begin{figure}
\includegraphics[width=\textwidth]{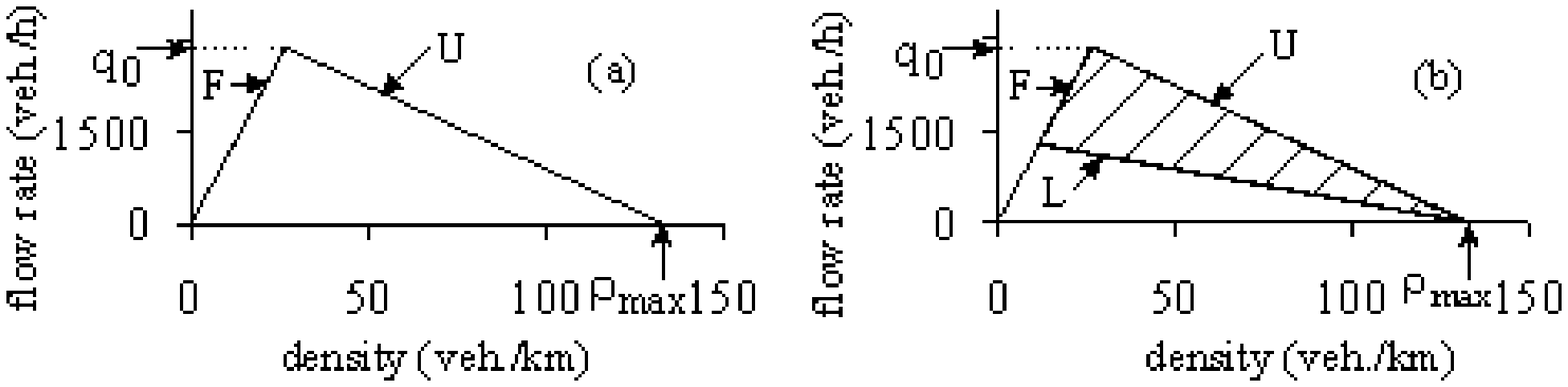}
\caption{Hypothetical spatially homogeneous and time-independent states (steady states): (a) -
for the initial NaSch CA-model~\protect\cite{NaSch}; 
(b) - the 2D-region for  the steady states for the KKW-1-model
(Table \ref{KKW-1} and Table \ref{parameters}) is the same as
in~\protect\cite{KKl}.  
 \label{NaSch_Diagram}  }
\end{figure}

The density $\rho$ and the flow rate $q$ are related to the gap $g$
and the speed $v$ by  
\begin{equation}
\rho = 1/(x_{\ell} - x)= 1/(g + d), \quad q = \rho v = v/(g+d).
\label{Wo:rho}
\end{equation} 
Because $v$ and $g$ are integer in CA-models, the steady states do not
form a continuum in the flow-density plane as they do in~\cite{KKl}.
However, the inequalities of (\ref{2D}) define a two-dimensional region in the
flow-density plane, in which steady states exist. 
As in~\cite{KKl} it is limited by three boundaries
(Figs.~\ref{NaSch_Diagram} (b) and~\ref{FlowDensity} (a, b)), 
the upper line $U$, the lower curve $L$, and the left line $F$. The
parameters of the lines $F$ and $U$ are chosen to 
be the same for all CA models under consideration. 
Note that without the lower boundary $L$, the lines $F$ and $U$ 
constitute the fundamental diagram of the NaSch CA-model
(Fig.~\ref{NaSch_Diagram} (a))~\cite{NaSch}. 

The left boundary $F$ is given by $q= \rho v_{\rm free}$. This is free
flow, where the flow rate $q$ is not restricted by safety-requirements.
On the upper boundary $U$ the flow rate is determined by the safe
speed $v_{\rm s}$. For example, inserting (\ref{safe}) and (\ref{Wo:rho}) it is given by
\begin{equation}
q = \rho v_{\rm s} = (1-\rho d)/\tau.
\label{upper}
\end{equation}
The lower boundary $L$  is determined by the synchronization distance
$D$: A steady state with density $\rho$ and a speed $v < v_{\rm
  free}$ requires that $D(v) \geq 1/\rho$ with equality on the lower
boundary $L$. For example, using (\ref{D}) with $d_1 = d$ one obtains
(see Fig.~\ref{NaSch_Diagram}(b))
\begin{equation}
q = (1-\rho d)/k\tau.
\label{lower}
\end{equation}
In the second model (\ref{D2}) the lower boundary $L$ is a non-linear
curve (see Fig.\ref{FlowDensity}(a)):
\begin{equation}
q= \frac{\tilde\rho}{\tau} \left(\sqrt{1 + \frac{2}{\tilde\rho}(1-\rho d)}
 \ - \ 1\right), \quad {\rm with} \quad 
\tilde\rho = \frac{\rho \tau^2 a}{\beta}.
\label{lowerD2}
\end{equation}
This curve has the upper line $U$ as a tangent at $\rho = \rho_{\rm max} =
1/d$. As will be shown below, this allows to reproduce qualitatively
the diagram of congested patterns of three-phase traffic theory with
simpler fluctuations than what is needed in the case of the
linear synchronization distance (\ref{D}) with $d_{1}=d$.
However, if the parameter $d_{1}$ is chosen smaller than $d$ in
(\ref{D}), the line $L$ intersects 
the line $U$ before the jam density $\rho_{\rm max}$ is reached
(Fig. \ref{FlowDensity}(b)). In this case, if the difference $d-d_{1}$
is chosen in a proper way, the
fluctuations in the model with linear synchronization distance (\ref{D}) 
may be as simple as for the non-linear $D$ (\ref{D2}), in order to
lead to qualitatively the same features. We also did simulations (not
shown in this paper), where we replaced $d$ in (\ref{D2}) by $d_1 <
d$: The qualitative results remain unchanged.

\subsection{Fluctuations of acceleration and deceleration}

In order to show the power of the basic model
(\ref{next}), (\ref{next1})~\cite{KKl},
the remaining model specifications (free and safe speeds,
fluctuations) will be the same as introduced in different
Nagel-Schreckenberg CA-models in the fundamental diagram 
approach~\cite{NaSch,Schreckenberg,Barlovic,Knospe2000,Knospe2001,Nagel,Brilon}
(with a slightly more general modeling of fluctuations).
Nevertheless, it will be shown that  all features of congested patterns 
which spontaneously occur upstream of the on-ramp
as well as of their evolution (when the flow rate to the on-ramp is changing)
are different in our CA-models. This proves that the
basic rules (\ref{next}), (\ref{next1})~\cite{KKl}
place our CA-models in the class belonging to the three-phase traffic theory.

In particular, as in the NaSch 
CA-models~\cite{NaSch,Schreckenberg,Barlovic,Knospe2000,Knospe2001,Nagel},
the accelerations and decelerations are stochastic. They are
implemented like in~\cite{Barlovic,Knospe2000}: 
{\it In a first step}, a preliminary vehicle speed of each vehicle
$\tilde v_{n+1}$ is
\begin{equation}
\tilde v_{n+1}=v_{n+1},
\label{dynamic}
\end{equation}
where $v_{n+1}$  is calculated based on the system of the dynamical
equations (\ref{next}) -- (\ref{next1}). 
 {\it In a second step}, a fluctuation $a\tau\eta_{n}$ (to be specified
 below) is added to the value $\tilde v_{n+1}$ calculated from the first step.
{\it Finally} the speed $v_{n+1}$ at the time $n+1$ is calculated by
\begin{equation}
v_{n+1}=\max(0, \min({\tilde v_{n+1}+a\tau\eta_{n}, v_{n}+a\tau, v_{\rm free}, v_{{\rm s},n}})).
\label{final}
\end{equation}
This means, that the stochastic contribution $a\tau\eta_n$ may neither lead to a speed
smaller than zero, nor to a speed larger than what the deterministic
acceleration $a$ would give, taking the limitations by  $v_{\rm
  free}$ and $v_{{\rm s},n}$ into account.

We implement the fluctuation $\eta_{n}$ in (\ref{final}) as
\begin{equation}
\eta_{n}=\left\{
\begin{array}{ll}
-1 &  \textrm{if $r< p_{\rm b}$}, \\
1 &  \textrm{if $p_{\rm b}\leq r<p_{\rm b}+p_{\rm a}$}, \\
0 &  \textrm{otherwise}
\end{array} \right.
\label{noise}
\end{equation}
where $r$ denotes a random number uniformly distributed between 0 and
1. This is a generalization of the random deceleration (with
probability $p_{\rm b}$) in NaSch cellular automata 
models~\cite{Knospe2000,Knospe2001}, because with probability $p_{\rm a}$
also a random acceleration can occur. $p_{a}+p_{b}\leq 1$ must be fulfilled.
In a different way and for a different purpose a random acceleration
was also introduced in CA-models by Brilon and Wu~\cite{Brilon}. 
As in a NaSch CA-model in the fundamental diagram 
approach~\cite{Barlovic}, the probability $p_{\rm b}$ in (\ref{noise}) is taken as 
a decreasing function of the vehicle speed $v_{n}$: 
\begin{equation}
p_{\rm b}(v_{n})=\left\{
\begin{array}{ll}
p_{0} & \textrm{if $v_{n}=0$} \\
p &  \textrm{if $v_{n}>0$}.
\end{array} \right.
\label{prob_b}
\end{equation}
where $p$ and $p_{0}>p$ are constants. This corresponds to
 the slow-to-start rules 
first introduced by Takayasu and Takayasu~\cite{TaTa}
and later used in the NaSch CA model by
Barlovic {\it et al.}~\cite{Barlovic}: 
Vehicles escape at the downstream front of a wide moving jam 
with the mean delay time $\tau_{\rm del}=\tau/(1-p_{0})$.
As in~\cite{Knospe2001} this 
provides the jam propagation through free and synchronized flows 
with the same velocity $v_{\rm g}$ of the downstream jam front that corresponds
 to a qualitative theory and to the related formula
$v_{\rm g}=-1/(\rho_{\rm max}\tau_{\rm del})$ from~\cite{Kerner1998}. $\rho_{\rm
 max}=1/d$ is the density inside the jam.

The probability $p_{\rm a}$ of the random acceleration in (\ref{noise})
is also taken as a decreasing function of the vehicle speed $v_{n}$: 
\begin{equation}
p_{\rm a}(v_{n})=\left\{
\begin{array}{ll}
p_{\rm a 1} & \textrm{if $v_{n} <   v_{\rm p}$} \\
p_{\rm a 2} & \textrm{if $v_{n}\geq v_{\rm p}$}.
\end{array} \right.
\label{prob_a}
\end{equation}
where $v_{\rm p}$,  $p_{\rm a1}$ and $p_{\rm a2}<p_{\rm a1}$ are constants.
This simulates  
the effect that  the vehicle moving at low speed in the dense flow tends
to close up to the leading one. Indeed, 
according to (\ref{next})-(\ref{D}), (\ref{final}), (\ref{noise}),
if the probability $p_{\rm a}$ of the acceleration is high,
the effect of adapting one's speed to the speed of the leading vehicle
is weak: With probability $p_{\rm a}$ the vehicle  
does not reduce its speed, and it may do so 
until it reaches the minimal safe gap.
Note that the tendency to minimize the space gap at low speed 
can lead in particular to 
the 'pinch' effect in synchronized traffic flow~\cite{Kerner2001,Kerner1998}, i.e., to 
the self-compression of the synchronized flow at lower vehicle speed with
the  spontaneous emergence of moving jams.

The tendency to minimize the space gap at low speed is automatically
built into our models, if the lower boundary $L$ approaches the upper
line $U$ as in Fig.\ref{FlowDensity}(a) and (b), because then the
synchronization distance is no longer larger than the security gap at
small speed. Indeed it 
turns out that the speed dependence (\ref{prob_a}) of $p_{\rm a}$
is not required in these cases for a realistic modelling. Therefore we choose
the probability $p_{\rm a}$ of acceleration in (\ref{noise})
to be a constant in the model variants of Fig. \ref{FlowDensity}.

\subsection{Cellular automata models with
cruise control within three-phase traffic theory \label{Cruise} }

Nagel and Paczuski~\cite{Nagel1995} proposed a variant of the NaSch CA-model
where fluctuations are turned off for the vehicle speed $v_{n}=v_{\rm free}$.
This variant has been called the NaSch CA-model
with {\it cruise control}~\cite{Helbing,Nagel1995}.

In the case of such cruise control, i.e.,
when fluctuations are turned off for the vehicle speed $v_{n}=v_{\rm free}$,
simpler CA models  can be used, if either the synchronization distance
is the non-linear one, (\ref{D2}) (Fig.~\ref{FlowDensity} (a)), or if in
the synchronization distance (\ref{D}) $d_{1}<d$ 
(Fig.~\ref{FlowDensity} (b)).

In these cases, the CA-models with cruise control
within three-phase traffic theory consist again of the formulas
(\ref{next})-(\ref{next1}) and (\ref{D2}) (or (\ref{D}) where $d_{1}<d$).
The formula (\ref{final}) for the incorporation of fluctuations into the
final value of the speed $v_{n+1}$ at time $(n+1)\tau$ is also valid,
but can be simplified, if only deceleration noise like in the
NaSch-type cellular automats~\cite{Barlovic,Knospe2000,Knospe2001} 
is implemented:
\begin{equation}
\eta_{n}=\left\{
\begin{array}{ll}
-1 &  \textrm{if $r< p_{\rm b}$}, \\
0 &  \textrm{otherwise}
\end{array} \right.
\label{noise_}
\end{equation}
with the probability~\cite{Nagel1995,Barlovic} 
\begin{equation}
p_{\rm b}(v_{n})=\left\{
\begin{array}{ll}
p_{0} & \textrm{if $v_{n}=0$} \\
p &  \textrm{if $0<v_{n}<v_{\rm free}$} \\
0 & \textrm{if $v_{n}=v_{\rm free}$}. 
\end{array} \right.
\label{prob_b_}
\end{equation}
Then $\tilde v_{n+1} + a\tau\eta_{n} \leq \tilde v_{n+1}$ so that
(\ref{final}) can also be written in the simpler form
\begin{equation}
v_{n+1}=\max(0, \tilde v_{n+1}+a\tau\eta_{n}).
\label{final_}
\end{equation}

\begin{table}
\caption{Definition of four CA-models KKW-1 -- KKW-4}
\label{KKW-1}
\begin{center}
\begin{tabular}{|l|}
\hline
\multicolumn{1}{|c|}{Dynamical part of all KKW-models}\\
\hline\\
$\tilde v_{n+1}=\max(0, \min(v_{\rm free},  v_{{\rm s},n}, v_{{\rm
    c},n})), \qquad g_{n}=x_{\ell, n}-x_{n}-d$,\\
$v_{{\rm s},n} = g_{n}/\tau, \qquad
v_{{\rm c},n}=\left\{\begin{array}{ll}
v_{n}+a\tau &  \textrm{for \, $g_{n} > D_{n}-d$},\\
v_{n}+a\tau \sign(v_{\ell,n}-v_{n}) &  
\textrm{for \, $g_{n} \leq D_{n}-d$}, \\
\end{array} \right.$ \\
$v_{\rm free}$, $d$, $\tau$ and $a$ are constants. \\\\
\hline
\multicolumn{1}{|c|}{Stochastical part of all KKW-models}\\
\hline\\
$v_{n+1}=\max(0, \min({\tilde v_{n+1}+a\tau\eta_{n}, v_{n}+a\tau,
  v_{\rm free}, v_{{\rm s},n}}))$,\quad
$x_{n+1}= x_{n}+v_{n+1}\tau$, \\
\multicolumn{1}{|c|}{
$\eta_{n}=\left\{
\begin{array}{ll}
-1 &  \textrm{if $r< p_{\rm b}$}, \\
1 &  \textrm{if $p_{\rm b}\leq r<p_{\rm b}+p_{\rm a}$}, \\
0 &  \textrm{otherwise}. \\
\end{array} \right.$ }\\\\
\hline
\multicolumn{1}{|c|}{Specifications of synchronization distance
  $D_{n}$ and noise $\eta_{n}$}\\
\hline
\multicolumn{1}{|c|}{for CA-model KKW-1 (cf. Fig.\ref{NaSch_Diagram}(b)):}\\\\
\multicolumn{1}{|c|}{
$D_{n}=d + kv_{n}\tau, \quad  p_{\rm b}(v_{n})=\left\{\begin{array}{ll}
                                      p_{0} & \textrm{if $v_{n}=0$} \\
                                      p &  \textrm{if $v_{n}>0$}
                                      \end{array} \right., 
                   \quad  p_{\rm a}(v_{n})=\left\{\begin{array}{ll}
                          p_{\rm a 1} & \textrm{if $v_{n} <   v_{\rm p}$} \\
                          p_{\rm a 2} & \textrm{if $v_{n}\geq v_{\rm p}$}
                          \end{array} \right.$,
}\\
constant parameters: $k$, $p_{0}$, $p$, 
$p_{\rm a 1}$, $p_{\rm a 2}$, $v_{\rm p}$. \\
\hline
\multicolumn{1}{|c|}{for CA-model KKW-2 (cf. Fig.\ref{FlowDensity}(a)):}\\\\
\multicolumn{1}{|c|}{
$D_{n}=d + v_{n}\tau + \beta v_{n}^2/2a, \qquad
p_{\rm b}(v_{n})=\left\{
\begin{array}{ll}
p_{0} & \textrm{if $v_{n}=0$} \\
p &  \textrm{if $v_{n}>0$}\\
\end{array} \right.$,
}\\
constant parameters: $\beta$, $p_{0}$, $p$, $p_{\rm a}$. \\
\hline
\multicolumn{1}{|c|}{for CA-model KKW-3 (cf. Fig.\ref{FlowDensity}(a)):}\\\\
\multicolumn{1}{|c|}{
$D_{n}=d + v_{n}\tau + \beta v_{n}^2/2a, \qquad
p_{\rm b}(v_{n})=\left\{
\begin{array}{ll}
p_{0} & \textrm{if $v_{n}=0$} \\
p &  \textrm{if $0<v_{n}<v_{\rm free}$}\\
0 &  \textrm{if $v_{n}=v_{\rm free}$}
\end{array} \right.$,
}\\
constant parameters: $p_{\rm a}=0$, $\beta$, $p_{0}$, $p$. \\ 
\hline
\multicolumn{1}{|c|}{for CA-model KKW-4 (cf. Fig.\ref{FlowDensity}(b)):}\\\\
\multicolumn{1}{|c|}{
$D_{n}=d_1 + kv_{n}\tau, \qquad
p_{\rm b}(v_{n})=\left\{
\begin{array}{ll}
p_{0} & \textrm{if $v_{n}=0$} \\
p &  \textrm{if $v_{n}>0$}
\end{array} \right.$,
}\\
constant parameters: $d_1<d$, $k$, $p_{0}$, $p$, 
$p_{\rm a}$. \\
\hline
\end{tabular}
\end{center}
\end{table}
\vspace{1cm}

\subsection{Summary of the new models and their parameters \label{Summary} }

In the following sections we shall discuss the congestion patterns obtained
from simulations of four CA-models belonging to the class of
three-phase traffic theory as introduced above. For easier reference 
we specify them in Table \ref{KKW-1} with the abbreviations KKW-1 to KKW-4.  
In addition we provide two tables containing a list of symbols (Table
\ref{symbols}) and typical values for the parameters (Table \ref{parameters}). 

A few remarks on the time and space discretization are in order.
The usual choice of the time step $\tau$ in CA-models of traffic is
$\tau=1s$~\cite{Ch,NaSch}. 
Like in~\cite{Knospe2000} we use a small-scale discretisation of space: 
The length of cells is chosen equal to $\delta x=0.5 \ m$. 
This leads to a speed discretisation in units of $\delta v = 1.8 \ km/h$. 
Hence two vehicles are only considered as moving with different speeds, 
if this difference is equal to (or larger) than $\delta v$. In the 
model~\cite{KKl} with continuous changes in the vehicle speed, two
vehicle speeds are considered as different,  if their difference exceeds a
much smaller  value $\delta v=10^{-6} \ m/s$. This is one of  the reasons why
fluctuations in cellular automata are in general stronger than in the
corresponding continuum model. 

For all CA-models within the three-phase-traffic theory investigated
here we chose the probability $p_{0}=0.425$.  This corresponds 
to a velocity $v_{\rm g}=-15.5 \ km/h$ of the downstream front of a
wide moving jam  
and an outflow $q_{\rm out}=1810 \ {\rm vehicles/h}$ from a wide moving jam.  
Note that the flow rate $q_{\rm out}$ refers to 
the case, where the vehicles reach the maximum vehicle speed $v=v_{\rm free}$,
after they have escaped from the jam. 

For the sake of comparison with other traffic models,
we choose occasionally in Sec.~\ref{Dis} model parameters deviating
from the ones given in Table~\ref{parameters}. In this case the
parameter values are given in the related figure captions.

\begin{table}
\caption{List of symbols}
\label{symbols}
\begin{tabular}{|l|l|}
\hline
$\tau$ & time discretization interval \\
$\delta x$ & space discretization length\\
$\delta v=\delta x/\tau$ & speed discretisation unit \\
$n=0, 1, 2, ...$ & number of time steps \\
$\tilde v_{n}$ & vehicle speed at time step $n$ without fluctuating part \\
$v_{n}$ & vehicle speed at time step $n$ \\
$v_{\ell,n}$ & speed of the leading vehicle at time step $n$ \\
$v_{{\rm s}, n}$ & safe speed at time step $n$ \\
$v_{\rm free}$ & maximal speed (free flow) \\
$x_{n}$ & vehicle position at the time step $n$ \\
$x_{\ell,n}$ & position of the leading vehicle at time step $n$\\
$d$  & vehicle length \\
$g_{n}=x_{\ell,n}-x_{n}-d$ & gap (front to end distance) at time step $n$ \\
$D_{n}$ & synchronization distance at time step $n$ \\
$a$ & vehicle acceleration \\
$b$ & vehicle deceleration \\
$\eta_{n}$ & speed fluctuation at time step $n$, \\
$p_{\rm a}$ & probability of vehicle acceleration \\
$p_{\rm b}$ & probability of vehicle deceleration \\
$r$ & random number uniformly distributed between $0$ and $1$ \\
$q_{0}$ & maximum flow rate in steady states \\
$q_{\rm max}$ & maximum flow rate in free flow ($v=v_{\rm free}$) \\
$\rho_{\rm max}=1/d$ & density of traffic jam\\
$v_{\rm g}$ & velocity of downstream front of wide moving jam \\
$q_{\rm out}$ & flow rate in the outflow from a wide moving jam\\
$\rho_{\rm min}$ & density in free flow related to the flow rate
                   $q_{\rm out}$ \\
$t_{0}$ & time at which vehicle influx from on-ramp is switched on \\
$q^{\rm (pinch)}_{\rm lim}$ & limit flow rate in pinch region of general pattern \\
$q^{\rm (bottle)}_{\rm out}$ & outflow from a congestion pattern at an on-ramp \\
\hline\hline
\end{tabular}
\end{table}
\vspace{1cm}

\begin{table}
\caption{Model parameters and characteristic values}
\label{parameters}
\begin{tabular}{|c|c|}
\hline
\multicolumn{2}{|c|}{} \\
\multicolumn{2}{|c|}{Common parameters and values for all KKW-models}\\
\hline
discretization units &
$\tau     = 1 \ s$,
$\delta x = 0.5 \ m$, \\
& $\delta v = \delta x/\tau = 1.8 \ km/h$,
$a = \delta v/\tau = 0.5 \ m/s^2$\\
\hline
model parameters &
$v_{\rm free} = 108 \ km/h = 60 \ \delta v$, 
$d= 7.5 \ m = 15 \ \delta x$, \\ & 
$p_{0}= 0.425$ \\
\hline  
model results  &
$v_{\rm g}= - 15.5 \ km/h$, $q_{\rm out}=1810 \ {\rm vehicles}/h$, \\ 
& 
$\rho_{\rm min}=16.76 \ {\rm vehicles}/km$,  
$q_{0}= 2880 \ {\rm vehicles}/h$ \\
\hline
\multicolumn{2}{|c|}{} \\
\multicolumn{2}{|c|}{CA model KKW-1}\\
\hline
general model parameters &
$k=2.55$, $v_{\rm p}=50.4 \ km/h= 28 \ \delta v$,  $p_{\rm a 1}=0.2$ \\
\hline
parameter-set I:   &  \\ 
model parameters & $p=0.04$, $p_{\rm a 2}=0.052$ \\ 
model results & $q_{\rm max}\approx 2400 \ {\rm vehicles}/h$, $q^{\rm
  (pinch)}_{\rm lim}\approx  1150 \ {\rm vehicles}/h$ \\
\hline
parameter-set II:  &   \\
model parameters & $p=0.055$, $p_{\rm a 2}=0.085$ \\
model results & $q_{\rm max}\approx  2630 \ {\rm vehicles}/h$, 
$q^{\rm (pinch)}_{\rm lim}\approx  1000 \ {\rm vehicles}/h$ \\
\hline
\multicolumn{2}{|c|}{} \\
\multicolumn{2}{|c|}{CA model KKW-2 } \\
\hline
model parameters & $p=0.04$, $p_{\rm a}=0.052$, $\beta= 0.05$ \\
model results & $q_{\rm max}\approx  2400 \ {\rm vehicles}/h$, 
$q^{\rm (pinch)}_{\rm lim}\approx  1150 \ {\rm vehicles}/h$ \\
\hline
\multicolumn{2}{|c|}{} \\
\multicolumn{2}{|c|}{CA model KKW-3 } \\
\hline
model parameters & $p=0.04$,  $\beta= 0.05$ \\
model results & $q_{\rm max}\approx  1460 \ {\rm vehicles}/h$, 
$q^{\rm (pinch)}_{\rm lim}\approx  1150 \ {\rm vehicles}/h$ \\
\hline
\multicolumn{2}{|c|}{} \\
\multicolumn{2}{|c|}{CA model KKW-4 } \\
\hline
model parameters & $d_{1}=2.5 \ m = 5 \ \delta x$, $k=2.55$, $p=0.04$,
$p_{\rm a}=0.052$ \\ 
\hline
\end{tabular}
\end{table}
\vspace{1cm}

\section{Congested patterns on a homogeneous one-lane road
\label{Homo} }

All CA-models introduced here (Table \ref{KKW-1}) show qualitatively
similar results on the homogeneous one-lane road. Therefore, only
numerical results of a simulation of model KKW-1 
(Fig.~\ref{NaSch_Diagram} (b)) will be presented in this section.

For simulations of congested patterns on a homogeneous one-lane road, 
cyclic boundary conditions have been used.
The one-lane homogeneous road has the length $60000$ cells ($30 \ km$). 
We checked that all qualitative results remain the same, if
open boundary conditions are used and the length of the road is large enough.

\subsection{Complex dynamics of synchronized flow \label{Syn_Sec} }

The features of spatio-temporal pattern formation on a homogeneous
road (i.e. a road without bottlenecks or on-ramps) are largely the
same for the KKW-models considered here and the continuum model of
Kerner and Klenov~\cite{KKl}.  However, due to stronger fluctuations
in the CA-models, the dynamics of perturbations is somewhat different.

(i) If the initial flow rate $q_{\rm in}$ does not exceed a value
$q_{\rm max}<q_{0}$, 
\begin{equation}
0<q_{\rm in}<q_{\rm max}, 
\label{Free}
\end{equation}
the fluctuations do not perturb the speed of a vehicle, which
  initially has maximal speed $v_{\rm free}$, significantly
(Fig.~\ref{Cycle} (a-c)). In this case, the fluctuations lead to 
changes in the distances between vehicles, i.e. to a change in the vehicle density 
and hence to a change in the flow rate (black points $F$ in Fig.~\ref{Cycle} (c)).

(ii) However, if
\begin{equation}
q_{\rm max}\leq q_{\rm in} \leq q_{0}
\label{FS_range} 
\end{equation}
the model fluctuations lead to an occurrence of inhomogeneous 
and non-stationary synchronized flow states where the vehicle speed is
lower than  $v_{\rm free}$ (Fig.~\ref{Cycle} (d-f)). According to Kerner's
hypothesis about continuous spatio-temporal transitions between different
states of synchronized flow in
three-phase traffic theory~\cite{Kerner1998A,Kerner1999B,Kerner1999C}
this behaviour corresponds to a complex motion within the 2D-region in
the flow-density plane, where steady states exist (open circles $S$ in 
Fig.~\ref{Cycle} (f)).  Similar continuous spatio-temporal transitions
between different states of synchronized flow in agreement with
Kerner's hypothesis have recently also been found in a different CA
model by Fukui {\it et al.}~\cite{Fukui}. 

These inhomogeneous synchronized flow states are the result of
many independent local transitions at different road locations. 
As they cause a reduction of the initial maximal vehicle speed,
they look similar to the F$\rightarrow$S-transitions
on the two-lane road which have been studied in~\cite{KKl}. However,
in~\cite{KKl} the F$\rightarrow$S-transition was in general caused by
an external local perturbation, which led to the
formation of a local region of synchronized flow
(see Fig. 1 (c) in~\cite{KKl}). 
By contrast, in the CA models under consideration the
local transitions were induced by the intrinsic model fluctuations
in a wide range of densities given by (\ref{FS_range}). No
external local perturbation was applied.
As a consequence, a complex inhomogeneous spatio-temporal pattern
of synchronized flow appears everywhere on
the road instead of the local region
triggered by a perturbation in the continuum
model (compare Fig.~\ref{Cycle} (d)
in this article with Fig. 1 (c) in~\cite{KKl}).

(iii) A homogeneous initial state with vehicle speed lower than
$v_{\rm free}$,  which in the absence of fluctuations would
belong to the steady states within the 2D-region of the flow-density
plane,  remains a synchronized flow state for a long time. However, as
in (ii) the evolution of these synchronized flow states shows
a complex spatio-temporal behaviour of all traffic flow variables
(Fig.~\ref{CycleSyn}) due to the intrinsic model fluctuations.

\subsection{Emergence of wide moving jams  \label{Emergence} }

The emergence of wide moving jams (Fig.~\ref{CycleJam})
shows qualitatively the same  features as in~\cite{KKl}.

(i) In particular, as in~\cite{KKl}, moving jams do {\it not} occur spontaneously,  
if the initial state with maximal vehicle speed $v=v_{\rm free}$ lies
within the range of flow rates (\ref{Free}), where the maximal
speed can be maintained.  

For a subset of these states, those with a density above a threshold 
for the wide moving jam formation, $\rho_{\rm min}$
(Figs.~\ref{Cycle} (c, f) 
and~\ref{CycleJam} (c)), wide moving jams can be induced, but only by a
very strong local perturbation.
For the case when traffic flow with the maximal vehicle speed $v=v_{\rm free}$
is formed downstream of a wide moving jam, the density  $\rho_{\rm
  min}$ is related to the flow rate in the outflow of the wide moving
jam, $q_{\rm out}$.  

The velocity of the downstream front of the wide moving jam $v_{\rm g}$ 
is a characteristic, i.e., unique, predictable and reproducible parameter 
which is a constant for given model
parameters. This velocity together with
the threshold point, $(\rho_{\rm min}, q_{\rm out})$, determines the characteristic 
line $J$ for the downstream front of a wide moving jam (the line $J$ in
Figs.~\ref{Cycle} (c, f), \ref{CycleSyn}(c) and \ref{CycleJam}(c)).  

The strength of the perturbation needed to trigger a wide moving
jam is highest for densities close to $\rho_{\rm min}$. Then it is not
enough that a vehicle is forced to stop (the maximal amplitude of a perturbation), 
but this stop must be maintained for some time (about 2-3 minutes at
$\rho_{\rm min}$)  for a wide moving jam to nucleate
in an initial traffic flow with maximal speed $v=v_{\rm free}$.
Note that for initial densities which are only slightly higher than $\rho_{\rm min}$
a wide moving jam often spontaneously disappears
due to the high fluctuations of the outflow from the jam in the KKW-1 model.

(ii) As in~\cite{KKl}, the line $J$ determines the threshold of the wide
moving jam excitation in synchronized flow: All densities 
in steady synchronized flows related to the line $J$ are
threshold densities
with respect to the jam formation (the S$\rightarrow$J-transition).
At a given speed, 
the higher the density, the lower is the critical amplitude 
$\delta v_{c}$ of a local perturbation
for the S$\rightarrow$J-transition: The critical amplitude $\delta v_{c}$ 
for the S$\rightarrow$J-transition reaches its maximum value at 
the threshold density. At a given difference between an initial
density and the threshold density, the lower the initial speed, the
lower the critical amplitude  $\delta v_{c}$  is. 

However, for synchronized flow states which lie in the vicinity of the
safe speed (in the vicinity of the line $U$ in
Figs.~\ref{NaSch_Diagram} (b) and~\ref{FlowDensity}) the strong
intrinsic model fluctuations in the KKW-models (Table \ref{KKW-1})
lead to the S$\rightarrow$J-transition without the need of any
external local perturbation.

(iii) The latter result allows a simulation of the spontaneous
emergence of wide moving jams (Fig.~\ref{CycleJam} (a-c)). 
In the initial state all vehicles move with the maximal speed $v=v_{\rm free}$.
For a flow rate in the interval (\ref{FS_range}),
the initial  vehicle speed $v=v_{\rm free}$
can not be maintained for  a long time: Due to local transitions
the vehicle speed decreases and states of synchronized
flow are formed at some locations on the road (Fig.~\ref{CycleJam} (b, c) at
$x=14 \ km$) as already described in (ii) in Sect.~\ref{Syn_Sec}.
In these synchronized flow states, model fluctuations grow leading
to the {\it spontaneous} emergence of wide moving jam (Figs.~\ref{CycleJam} (a) and
(b, c) at $x=8 \ km$). 
For the model parameters used in Fig.~\ref{CycleJam} the spontaneous
emergence of a wide moving jam can also occur for an initial state of
synchronized flow with speed $v_{\rm in}<v_{\rm free}$.

Note that the fluctuation parameters $p$ and $p_{\rm a2}$ of both,
random deceleration and random acceleration, are larger than in
Fig.~\ref{Cycle} and \ref{CycleSyn}, where no spontaneous emergence of wide moving
jams from
synchronized flow states had been observed within the simulation time.
Fast drivers are more ``nervous'' in Fig.~\ref{CycleJam} than in the previous
figures. However, this is not the only reason for the
emergence of wide moving
jams in the present example. This can be seen by
comparing the average values of the random contribution $\eta$ to the 
speed for speeds larger than $v_{\rm p}=50.4 \ km/h$: While in
the previous Figures $\langle \eta \rangle = p_{\rm 
  a2} - p = 0.012$, it is here more than twice as large: $\langle \eta
\rangle = 0.03$. A positive value of $\langle \eta \rangle$ means that
the drivers are biased towards stochastic acceleration rather than
deceleration. A stronger bias implies a higher delay time in the
vehicle deceleration. We will come back to the question, how this
makes wide moving jams more likely, below, in Sec. \ref{4.2}(ii).

(iv) Thus, as in the model of Kerner and Klenov~\cite{KKl}, in an
initial traffic flow with the maximal speed $v=v_{\rm free}$ model
fluctuations can cause the spontaneous occurrence of synchronized
flow, but not the spontaneous emergence of wide moving jams.  The
latter was only found in the KKW-models (Table \ref{KKW-1}), once
synchronized flow was established.  That synchronized flow states
should occur first and only later the spontaneous wide moving jam, is a
common feature of three-phase-traffic theory and agrees with empirical
observations~\cite{Kerner1999A,Kerner1998}.  Thus, in the KKW-models
within the three-phase traffic theory the diagram of congested
patterns on the homogeneous road is qualitatively different from those
in other approaches~\cite{Helbing,Ch}.

\section{Congested patterns at on-ramps}

\subsection{Model of on-ramp \label{On-ramp} }

In this section, for all simulations of congested patterns at an on-ramp a
one-lane road of $100 \ km$ length ($200000$ cells) with open boundary
conditions is used. The reference point $x=0$ is placed at the distance $20 \
km$ from the end of the road, so that it begins at the coordinate $x=-80 \
km$. The on-ramp starts at the point $x=16 \ km$ ($32000$ cells) and
its merging area was  $0.3 \ km$ long ($600$ cells).

For simulation of the on-ramp two consecutive vehicles on the main
road within the on-ramp area are chosen randomly, their coordinates
being denoted by $x^{+}>x^{-}$. The entering vehicle is placed in the
middle point between them at the coordinate
$x_{n}=[(x^{+}+x^{-}+1)/2]$ (here $[ \ ]$ denotes the integer part),
taking the speed of the leading vehicle $v^{+}$~\cite{Tomer}.  In
addition it was required that the distance between the two vehicles on
the main road should  exceed some value 
\begin{equation}
dx^{\rm (min)}_{\rm on}=\lambda v^{+} +2d, 
\label{lambda}
\end{equation}
where $\lambda$ was chosen to be equal to $0.55$, 
if not stated explicitely otherwise.

Our numerical investigations have shown that the main qualitative
features of congested patterns at the on-ramp and of the related
diagrams do not change, if instead of these simple
rules for a vehicle squeezing in from the on-ramp to the road 
more sophisticated lane changing rules are used.
It is important, however, that on the one hand
the model of the on-ramp allows a gradual change of the flow rate to the
on-ramp $q_{\rm on}$ from nearly zero to a relatively large (but for
traffic flow relevant) value, and on the other hand does not introduce large
additional 
speed fluctuations, when a vehicle from the on-ramp
merges with the traffic flow on the main road.

\subsection{Diagrams of congested patterns at on-ramps}
\label{4.2}

A diagram of congested patterns at the on-ramp represents regions of
spontaneous occurrence of congested patterns upstream of the on-ramp
at different values of the initial flow rate to the on-ramp $q_{\rm
  on}$ and the initial flow rate on the one-lane road $q_{\rm in}$
upstream of the on-ramp.  As will be discussed in more detail below,
these regions in principle depend on how long one waits, i.e. in how
far one samples rare events. In practice a typical waiting time much
longer than one hour has little meaning, as real traffic is not a
stationary stochastic process on large time scales.

We found that the CA-models under consideration (Table \ref{KKW-1})
essentially show the same diagram of congested patterns which
has been predicted for the three-phase traffic theory~\cite{Kerner2002B}
and previously obtained in the continuous microscopic model 
\cite{KKl}. However, for some parameter values there are interesting
peculiarities in the KKW-models. 
 
(i) 
In Fig.~\ref{Diagram1} (a) and (b) the diagram of congested patterns
at an on-ramp is shown for two different sets of the parameters
of the KKW-1 model (Table \ref{KKW-1})  
(parameter-set I and parameter-set II in Table \ref{parameters},
respectively). Although the KKW-1 model 
is considerably simpler than the model of Kerner and Klenov studied
in~\cite{KKl}, the main features of the diagram (Fig.~\ref{Diagram1}
(a)) and the related congested patterns which spontaneously occur
upstream of the on-ramp (Figs.~\ref{Mesh_Patterns1},
~\ref{Patterns1_GP} and~\ref{Patterns1_WSP}) are qualitatively the
same as in~\cite{KKl}.

There are two
main boundaries $F^{(B)}_{S}$ and $S^{(B)}_{J}$ on the diagram
(Fig.~\ref{Diagram1}).
The limit point of the boundary $F^{(B)}_{S}$ at $q_{\rm on}=0$
is related to
the maximum flow rate in free flow where $q_{\rm in}=q_{\rm max}$.
The explanation of the limit point $q_{\rm in}=q_{\rm max}$ is very simple:
On the homogeneous one-lane road (i.e. without the on-ramp) 
synchronized flow occurs spontaneously, if the flow rate exceeds
$q_{\rm max}$ (range (\ref{FS_range})).
Therefore, synchronized patterns should spontaneously occur
upstream of the on-ramp for $q_{\rm in}=q_{\rm max}$ already for
vanishing  flow rate  to the on-ramp $q_{\rm on}\rightarrow 0$. 

Below and left of the boundary $F^{(B)}_{S}$ free flow occurs.
Between the boundaries $F^{(B)}_{S}$ and $S^{(B)}_{J}$ different
synchronized flow patterns (SP) occur upstream of the on-ramp,
without wide moving jams being observed.

Right of the boundary $S^{(B)}_{J}$ wide moving jams spontaneously
emerge in synchronized flow which has been formed upstream of the on-ramp.
The only difference compared to the results
in~\cite{KKl} is that in Fig.~\ref{Diagram1} (a)
there is no region where the moving synchronized flow pattern (MSP) occurs
(about a possible occurrence of MSP see below).
The diagram  in Fig.~\ref{Diagram1} (a) is in accordance
with general features of the diagrams of congested
patterns at the on-ramp on a one-lane road which was postulated from
qualitative considerations in~\cite{Kerner2002B}.

A few technical remarks about the determination of the boundary 
$F^{(B)}_{S}$ are in order: 
After the on-ramp has been switched on at $t=t_{0}$, there is a
transient, before the congested patterns are established
upstream of the on-ramp, and one needs some criterion to detect the
transition from free to synchronized flow. The criterion we used is
that the speed drops below 
some threshold value $V_{\rm FS}$
and stays low for at least $4 \ min$.  The value of $V_{\rm FS}$ is
chosen equal to $80 \ km/h$. 

The delay time, until this criterion of the transition from free flow to synchronized flow
is fulfilled,
is marked as $T_{\rm FS}$ in
Fig.~\ref{FS_Fig}. This delay time can be rather short ($\approx 1-2 \
min$) for large values of $q_{\rm on}$ (Fig.~\ref{FS_Fig} (a)) and
increases for decreasing $q_{\rm on}$ (Fig.~\ref{FS_Fig} (b)). 
For small values of $q_{\rm on}$ the speed returns quickly to high
values, whenever it happened to drop below  
$V_{\rm FS}$.
Then one would not speak of a transition into synchronized flow any more
(Fig.~\ref{FS_Fig} (c)). This behaviour of $T_{\rm FS}$ has been used 
to determine the boundary  $F^{(B)}_{S}$. Scanning the $(q_{\rm
  on},q_{\rm in})$-plane on a grid the leftmost points were determined, where
$T_{\rm FS}$ is still smaller or equal $30 \ min$ (Fig.~\ref{FS_Fig} (d)).
Similarly, a point on the  boundary $S^{(B)}_{J}$ is found as 
the leftmost point ($q_{\rm on}, q_{\rm in}$),  where a wide moving jam emerges
in synchronized flow within $60 \ min$. Thus, the quantitative positions
of the boundaries $F^{(B)}_{S}$ and $S^{(B)}_{J}$
depend on the chosen time intervals.
However, the qualitative forms of these boundaries are independent of the
choice of the time intervals, provided they are high enough.

(ii) The diagram of congested patterns at the on-ramp in
Fig.~\ref{Diagram1} (b) is obtained for the same model KKW-1
(Table~\ref{KKW-1}) as in (a), however with the fluctuation parameters
parameter-set II in Table~\ref{parameters}. As pointed out already in
connection with Fig.~\ref{CycleJam}, this means that
the delay time of vehicle deceleration was increased
compared to Fig.~\ref{Diagram1} (a).  This has two effects: 
First, the region between the boundaries $F^{(B)}_{S}$ and $S^{(B)}_{J}$,
where SP occur without spontaneous wide moving jam formation, is reduced. 
Second, the boundaries $F^{(B)}_{S}$ and $S^{(B)}_{J}$ merge 
in the limit point $q_{\rm in}=q_{\rm max}$, where $q_{\rm on}\rightarrow 0$.
This means that in the CA-model with higher delay time of vehicle deceleration 
wide moving jams can spontaneously occur  in synchronized flow
already at extremely low flow rates to the on-ramp in the vicinity of the point 
$q_{\rm in}=q_{\rm max}$.

This can be explained by realizing that
in the present model, an increase in the delay time of vehicle
deceleration makes the effect of speed adjustment within the
synchronization distance weaker. As a result,  it becomes more likely
that a vehicle closes up to the leading one, i.e. 
reaches the minimal safe gap $g_{n}$, so that model fluctuations
more easily get amplified to cause a wide moving jam.

(iii) In model KKW-2 (non-linear synchronization distance $D(v)$) 
the diagram of congested patterns at the on-ramp (Fig.~\ref{Diagram2} (a)) 
and the related congested patterns upstream of the on-ramp
(Figs.~\ref{Mesh_Patterns2}, ~\ref{Patterns2_GP}, ~\ref{Patterns2_WSP})
possess the same qualitative features as 
discussed in Fig.~\ref{Diagram1}.

\begin{figure}
\includegraphics[width=\textwidth]{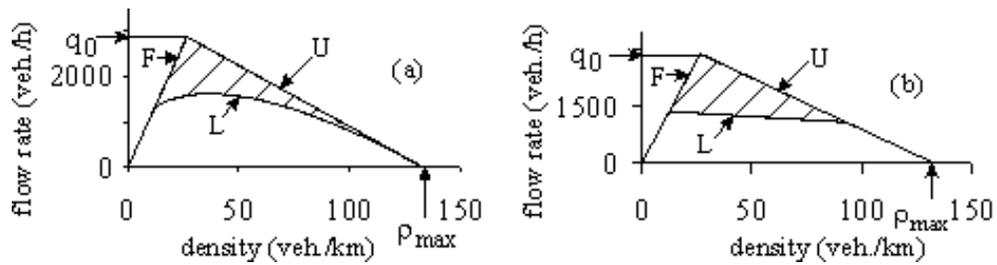}
\caption{2D-regions supporting steady states in the flow-density
plane for two of the KKW-models (Table \ref{KKW-1} and Table
\ref{parameters}): (a) - for KKW-2-model (non-linear dependence
of the synchronization distance (\ref{D2}) on the vehicle speed) 
(b) - for KKW-4-model (linear synchronization distance (\ref{D}) with
$d_{1}<d$).
 \label{FlowDensity}  }
\end{figure}

\begin{figure}
\includegraphics[width=\textwidth]{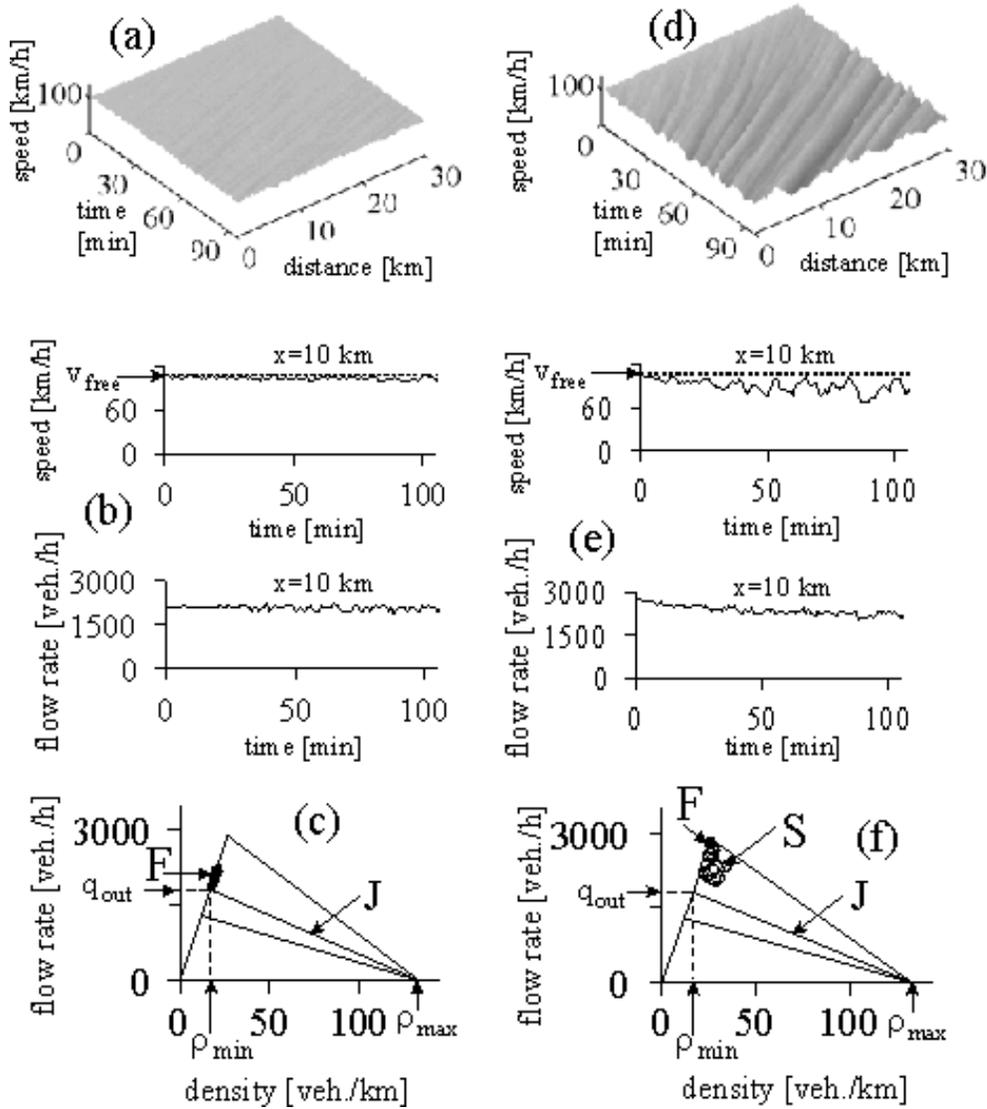}
\caption{Traffic patterns on a homogeneous one lane road with
  periodic boundary conditions for KKW-1-model (parameter-set I of
  Table \ref{parameters}).  (a, b, c) Perturbation of free flow at
  moderate flow rate ($q = 2160 \ {\rm vehicles/h}  < q_{\rm max}
  \approx 2400 \ {\rm vehicles/h}$) and (d, e, f) at high flow rate
  ($q=2842 \ {\rm vehicles/h} > q_{\rm max}$).
(a, d) - vehicle speed as function of time and distance
  (distance increases in downstream direction);
(b, e) - vehicle speed and flow rate
at the location $x=10 \ km$ as functions of time
(one minute averaged data of virtual detectors);
(c, f) -  data in the flow-density plane which correspond
to (b, e), respectively. 
\label{Cycle} }
\end{figure}

\begin{figure}
\includegraphics[width=\textwidth]{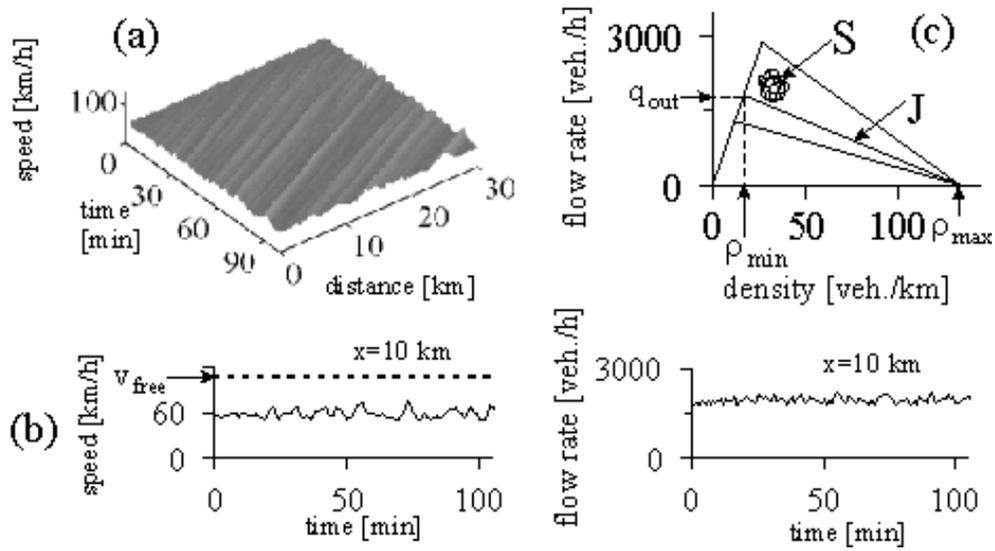}
\caption{Synchronized flow behaviour on  a homogeneous one lane road
  with periodic boundary conditions for KKW-1-model (parameter-set I of
  Table \ref{parameters}).
(a) - vehicle speed as a function of time and distance;
(b) - vehicle speed (left)  and  flow rate (right)
at a fixed location ($x=10 \ km$) as functions of time
(one minute averages);
(c) -  data in the flow-density plane which correspond to (b).
Initial free flow $q_{\rm in}=1800 \ {\rm vehicles/h}$, 
initial speed $v_{\rm in}=54 \ km/h$ ($v_{\rm in}=15 \ m/s$).
\label{CycleSyn} }
\end{figure}

\begin{figure}
\includegraphics[width=\textwidth]{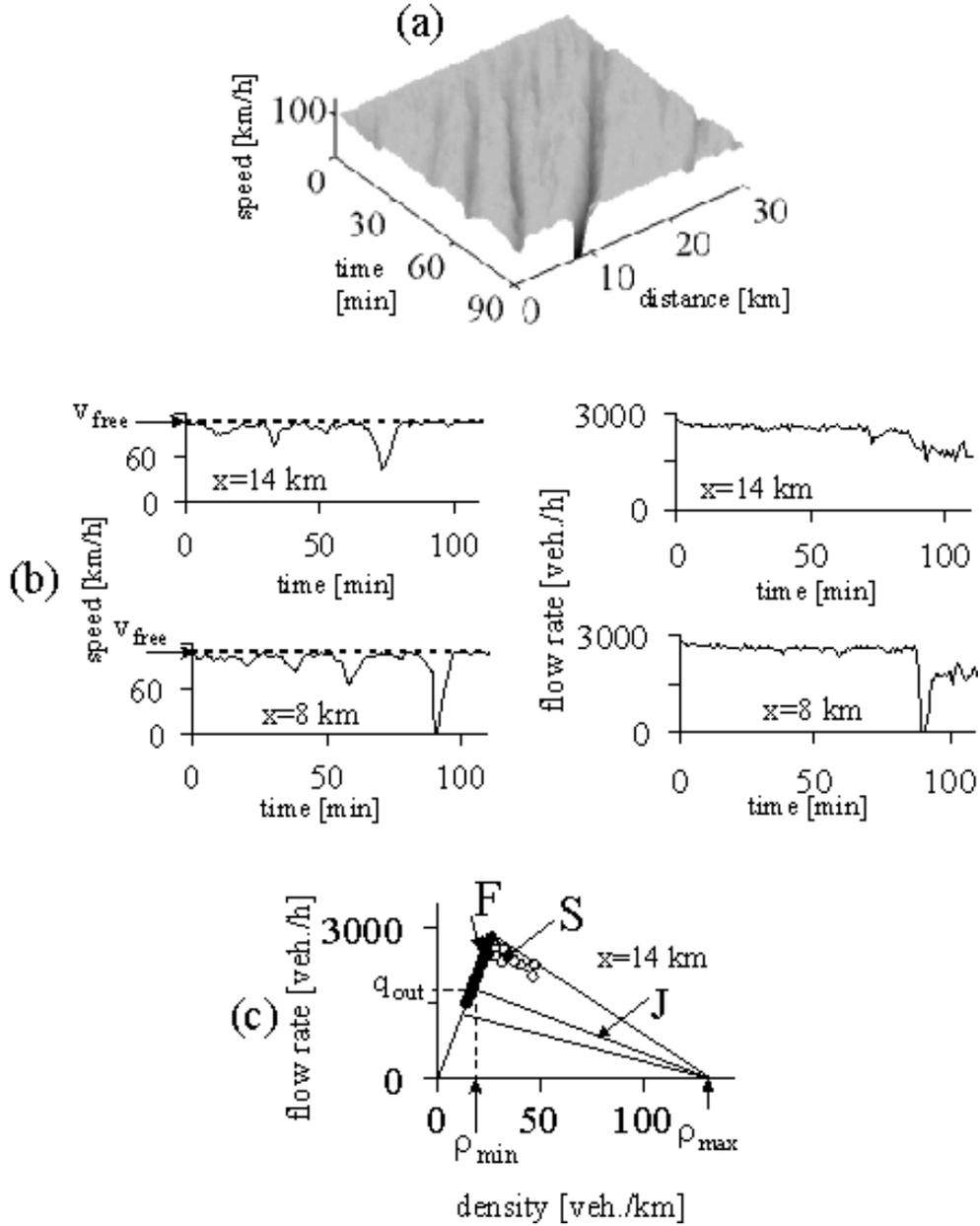}
\caption{Wide moving jam formation on a homogeneous one-lane road with periodic
  boundary conditions for KKW-1-model with parameter-set II of
  Table \ref{parameters}, where the
  probabilities of random deceleration, $p$, and random acceleration
  at high speed, $p_{a2}$, are larger than in Fig.~\ref{Cycle}. 
  (a) - vehicle speed as function of time and distance; 
  (b) - vehicle speed and  flow rate at the locations $x=14 \ km$ and 
  $x=8 \ km$ as functions of time (one minute averages); 
  (c) - data in
  the flow-density plane which correspond to (b) at the location $x=14
  \ km$.  Initial free flow rate $q_{\rm in}=2842 \ {\rm vehicles/h}$
  is larger than $q_{\rm max}=2634 \ {\rm vehicles/h}$ (cf.
  (\ref{Free}) and (\ref{FS_range})).  In (c) black points are related
  to the states of free flow with speed $v$ close to $v_{\rm free}$
  (the points $F$), and circles are related to states of synchronized
  flow (the points $S$).  
\label{CycleJam} }
\end{figure}

\begin{figure}
\includegraphics[width=0.7\textwidth]{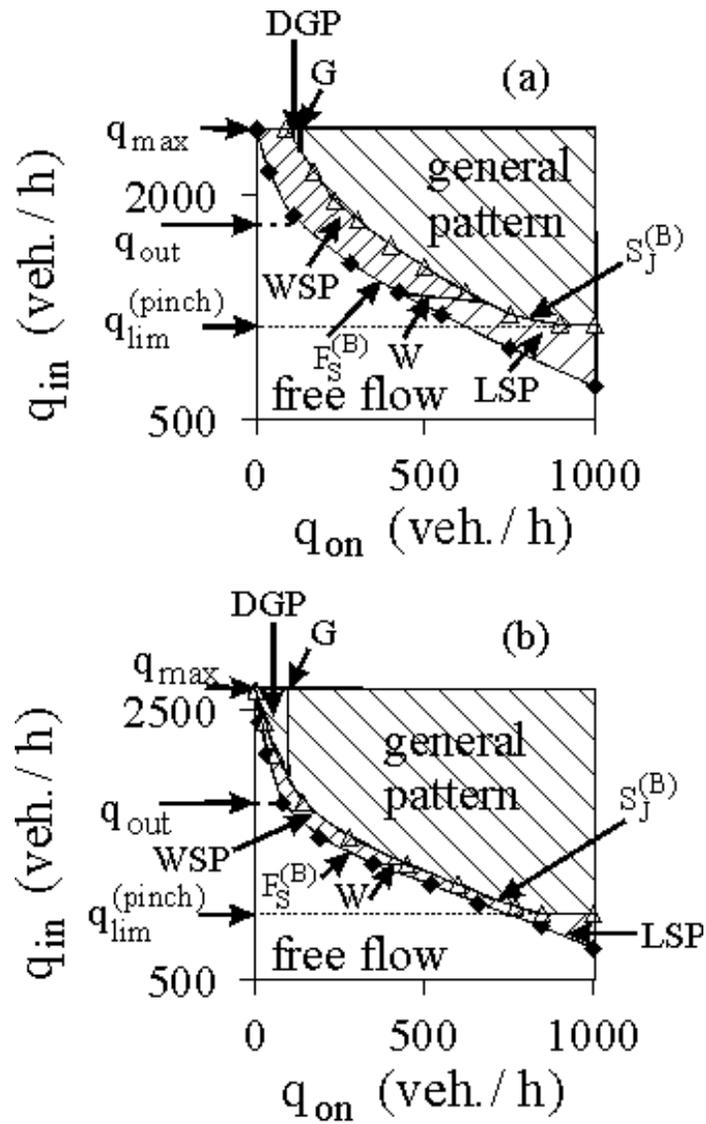}
\caption{Diagrams of congested patterns at the on-ramp for the KKW-1-model.
  (a) - Parameter-set I of Table \ref{parameters} as in Fig.~\ref{Cycle}.
  (b) - Parameter-set II of Table \ref{parameters} as in Fig.~\ref{CycleJam}.
  GP - general pattern, DGP - dissolving general pattern, WSP -
  widening synchronized flow pattern, LSP - localized synchronized
  flow pattern.
 \label{Diagram1}  }
\end{figure}

There is a peculiarity of the diagram of congested patterns
(Fig.~\ref{Diagram2} (b)) for model KKW-3 (cruise control). 
In this case, there are no model fluctuations at the maximal speed
$v=v_{\rm free}$. Thus, if the flow rate to the on-ramp $q_{\rm on}=0$
and the initial state is related to the  maximal speed $v=v_{\rm
  free}$, synchronized flow with lower vehicle speed can not appear 
{\it spontaneously} up to the top flow rate $q_{\rm in}=q_{0}$ (see
Fig.~\ref{Diagram2} (b) and Fig.~\ref{FlowDensity} (a)).

However, already at an extremely low flow rate of $q_{\rm
  on}\approx 1-2 \ {\rm vehicles}/h$
these synchronized flow states with lower speed spontaneously appear
upstream of the on-ramp,
because of a small disturbance of the initial flow at the on-ramp.
Moreover, the flow rate $q_{\rm in}=q_{\rm max}$ at which this effect occurs
can be noticeably lower than $q_{\rm out}$ in this case (Fig.~\ref{Diagram2} (b)).
Therefore, at all flow rates $q_{\rm in}$ within the range
\begin{equation}
q_{\rm max}\leq q_{\rm in}\leq q_{0}
\end{equation}
 SP occur spontaneously at the on-ramp 
for  $q_{\rm on}$ as small as $1-2 \ {\rm vehicles}/h$
(Fig.~\ref{Diagram2} (b)).

We found that the KKW-4 model, which differs from KKW-2 only by 
having the synchronization distance which leads to 
Fig.~\ref{FlowDensity} (b) rather than Fig.~\ref{FlowDensity} (a),
can show qualitative the same diagram of congested patterns
as obtained for model KKW-2 (Fig.~\ref{Diagram2} (a)). 
Likewise, if one replaces the non-linear synchronization distance in
the cruise control model KKW-3 by formula (\ref{D}) with $d_{1}<d$ (as
in KKW-4), the  diagram of congested patterns can remain
qualitative the same as Fig.~\ref{Diagram2} (b).

To explain these results, first recall that in all these KKW-models 
the probability $p_{\rm a}$ (\ref{prob_a}) was independent of the
vehicle speed, in contrast to the previously considered KKW-1 model.
The fact that this simplification of acceleration noise nevertheless
allows us  to simulate the qualitatively correct pattern formation in
the three-phase 
traffic theory can be traced back to the difference in the 2D-regions of
the steady states in the flow density plane for these CA-models:

For {\it low vehicle speeds} the  boundary $L$ for these models 
is either tangential to the boundary $U$ (Fig.~\ref{FlowDensity}(a)),
or even coincides with it (Fig.~\ref{FlowDensity}(b)). On the other hand, 
the numerical study of the KKW-models shows that,
if the vehicle speed in synchronized flow  is very close to the safe
speed, i.e. for synchronized states close to the boundary $U$, fluctuations easily 
lead to the emergence of a wide moving jam.
As already mentioned, the purpose of introducing acceleration noise in our models 
is a simulation of the pinch effect, where moving jams  emerge spontaneously. 
Both models in Fig.~\ref{FlowDensity} are sufficiently sensitive to
fluctuations at low speeds that the pinch effect can be achieved with
a relatively small probability $p_{\rm a}$ independent of the vehicle speed.

In contrast, in the KKW-1 model
the  boundary $L$ of the 2D-region 
of steady states is not close enough to the boundary $U$
even at low vehicle speeds  (Fig.~\ref{NaSch_Diagram} (b)). 
In particular, there are synchronized
 states with low vehicle speed, which lie below the line $J$. Therefore,   
the probability $p_{\rm a}$ had to be chosen higher at low speeds 
in order to enhance the likelihood that a driver comes close to the
 boundary $U$. 
Then the pinch effect is also obtained in this model,
 in accordance with empirical observations in~\cite{Kerner1998,Kerner2002B}.

\subsection{Synchronized flow patterns (SP)}

As in~\cite{KKl}, depending on the
parameters either the widening
synchronized flow pattern (WSP), or the moving synchronized flow pattern (MSP),
 or the localized synchronized flow pattern (LSP) can occur in our CA-models.

(i) As in~\cite{KKl}, WSP  occurs
at high initial flow rate on the road upstream of the on-ramp, $q_{\rm in}$,  
and low flow rate to the on-ramp, $q_{\rm on}$, 
between the boundaries $F^{(B)}_{S}$ and $S^{(B)}_{J}$.
The downstream front of WSP is localized at the on-ramp. The upstream
front is continuously moving upstream, so that the
width of WSP is gradually increasing in time (Figs.~\ref{Mesh_Patterns1} (d, e),
~\ref{Patterns1_WSP} (a, b),
~\ref{Mesh_Patterns2} (d), ~\ref{Patterns2_WSP},
~\ref{Mesh_Patterns3} (b) and~\ref{Patterns3_WSP}). 

(ii) Depending on the flow rates $q_{\rm in}$ and $q_{\rm on}$
the distribution of the vehicle speeds inside WSP can be related to 
states with nearly homogeneous speed
(Figs.~\ref{Mesh_Patterns1} (d)
and~\ref{Patterns1_WSP} (a, c)), or to non-homogeneous 
speed distributions, where sometimes non-stationary vehicle speed waves 
(propagating with different, negative and positive velocities)
can occur (Figs.~\ref{Mesh_Patterns1} (e)
and~\ref{Patterns1_WSP} (b, d)).

(iii) Note a peculiarity of the KKW-models:
The upstream front in WSP which separates synchronized flow 
downstream and free flow upstream moves with a relatively high velocity
($v^{(WSP)}_{\rm g}\approx -40$ km/h).
This non-realistic velocity is 
a consequence of simplicity of the models presented. 
In spite of this, the models give a realistic qualitative 
description of congested patterns and their evolution. 
More correct values for the front velocity have been found 
in the  microscopic model of Kerner and Klenov~\cite{KKl}. 
The other way to obtain realistic velocity of the front between free and
synchronized flows 
may be the use of a strongly non-uniform free flow upstream of the on-ramp.
In this case, there is a large spread of gaps between vehicles in free flow. 
As a result, the speed of the front is diminished due 
to the presence of vehicles with too small gaps between them.

(iv) Recall that  at very low  flow rate to the on-ramp, $q_{\rm on}$,
and high initial flow rate on the road upstream of the on-ramp, $q_{\rm in}$,
in the diagram derived in~\cite{KKl} there is a region ``MSP'', where the moving
synchronized flow pattern (MSP) spontaneously occurs.
In this case, after SP has emerged upstream of the on-ramp,  this SP
comes off the on-ramp and begins to move upstream.
In some cases, a new SP emerges at the on-ramp; this SP 
comes off the on-ramp later, and so on.

In contrast to the diagram of congested patterns in~\cite{KKl},
WSP can appear also at very low flow rate to the on-ramp $q_{\rm on}$
(Figs.~\ref{Diagram1}, \ref{Diagram2}). In other words, there is no
region ``MSP'' in our diagrams, which look like the one
postulated by Kerner~\cite{Kerner2002B}
based on a qualitative consideration of the three-phase traffic theory 
approach for a one-lane road. 
However, in \cite{Kerner2002B}
it was also mentioned that fluctuations may cause
MSP in the region, where WSP exists normally.
Apparently for this reason,
sometimes MSP appears at very low flow rate to the on-ramp $q_{\rm on}$
(close to the boundary $F^{(B)}_{S}$)
in the region of WSP   in the CA-models.

This effect is shown for the KKW-3 cruise-control-model  (Table~\ref{KKW-1})
in Figs.~\ref{Mesh_Patterns3} (d) and~\ref{MSP}, where MSP usually
occurs in the region marked ``WSP $\&$ MSP''  in the related diagram of  
congested patterns (Fig.~\ref{Diagram2} (b))
at lower flow rate to the on-ramp, $q_{\rm on}$. 
Within the region ``WSP $\&$ MSP'' (at a slightly higher flow rate
$q_{\rm on}$ than the one at which MSP occurs) 
a pattern which looks like a mixture of WSP and MSP can occur (Fig.~\ref{Mesh_Patterns3} (c)):
Near the on-ramp this pattern resembles MSP. However, upstream of the on-ramp
the pattern more and more transforms into WSP. 
Note another peculiarity of the KKW-3 model:
When WSP occurs in this model, the vehicle speeds and the flow rates in this WSP
are  usually related to points in the flow-density plane which are in the vicinity or lie on
the boundary $L$ in the flow-density plane (Fig.~\ref{FlowDensity} (a)),
which corresponds to the synchronization distance $D$
(see circles in Fig.~\ref{Patterns3_WSP} (b)).
Presumably, this behaviour is not a common feature of WSP, but due to
the simplicity of the model.

(v)
As in the diagram in~\cite{KKl}, at higher flow rate $q_{\rm on}$
and  lower flow rate $q_{\rm in}$ 
between the boundaries $F^{(B)}_{S}$ and $S^{(B)}_{J}$
the localized synchronized flow
pattern (LSP) occurs.
The downstream front of LSP is localized at the on-ramp. However,
the upstream front of LSP is not continuously moving upstream, so that  the
width of LSP remains spatially limited (Figs.~\ref{Diagram1}
and~\ref{Mesh_Patterns1} (f)). Note that the upstream front of LSP and therefore
the width of LSP can oscillate in time. 
Moreover, at flow rates close to the boundary $F^{(B)}_{S}$, fluctuations 
can cause random appearance and disappearance of LSP.  
The boundary which  separates the region of WSP 
(see Figs.~\ref{Diagram1} and~\ref{Diagram2}) from the region of LSP
is marked by the letter $W$ in the diagram of congested patterns.

\subsection{General patterns (GP)}

\begin{figure}
\includegraphics[width=\textwidth]{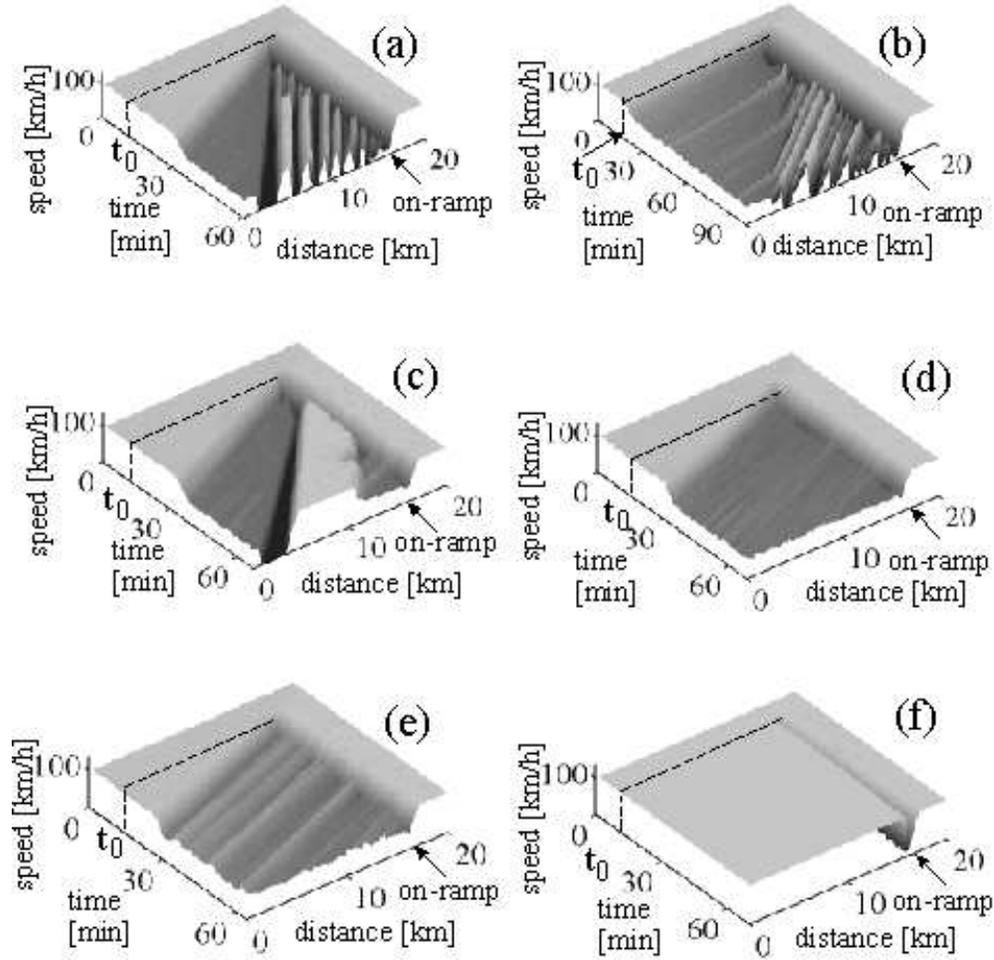}
\caption[]{Congested patterns at the on-ramp belonging to 
  Fig.~\ref{Diagram1} (a). (a) - General pattern (GP) at $q_{\rm
    in}>q_{\rm out}$, (b) - GP at $q_{\rm in}<q_{\rm out}$, (c) -
  dissolving general pattern (DGP), (d, e) - widening synchronized
  flow patterns (WSP), and (f) - localized synchronized flow pattern
  (LSP).  At $t_{0}=8 \ min$ flow from the on-ramp is switched on.
    Single vehicle data 
  are averaged over a space interval of $40 \ m$ and a time interval
  of $1 \ min$.  The flow rates $(q_{\rm on}, q_{\rm in})$ are: (a)
  (500, 2300), (b) (740, 1740), (c) (105, 2400), (d) (90, 2300), (e)
  (90, 2160), and (f) (760, 1080) ${\rm vehicles/h}$.
\label{Mesh_Patterns1}  }
\end{figure}

\begin{figure}
\includegraphics[width=0.95\textwidth]{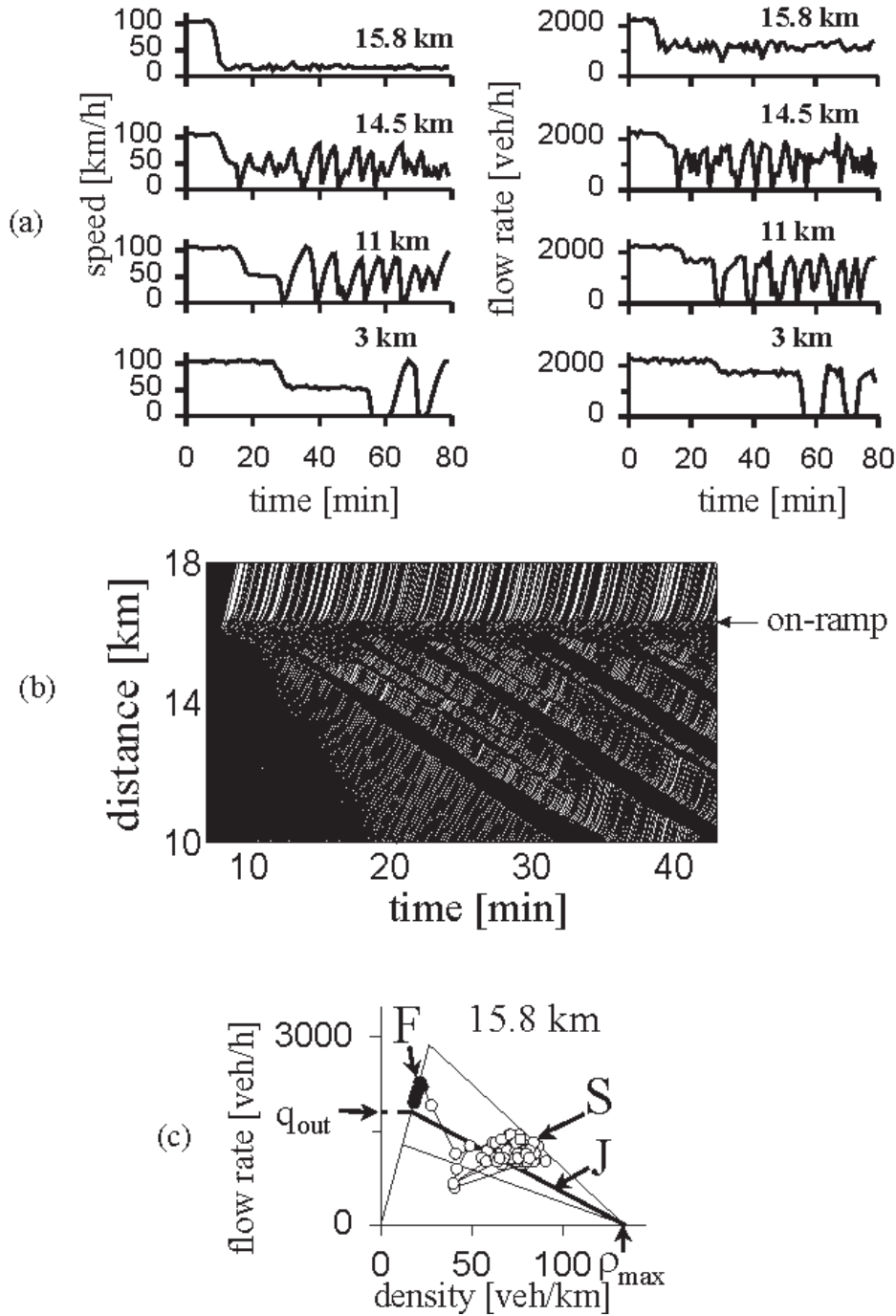}
\caption[]{The general pattern (GP) (KKW-1 model parameters as in
  Fig.~\ref{Mesh_Patterns1} (a)): 
  (a) - vehicle speed (left) and flow rate (right), 
  (b) - vehicle trajectories, 
  (c) - the corresponding data in the flow-density plane
  for the location $x=15.8 \ km$.  
  (a, c) - One minute averaged data
  of virtual detectors whose coordinates are indicated in the related
  figures. In (c) black points are related to the states of free flow
  with the speed $v$ close to the maximal one $v_{\rm free}$ (the
  points $F$) and circles are related to states of synchronized flow
  (the points $S$).  To show the spatio-temporal features of the GP
  clearly, only trajectories of every 6th vehicle are shown in (b).
  The dashed line in (b) shows the upstream front of the pattern which
  separates synchronized flow downstream from free flow upstream.
\label{Patterns1_GP} }
\end{figure}

\begin{figure}
\includegraphics[width=\textwidth]{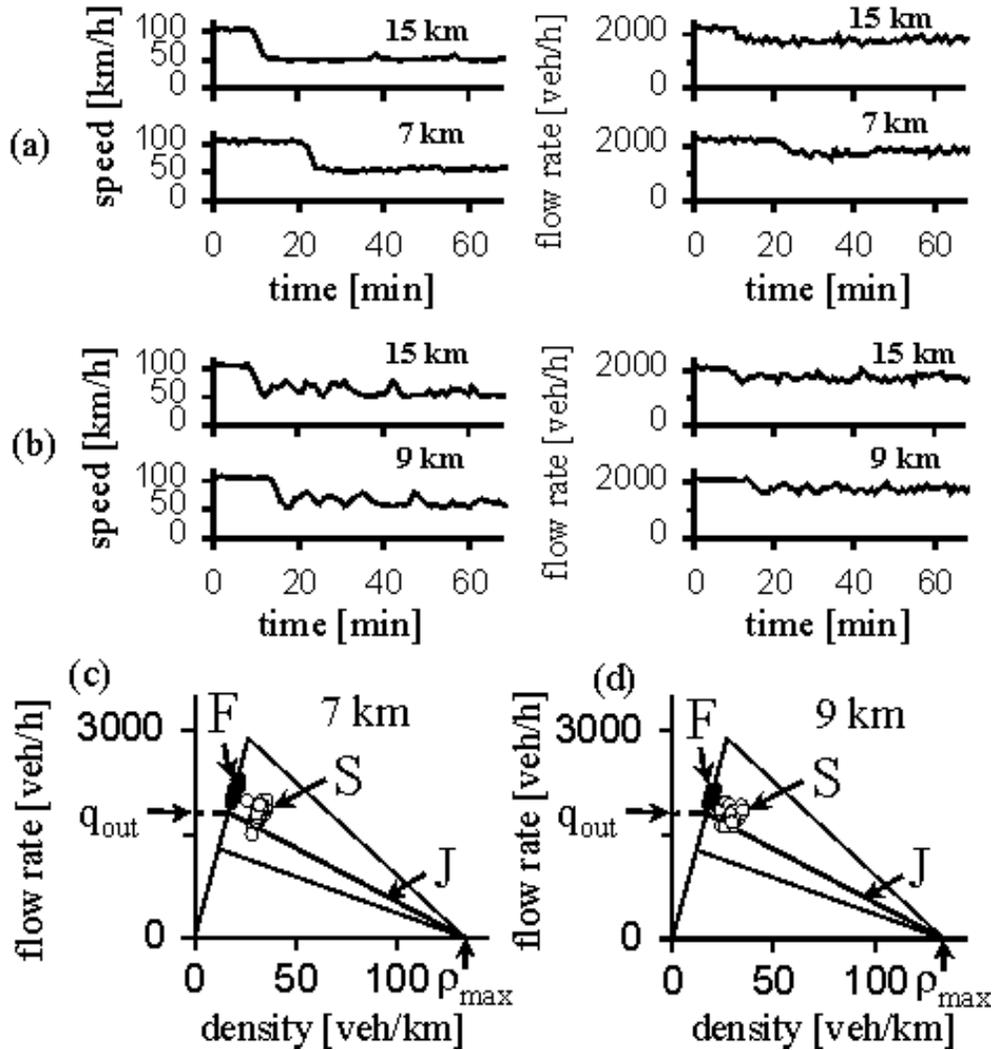}
\caption{The widening synchronized flow pattern (WSP):
  (a) - The vehicle speed (left) and the flow rate (right), (c) - the
  corresponding data on the flow-density plane for WSP shown in
  Fig.~\ref{Mesh_Patterns1} (d).  The similar plots (b), (d) are for
  WSP shown in Fig.~\ref{Mesh_Patterns1} (e).  One minute averaged
  data of virtual detectors whose coordinates are indicated in (a-d).
  In (c), (d) black points are related to the states of free flow with the
  speed $v$ close to the maximal one $v_{\rm free}$ (the points {\it
    F}) and circles are related to states of synchronized flow (the
  points {\it S}).  The KKW-1 model parameters for (a, c) are the same
  as in Fig.~\ref{Mesh_Patterns1} (d) and for (b, d) as in
  Fig.~\ref{Mesh_Patterns1} (e).
\label{Patterns1_WSP} }
\end{figure}

\begin{figure}
\includegraphics[width=\textwidth]{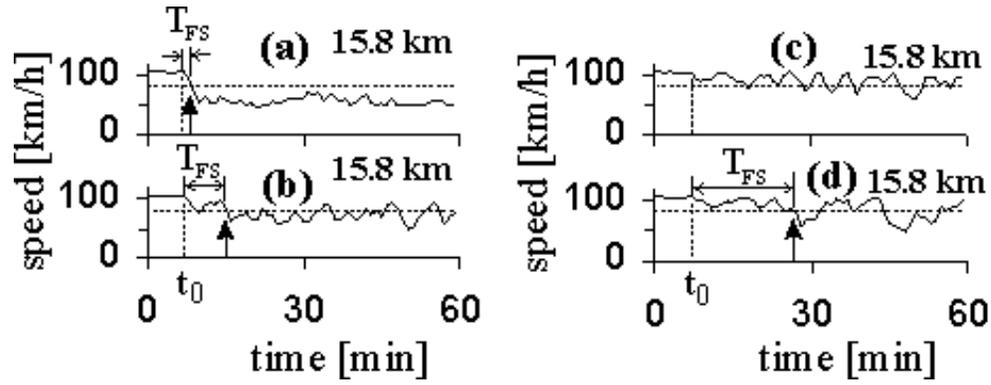}
\caption{Speed breakdown due to the F$\rightarrow$S-transition
  at the location
  $x=15.8 \ km$ ($0.2 \ km$ upstream of the begin of the on-ramp).
  Fixed flow rate $q_{\rm in} = 2000 \ {\rm vehicles}/h$. Flow
  rate $q_{\rm on}$ is: (a) $120$, (b) $90$, (c) $55$, and (d) $70$
  ${\rm vehicles/h}$.  Up-arrows mark the time $t_0 + T_{\rm FS}$, at
  which according to our criterion synchronized flow is detected: The
  vehicle speed drops below the level $80 \ km/h$ (dashed horizontal
  line) and remains low for more than $4 \ min$.  Simulations of the
  KKW-1-model (parameter-set I of Table~\ref{parameters}).
\label{FS_Fig} }
\end{figure}

\begin{figure}
\includegraphics[width=0.8\textwidth]{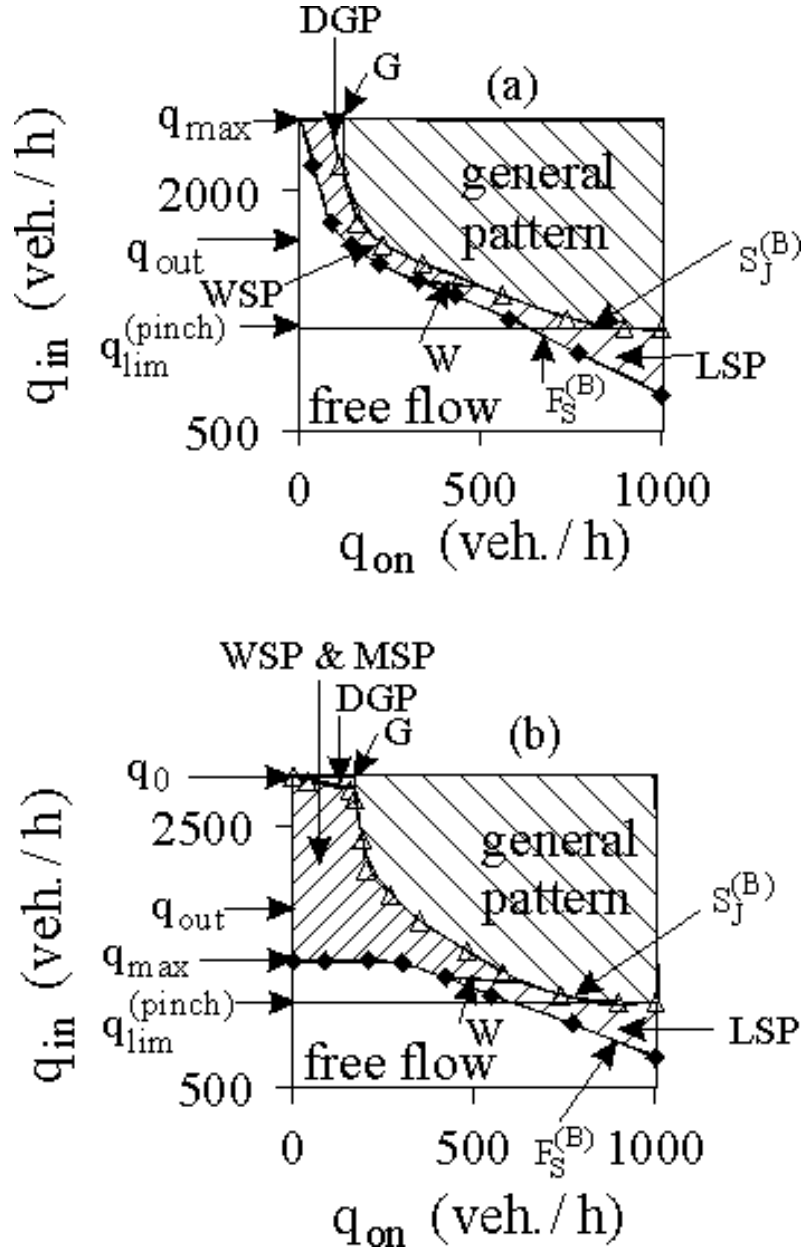}
\caption{Diagrams of congested patterns at the on-ramp for the
  KKW-models with non-linear dependence of the synchonization distance
  (\ref{D2}) on the vehicle speed  (Fig.~\ref{FlowDensity} (a)):
(a) - KKW-2-model, (b) - KKW-3-model  
(Table \ref{KKW-1} and Table \ref{parameters}).
GP - general pattern, 
DGP - dissolving general pattern, 
WSP - widening synchronized flow pattern, 
MSP - moving synchronized flow pattern, 
LSP - localized synchronized flow pattern.
Note that
for the KKW-2-model (a) and the KKW-3-model (b) the value $V_{\rm FS}$ discussed in Sec. \ref{4.2}(i) 
is also chosen equal to $80 \  km/h$. However for the KKW-3-model there is 
the exeption for the case, when $q_{\rm
  in}$ is close to the point of intersection of the curves $L$ and $F$
(Fig.~\ref{FlowDensity} (a)).  Near this point
the minimal speed in WSP can be greater than $80 \ km/h$, and the
threshold speed $V_{\rm FS}$ is chosen as the average between
$v_{\rm free}$ and this minimal speed.

 \label{Diagram2}  }
\end{figure}

\begin{figure}
\includegraphics[width=\textwidth]{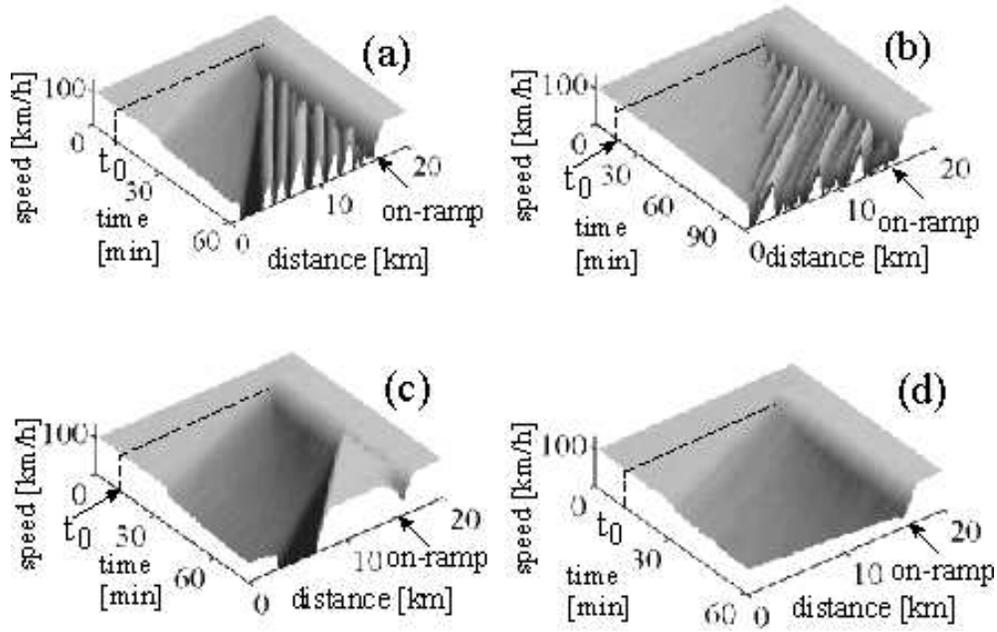}
\caption[]{Congested patterns at the on-ramp for the KKW-2-model of
  Fig.~\ref{Diagram2} (a):
(a) - General pattern (GP) at $q_{\rm in}>q_{\rm out}$, 
(b) - GP at $q_{\rm in}<q_{\rm out}$, 
(c) - dissolving general pattern (DGP),
(d) - widening synchronized flow pattern (WSP).
Single  vehicle data are averaged over a space interval of
$40 \ m$ and a time interval of $1 \ min$.
$t_{0}=8 \ min$. 
The flow rates $(q_{\rm on}, q_{\rm in})$ are:
(a) (500, 2250), 
(b) (800, 1650), 
(c) (110, 2400), and 
(d) (70, 2300) ${\rm vehicles/h}$.
\label{Mesh_Patterns2}  }
\end{figure}

\begin{figure}
\includegraphics[width=\textwidth]{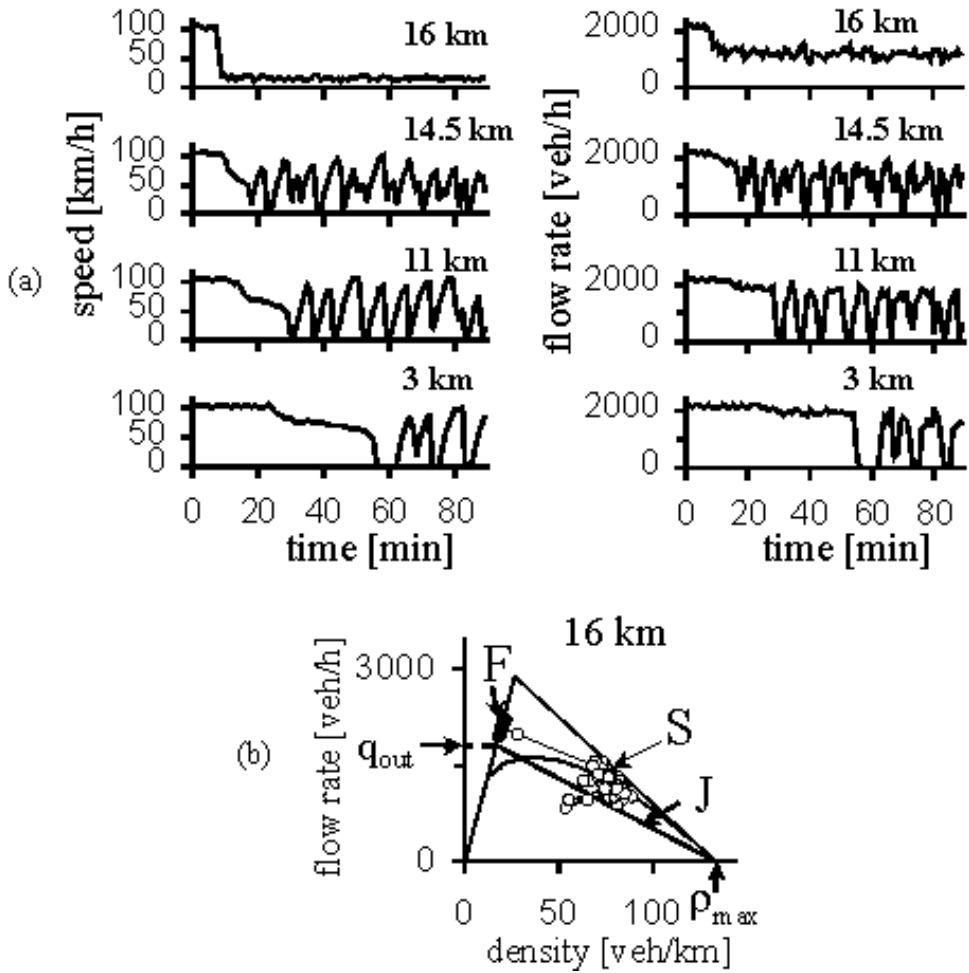}
\caption{The general pattern (GP) related to Fig.~\ref{Mesh_Patterns2}
  (a) (KKW-2-model): (a) - vehicle speed (left) and flow rate (right),
  (b) - corresponding data in the flow-density plane at the location
  $x=16 \ km$.  One minute averaged data of virtual detectors, whose
  coordinates are indicated in (a, b).  In (b) black points are
  related to the states of free flow with the speed $v$ close to the
  maximal one $v_{\rm free}$ (the points $F$) and circles are related
  to states of synchronized flow (the points $S$). 
\label{Patterns2_GP} }
\end{figure}

\begin{figure}
\includegraphics[width=\textwidth]{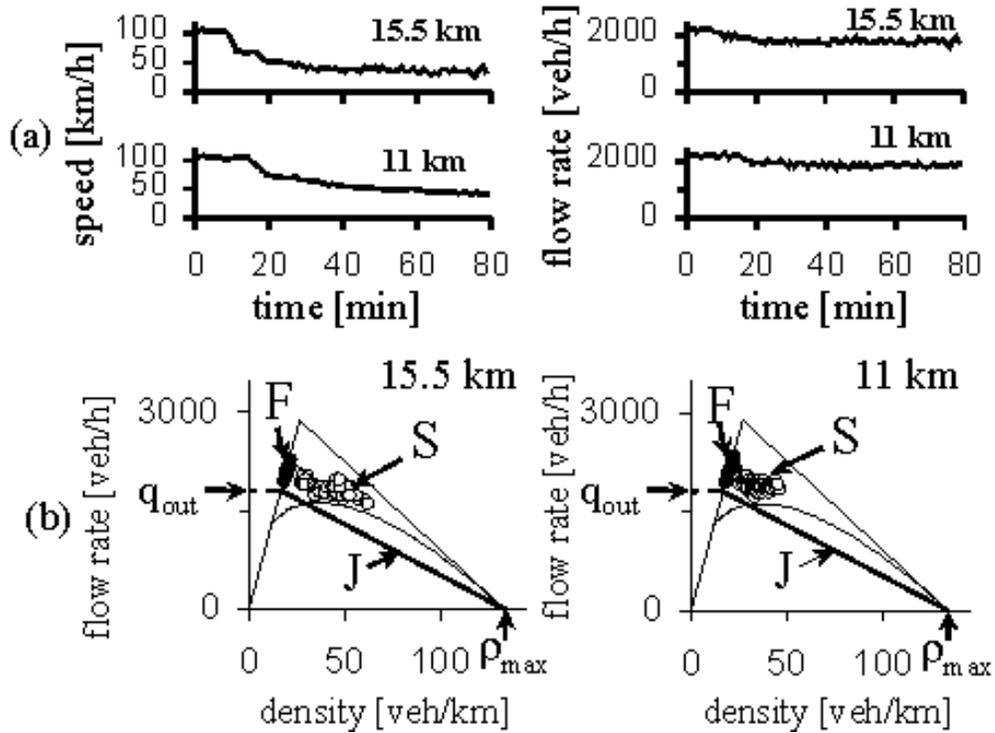}
\caption{The widening synchronized flow pattern (WSP) related to
  Fig.~\ref{Mesh_Patterns2} (d) (KKW-2-model): 
(a) - vehicle speed (left) and flow rate (right),
(b) - the corresponding data on the flow-density plane. 
One minute averaged data of virtual detectors, whose coordinates
are indicated in (a, b). In (b) black points are related to
the  states  of free flow   with the
 speed $v$ close to the maximal one $v_{\rm free}$ (the points $F$)
and circles are related to states of  synchronized flow 
(the points {\it S}). 
\label{Patterns2_WSP} }
\end{figure}

\begin{figure}
\includegraphics[width=\textwidth]{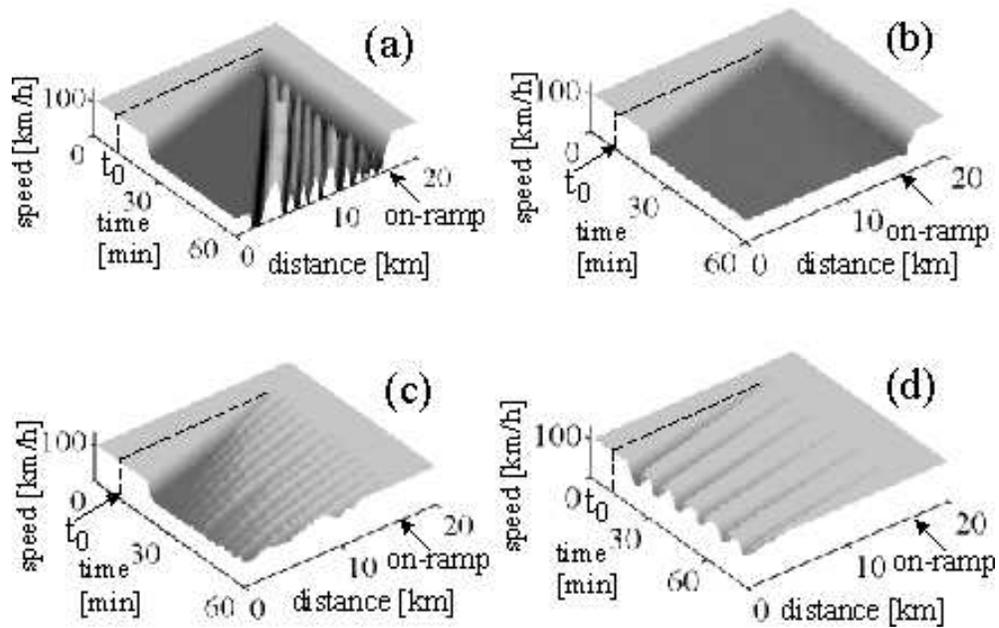}
\caption[]{Congested patterns at the on-ramp for the KKW-3-model of
  Fig.~\ref{Diagram2} (b):
(a) - General pattern (GP),
(b) - widening synchronized flow pattern (WSP).
(c) - widening synchronized flow pattern (WSP) at a lower value of the
      flow rate to the on-ramp than in (b), 
(d) - moving synchronized flow pattern (MSP). 
Single  vehicle data are averaged over a space interval of
$40 \ m$ and a time interval of $1 \ min$.
$t_{0}=8 \ min$. 
The flow rates $(q_{\rm on}, q_{\rm in})$ are:
(a) (480, 2300), 
(b) (120, 2160), 
(c) (15, 2160), and 
(d) (5, 2040) ${\rm vehicles/h}$.
\label{Mesh_Patterns3}  }
\end{figure}

\begin{figure}
\includegraphics[width=\textwidth]{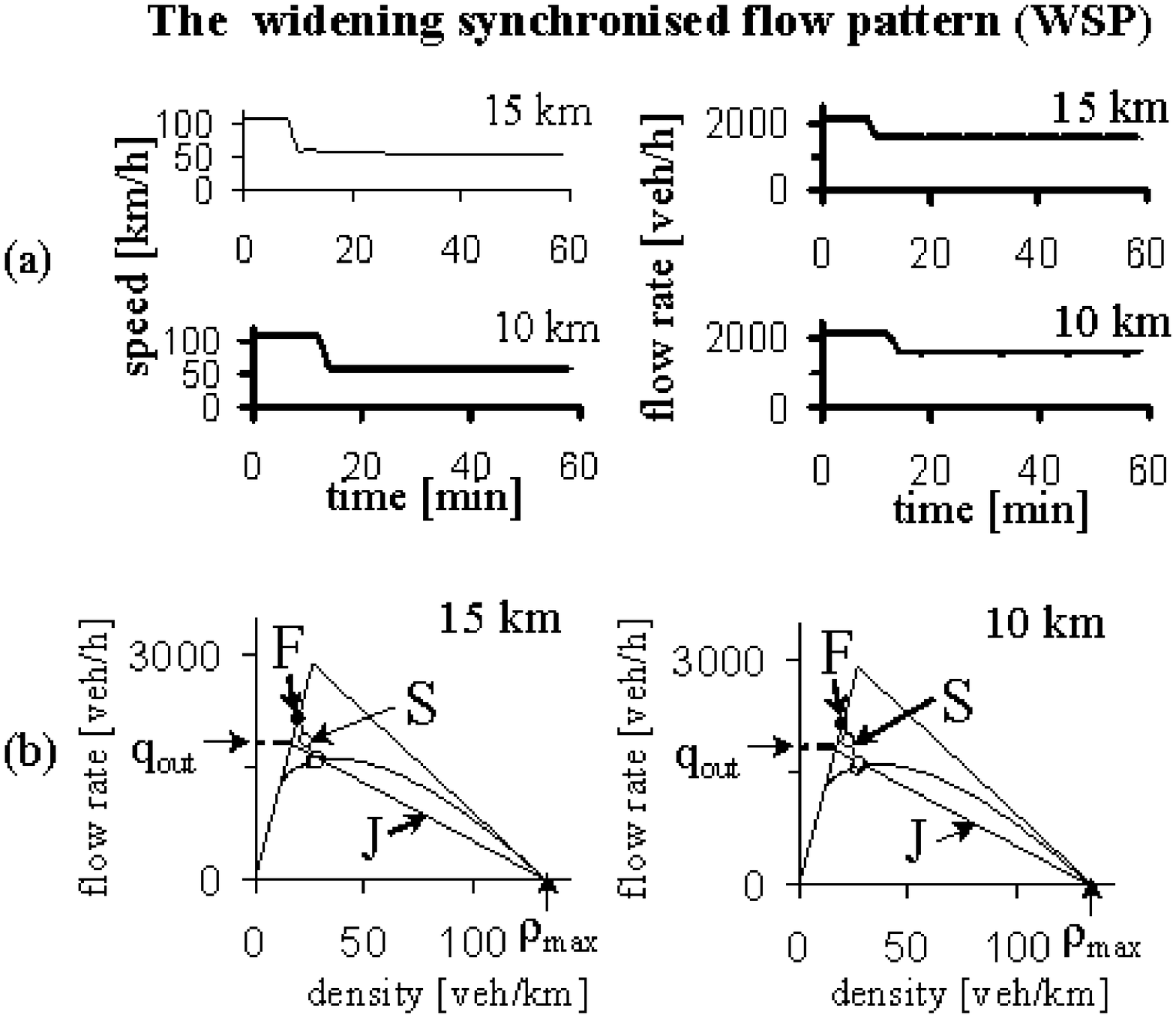}
\caption{The widening synchronized flow pattern (WSP) related to
  Fig.~\ref{Mesh_Patterns3} (b) (KKW-3-model): 
(a) - vehicle speed (left) and flow rate (right),
(b) - the corresponding data on the flow-density plane.
One minute averaged data of virtual detectors, whose coordinates
are indicated in (a, b). 
\label{Patterns3_WSP} }
\end{figure}

\begin{figure}
\includegraphics[width=\textwidth]{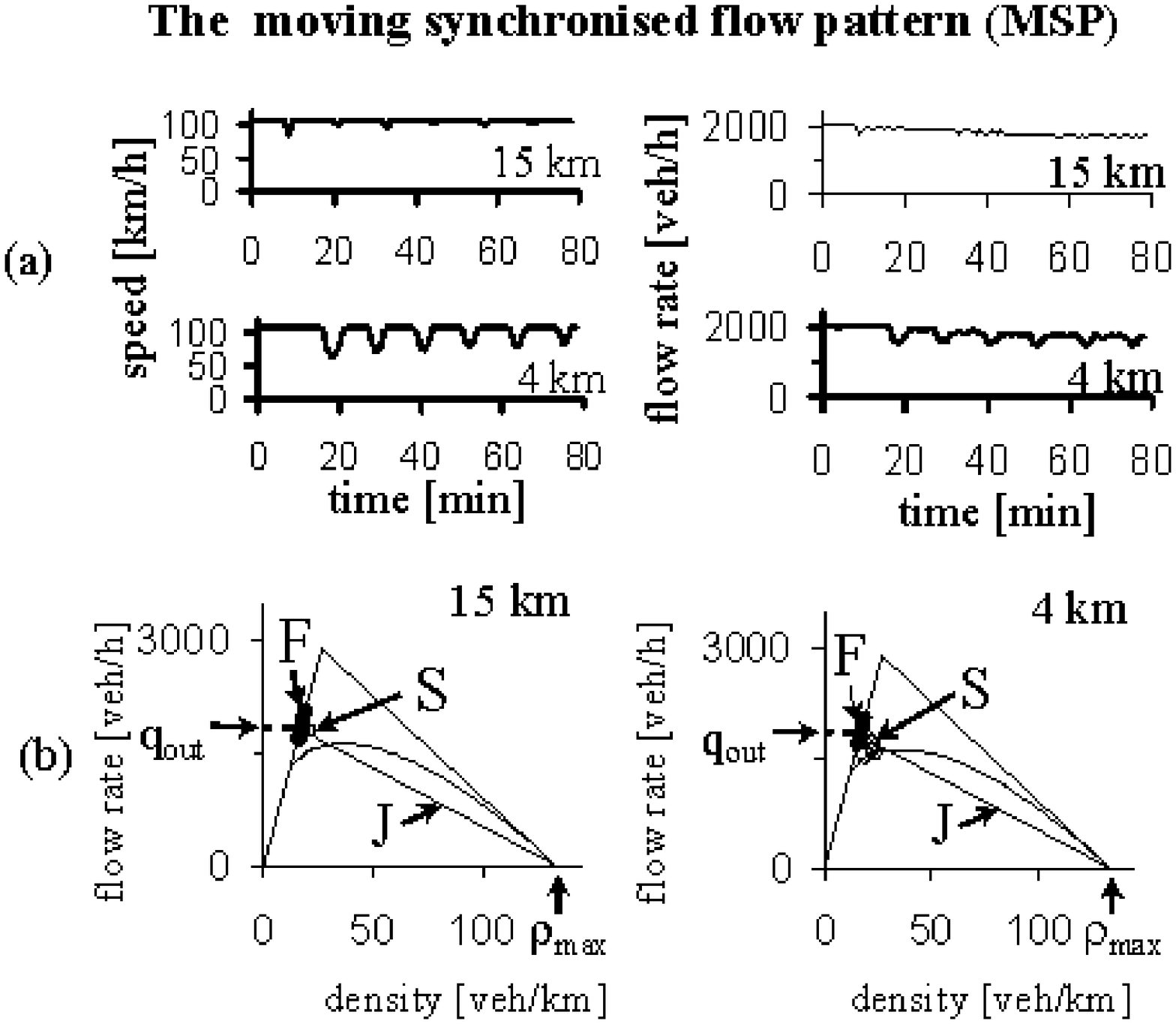}
\caption{The moving synchronized flow pattern (MSP) related to
  Fig.~\ref{Mesh_Patterns3} (d) (KKW-3-model): 
(a) -  The vehicle speed (left) and the flow rate (right),
(b) - the corresponding data on the flow-density plane.
One minute averaged data of virtual detectors, whose coordinates
are indicated in (a, b).
\label{MSP} }
\end{figure}

Right of the boundaries $S^{(B)}_{J}$
and $G$ in Figs.~\ref{Diagram1} and \ref{Diagram2} 
one finds the ``general pattern'' (GP).
It is a {\it self-maintaining} congested pattern,
where synchronized flow occurs upstream
of the on-ramp, and wide moving jams spontaneously emerge in this synchronized flow
(Figs.~\ref{Mesh_Patterns1} (a, b) and~\ref{Patterns1_GP},~\ref{Mesh_Patterns2} (a, b)
and~\ref{Patterns2_GP},~\ref{Mesh_Patterns3} (a)). 
In other words, in the GP wide moving jams are continuously generated
somewhere upstream of the on-ramp.
In the outflow of the wide moving jams either synchronized flow or free flow occurs.
GP in the KKW-models within the three-phase-traffic theory
have common features, which are very similar to those found in~\cite{KKl}:

(i)
If free flow occurs in the outflow of a wide moving jam,  then 
the mean velocity of the downstream jam front $v_{\rm g}$ and
the mean flow rate in the jam outflow $q_{\rm out}$ are
characteristic quantities of the model. They do not depend on initial conditions
and are the same for different wide moving jams. 
The mean velocity of the downstream front remains a characteristic 
parameter no matter, what the state of flow in the jam outflow is.

(ii)
If $q_{\rm in}>q_{\rm out}$, then it is obvious that
the width of the wide moving jam, which is furthest upstream, 
increases monotonously (Figs.~\ref{Mesh_Patterns1} (a),~\ref{Patterns1_GP} 
(b), and~\ref{Mesh_Patterns2} (a)).
If in contrast, $q_{\rm in}<q_{\rm out}$, the width of the most upstream wide moving jam
decreases  and this jam dissolves. This process of  wide moving jam dissolution
repeates for the next most upstream jam and so on
(Figs.~\ref{Mesh_Patterns1} (b) and 
and~\ref{Mesh_Patterns2} (b)).
Nevertheless, if the difference between $q_{\rm out}$ and $q_{\rm in}$
is not very large, 
the region of wide moving jams is widening upstream over time
(Figs.~\ref{Mesh_Patterns1} (b) and 
and~\ref{Mesh_Patterns2} (b)).
Thus, GP which is very similar to the
GP found in~\cite{KKl}
spontaneously occurs in all CA-models (Table \ref{KKW-1}) within
three-phase traffic flow theory under consideration.
However, there are some peculiarities of the KKW-models
which will be considered below.

(iii) 
The first peculiarity of the KKW-models considered here is linked to
the fact mentioned above that the  
 upstream front in WSP which separates synchronized flow 
downstream and free flow upstream moves with a very high (non-realistic) negative velocity.
Let us consider a case, in which the flow rate
$q_{\rm in}$ is high (it corresponds to a point
above the boundary $W$, Figs.~\ref{Diagram1} and~\ref{Diagram2})
and the flow rate $q_{\rm on}$ is related to a point 
right of the boundary $S^{(B)}_{J}$ in the diagram of congested patterns.
In this case, first WSP occurs which further
 transforms into the GP, i.e., wide moving jams 
spontaneously emerge inside the synchronized flow of the initial 
WSP (Figs.~\ref{Mesh_Patterns1} (a) and~\ref{Mesh_Patterns2} (a)).
However,
the upstream front of this initial WSP moves considerably faster
upstream than the fronts of any wide moving jam.
For this reason, the upstream front of the whole GP at any flow rate $q_{\rm in}$
is determined by this upstream front of synchronized flow 
rather than by the most upstream
wide moving jam. In Fig.~\ref{Patterns1_GP} this upstream front of the 
synchronized flow
is  marked by the dashed line.
Note that such a GP has been observed in empirical observations
(see Fig. 20 and Sec. V.A in~\cite{Kerner2002B}).
In all CA-models within the three-phase-traffic theory under consideration
GP possesses similar non-linear features 
(compare Fig.~\ref{Mesh_Patterns1} (a, b) with Figs.~\ref{Mesh_Patterns2} (a, b),
\ref{Mesh_Patterns3} (a) and Fig.~\ref{Patterns1_GP}
with Fig.~\ref{Patterns2_GP}).

(iv)
There is also some difference in the GP formation
in the  continuous model~\cite{KKl} and in the KKW-models
in the three-phase traffic theory.
This difference concerns
the boundary $G$ which separates the dissolving general pattern (DGP) and
the GP. 

Recall that DGP appears right of the boundary $S^{(B)}_{J}$
at the initial flow rate $q_{\rm in}>q_{\rm out}$.
In this case, after a wide moving jam 
has been formed in synchronized flow upstream of the on-ramp, the mean
flow rate in the jam outflow cannot exceed $q_{\rm out}$.
Thus, the initial condition $q_{\rm in}>q_{\rm out}$ is not fulfilled any more.
As a result, the GP transforms into DGP, where one
or several wide moving jams propagate upstream,
and  either free flow or one of SP occurs upstream of the on-ramp.
This behaviour is realized also in the KKW-models 
(Figs.~\ref{Mesh_Patterns1} (c) and~\ref{Mesh_Patterns2} (c)).

In~\cite{KKl} the boundary  $G$ intersects  the boundary $S^{(B)}_{J}$
in the point $q_{\rm in}=q_{\rm out}$.
In the CA-models under consideration, however, the boundary $G$ is shifted to the left
in the diagram of congested patterns, i.e., the boundary $G$ intersects 
 the boundary $S^{(B)}_{J}$
at some $q_{\rm in}>q_{\rm out}$
(Figs.~\ref{Diagram1} and~\ref{Diagram2}).

This behaviour may be explained by hysteresis effects or by the influence
of high amplitude fluctuations 
(see the related remark at the end of Sec. VII.B.1 in~\cite{Kerner2002B}).
Indeed, in comparison with the model in~\cite{KKl}
in the CA-models model fluctuations are very high.
High amplitude fluctuations occur also in the outflow of a wide moving jam.
This may explain why for $q_{\rm in}>q_{\rm out}$ right of the boundary $S^{(B)}_{J}$
still the general pattern rather than DGP may occur
at  considerably lower  flow rate to the on-ramp $q_{\rm on}$ than in the  model
in~\cite{KKl}.

(v) In some cases the GP like that shown in Fig.~\ref{Mesh_Patterns1}
(b), i.e., the GP where the most upstream jam is dissolved in the
course of time, can occur even if the initial flow rate $q_{\rm in}$
is slightly higher than $q_{\rm out}$. This is linked to the fact
mentioned in item (iii) above that the upstream front of synchronized
flow in the GP propagates upstream faster than the one of any wide
moving jam (see Fig.~\ref{Mesh_Patterns1} (a, b) and
Fig.~\ref{Patterns1_GP} (b), where this upstream front is marked by
the dashed line). Thus, upstream of the most upstream wide moving jam
in GP a synchronized flow is formed. The flow rate in this
synchronized flow, $q^{\rm (syn)}_{\rm in}$, is always lower than the
initial flow rate $q_{\rm in}$. The latter flow rate is realized
upstream of the upstream front of synchronized flow in GP
(Fig.~\ref{Mesh_Patterns1} (a, b)). Therefore, the flow rate
downstream of the upstream front of synchronized flow in GP, i.e., the
flow rate $q^{\rm (syn)}_{\rm in}$ is the incoming flow rate for the
most upstream wide moving jam rather than the initial flow rate
$q_{\rm in}$. It can occur that $q^{\rm (syn)}_{\rm in}$ is also lower
than $q_{\rm out}$. In this case the most upstream wide moving jam in
GP will be dissolved.

For the KKW-1 model (parameter-set I of Table~\ref{parameters})  the
maximal value of the flow rate $q_{\rm in}$  
at which GP of this type occurs is $1960 \ {\rm vehicles/h}$.  
For the KKW-2 model  this flow rate $q_{\rm in}$
 is very close to the flow rate $q_{\rm out}=1810 \ {\rm vehicles/h}$.
For the KKW-3 model this flow rate is $q_{\rm in}=2160 \ {\rm vehicles/h}$.

(vi) In empirical observations of GP at on-ramps, it has recently been
found~\cite{Kerner2002B} that GP possesses the following
characteristic feature: If the flow rate to the on-ramp $q_{\rm on}$
is high enough, the average flow rate in the pinch region $q^{{\rm
    (pinch)}}$ (averaged over a time interval which is considerably
larger than the time-distance between narrow moving jams emerging in
the pinch region of GP) reaches the limit flow rate $q^{{\rm
    (pinch)}}_{{\rm lim}}$. This means that the flow $q^{{\rm
    (pinch)}}$ does not decrease below $q^{{\rm (pinch)}}_{{\rm lim}}$
even if the flow rate $q_{\rm on}$ further increases.  This case is
called the ``strong'' congestion~\cite{Kerner2002B}. In the strong
congestion condition, GP can not exist if $q_{\rm in}<q^{{\rm
    (pinch)}}_{{\rm lim}}$~\cite{Kerner2002A,Kerner2002B}.  Note that
in the ``weak'' congestion condition which is realized at lower flow
rates $q_{\rm on}$ the flow rate $q^{{\rm (pinch)}}$ changes
noticeably when the flow rate $q_{\rm on}$ is changing.

As in the model~\cite{KKl}, both the strong congestion and the weak
one can be simulated in the KKW-models under consideration.
In particular, GP can not exist if $q_{\rm in}<q^{{\rm (pinch)}}_{{\rm lim}}$. Indeed,
at high flow rates to the on-ramp $q_{\rm on}$
the boundary $S^{(B)}_{J}$ transforms into a horizontal line at
$q_{\rm in}=q^{{\rm (pinch)}}_{{\rm lim}}$ (Fig.~\ref{Diagram1} and
\ref{Diagram2}). For the KKW-1 model 
(parameter-set I of Table~\ref{parameters}) we obtain  
$q_{\rm out}/q^{{\rm (pinch)}}_{{\rm lim}}\approx 1.57$.
This value is also approximately
in accordance with the empirical finding (see the empirical formula (4)
in~\cite{Kerner2002B}).

It should be noted that in the vicinity of this horizontal line on the
boundary $S^{(B)}_{J}$, after the GP has been formed (precisely, when
the related point in the flow-flow plane in Fig.~\ref{Diagram1} and
\ref{Diagram2} has been moved above and right of the boundary
$S^{(B)}_{J}$), the strong congestion in the pinch region occurs, and
the flow rate of the vehicles, which may actually squeeze to the road
from the on-ramp, can decrease in comparison with the initial flow rate
$q_{\rm on}$.  This effect has also occurred in the model~\cite{KKl}
at the related high initial flow rates $q_{\rm on}$.  Nevertheless,
the GP remains the GP after this decrease in the real $q_{\rm on}$.
This is due to a hysteresis effect in the flow rate to the on-ramp
$q_{\rm on}$ which accompanies the occurrence and the disappearance of
the GP when the flow rate $q_{\rm on}$ first increases and then
decreases, correspondingly.  However, the detailed investigation of
hysteresis effects is out of the scope of this paper and it will be
considered elsewhere.

\subsection{Probability of 
the breakdown phenomenon (the F$\rightarrow$S transition) at the on-ramp}

Let us first recall, how the breakdown phenomenon looks like in the
fundamental diagram approach.
From a numerical analysis of a macroscopic traffic flow 
model within the fundamental diagram approach Kerner and Konh\"auser
found in 1994 \cite{KK1994} that free flow is metastable with respect
to the formation of wide moving jams (F$\rightarrow$J transition), 
if the flow rate is equal to or higher than the outflow from a jam,
$q_{\rm out}$. The critical amplitude of a local perturbation in an
initially homogeneous free flow, which is needed for the
F$\rightarrow$J transition, decreases with increasing density: 
It is maximal at the threshold  
density $\rho = \rho_{\rm min}$, below which free flow is stable.
$\rho_{\rm min}$ is the characteristic density in the outflow from a wide
moving jam, i.e. when $q=q_{\rm out}$.
The critical amplitude becomes zero at some critical density
$\rho = \rho_{\rm cr}> \rho_{\rm min}$, above which free flow is linearly
unstable. Obviously the higher the amplitude of a random local 
perturbation the less frequent it is. Hence, the likelihood that the
F$\rightarrow$J transition occurs in a given time interval should
increase with density (or flow rate). 
The probability should be very small at the threshold density
$\rho_{\rm min}$ (at the threshold flow rate $q_{\rm out}$),
and it should tend to one at the critical density $\rho_{\rm cr}$
(at the related critical flow rate in free flow). 

In 1997  Mahnke {\it et al}~\cite{Mahnke1997,Mahnke1999} developed
a master equation approach for calculating the
probability of the F$\rightarrow$J transition on a homogeneous road
(i.e. without bottleneck). Based on this
approach  K{\"u}hne {\it et al}~\cite{Kuehne2002} confirmed that
the probability of the F$\rightarrow$J transition
in the metastable region is increasing with the flow rate in free
flow. They applied this result to explain the
breakdown phenomenon at a highway bottleneck.  
For a recent comprehensive discussion of
the breakdown phenomenon in CA-models and in the Krau{\ss} {\it et al.} model
in the fundamental diagram
approach see also \cite{Jost2002,JostNa}.
The theories in~\cite{KK1994,Mahnke1997,Mahnke1999,Kuehne2002,Jost2002,JostNa}
belong to the fundamental diagram approach.

In contrast to these results, in Kerner's three-phase traffic 
theory~\cite{Kerner1999A,Kerner1999B,Kerner1999C}  
it is postulated that metastable states of free flow decay into
synchronized flow (F$\rightarrow$S transition) rather than wide moving
jams (F$\rightarrow$J transition). Moving jams can emerge
spontaneously only in synchronized flow, i.e. after
a sequence of F$\rightarrow$S$\rightarrow$J transitions~\cite{Kerner1998}.
In particular, even the upper limit of free flow, $q_{\rm max}$, is related to
the F$\rightarrow$S transition: In this limit point the probability of
the F$\rightarrow$S transition should be equal to one whereas the
probability of  the emergence of a moving jam (F$\rightarrow$J
transition) should be very small. Thus, in this theory the breakdown
phenomenon in free traffic is related to 
the F$\rightarrow$S transition rather than to an emergence of moving jams.

In the three-phase traffic theory, it is also postulated that the
breakdown phenomenon at a highway bottleneck 
(i.e., due to an on-ramp) is due to a localized deterministic
perturbation at the bottleneck~\cite{Kerner2000A}.
Indeed, due to this perturbation the probability of the
F$\rightarrow$S transition should be considerably higher at the bottleneck   
than anywhere else. Thus, in this theory there is a spatially non-homogeneous
distribution of the probability of the breakdown phenomenon
(of the F$\rightarrow$S transition) on a road (per unit of space and time): At effective
bottlenecks on the road the probability of the breakdown phenomenon
should have maxima (see Fig. 1 (b) in~\cite{Kerner2000A}).
This explains, why the breakdown phenomenon  mostly occurs
at highway bottlenecks. 

In the following we confirm these hypotheses of three-phase traffic theory
for the KKW-models proposed in this paper.
In particular we find that a localized permanent perturbation indeed
occurs in the vicinity of the bottleneck (due to the on-ramp),
and that it triggers the breakdown of initially free flow
at the on-ramp rather than anywhere else.

As in~\cite{KKl}, at the boundary $F^{\rm (B)}_{\rm S}$ the 
F$\rightarrow$S transition occurs at the on-ramp.
Due to this transition the vehicle speed decreases sharply at the on-ramp.
A sharp drop of vehicle speeds at the on-ramp
(or at another bottleneck) is well-known from empirical observations
(e.g.,~\cite{Hall,Persaud}). 
Traffic engineers have called this effect "the breakdown phenomenon" in traffic flow.
The F$\rightarrow$S transition has the nature of such a breakdown phenomenon.

In~\cite{KKl} it was also shown that the F$\rightarrow$S transition
 is a first order phase transition: It reqires nucleation,
 i.e. the occurrence of a local perturbation in traffic flow whose amplitude exceeds 
some critical value. This critical amplitude is a
 decreasing function of the vehicle density in 
free flow (see the curve $F_{\rm S}$ in Fig. 1 (b) in~\cite{KKl}). 
The higher the  amplitude of a random perturbation (fluctuation) the less likely
 it is, with a  probability distribution that for very general reasons can be
 assumed to decay exponentially for large amplitudes. Therefore we expect that 
the probability that the F$\rightarrow$S transition happens within a given time
 interval increases exponentially with the vehicle density in free flow. 

Such behaviour of the probability of the breakdown phenomenon, i.e. the
F$\rightarrow$S transition, has indeed been observed empirically by Persaud {\it et
  al.}~\cite{Persaud}. 
In the case of the on-ramp, however, there is already a local permanent
non-homogeneity, 
which occurs due to the squeezing of vehicles into the main road. Corresponding
to~\cite{Kerner2000A} 
this should explain why the F$\rightarrow$S transition occurs at the on-ramp
with a considerable  higher probability than
away from the on-ramp at the same flow rate. 

This local permanent (deterministic) perturbation at on-ramp
determines the character of the boundary
$F^{(B)}_{S}$~\cite{Kerner2002B}.  The higher the flow rate to the
on-ramp $q_{\rm on}$ is, the higher is the amplitude of this permanent
perturbation. Therefore, the higher the flow rate to the on-ramp
$q_{\rm on}$ is, the lower is the flow rate $q_{\rm in}$ on the main road
upstream of the on-ramp, at which the related critical amplitude
occurs at the bottleneck: This may explain the negative slope of the
curve $F^{\rm (B)}_{\rm S}$ in the flow-flow plane in
Fig.~\ref{Diagram1}(a).

However, a real local perturbation which leads to the F$\rightarrow$S transition
at the on-ramp has always also a random component, i.e.,  the real local
perturbation should consist of two components: 
(a) a permanent perturbation, the amplitude of which is the higher, the higher the
flow rate to the on-ramp  $q_{\rm on}$ is, and (b) a random component. The latter component
should lead to the F$\rightarrow$S transition at the
on-ramp with some probability also, if the flow rate upstream of the
on-ramp, $q_{\rm in}$, and the flow rate to the on-ramp, $q_{\rm on}$,
belong to points in the flow-flow plane in
Fig.~\ref{Diagram1}(a) which lie to the {\it left} of the boundary  $F^{\rm
  (B)}_{\rm S}$, i.e. still in the free flow region. 
This probability should increase, if the flow rate $q_{\rm sum}=q_{\rm
  in}+q_{\rm on}$ approaches the boundary $F^{\rm (B)}_{\rm S}$. 
If these assumptions are correct, then 
the probability of  the F$\rightarrow$S transition at the
on-ramp must grow, if the flow rate upstream of the on-ramp $q_{\rm in}$  increases
at a given constant flow rate to the on-ramp $q_{\rm on}$,
which can be related to the results of empirical observations~\cite{Persaud}.

\begin{figure}
\includegraphics[width=\textwidth]{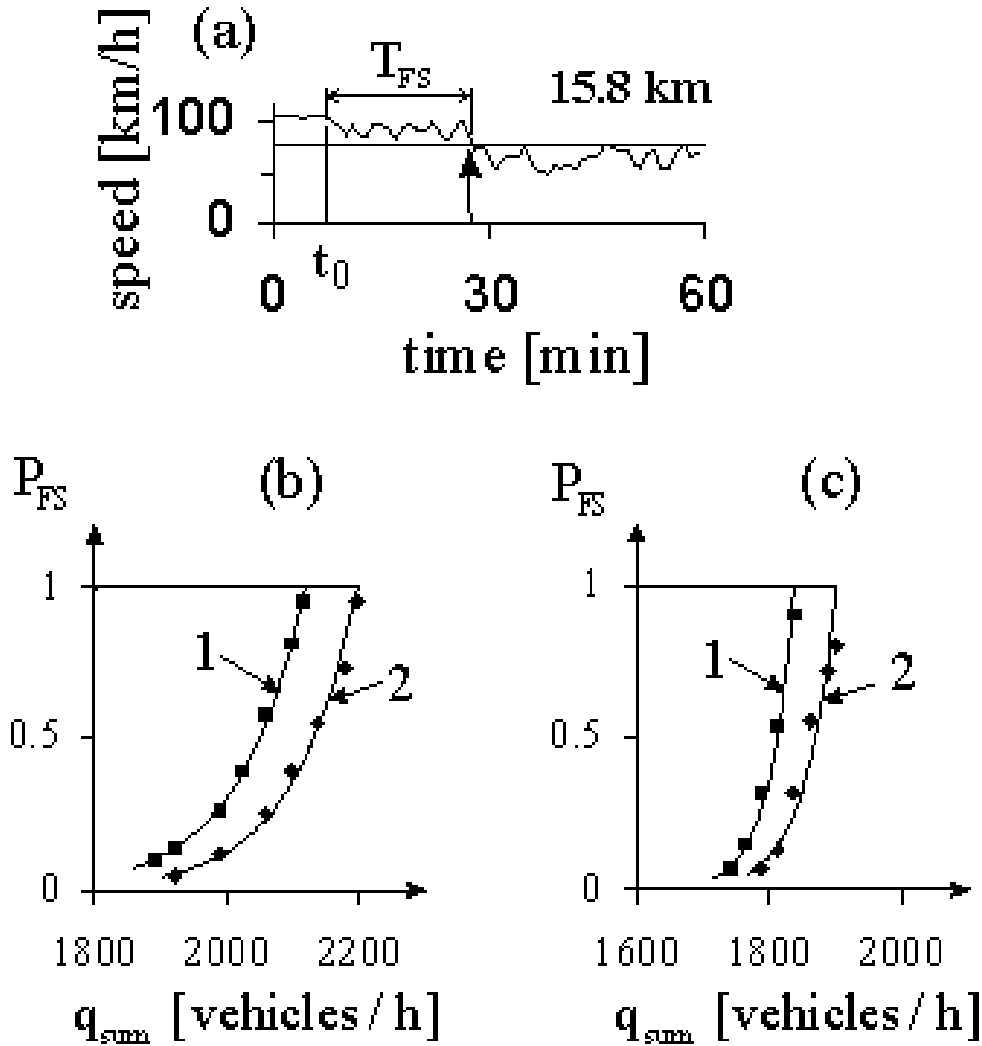}
\caption{Probability of breakdown phenomenon at the on-ramp
for the KKW-1-model (parameter-set I of Table \ref{parameters}):
(a) - Time dependence of the vehicle speed, when the
F$\rightarrow$S transition occurs at the on-ramp
(the arrow marks the F$\rightarrow$S transition).
Data are one minute averages of a virtual detector located at $x=15.8
\ km$ ($200 m$ upstream of  
the start of the on-ramp merging area). The dashed line
shows the speed level  $80 \ km/h$; the characteristic duration of a
sharp decrease in the vehicle speed  (this "breakdown" is marked by
the arrow) is about 1- 2 min in agreement with empirical observations.
$(q_{\rm on}, q_{\rm in})$ have the following values: (200, 1660) {\rm vehicles/h}.
 (b, c) - The probability 
$P_{\rm FS}$ that the F$\rightarrow$S transition occurs at the on-ramp
within $T_{0}=$30 min (curve 1) or already within $T_{0}=$15 min (curve 2),
after the on-ramp inflow was switched on (at $t_{0}=8 \ min$),
versus the traffic demand upstream of the on-ramp, $q_{\rm sum}=q_{\rm
  in}+q_{\rm on}$. Results are shown 
for two different flow rates to the on-ramp, 
$q_{\rm on}=60 \ {\rm vehicles/h}$ in (b) and
$q_{\rm on}=200 \ {\rm vehicles/h}$ in (c).
The following criterion that a F$\rightarrow$S transition has occurred is
used: The vehicle speed just
 upstream of the on-ramp drops below the level  $80 \ km/h$ 
and then remains at nearly the same low level for
 more than 
$4 \ min$
(cf. (a)). The probabilities were obtained from $N=40$ independent
runs.
\label{Probability}}
\end{figure}

To study the probability of the F$\rightarrow$S transition at the
on-ramp (Fig.~\ref{Probability}) in the KKW-1 model
(parameter-set I of Table~\ref{parameters}) a large
number of runs of the same duration $T_{0}$ has been studied for given
flow rates $q_{\rm sum}$ and $q_{\rm on}$. At the beginning of each
run there was free flow at the on-ramp. For each run it was checked,
whether the F$\rightarrow$S transition at the on-ramp occurred within
the given time interval $T_{0}$ or not. The result of these
simulations is the number of realizations $n_{\rm P}$ where the
F$\rightarrow$S transition at the on-ramp had occurred in comparison
with the number of all realizations $N_{\rm P}$. Then
\begin{equation}
P_{\rm FS}=n_{\rm P}/N_{\rm P} 
\end{equation}
is the probability that the F$\rightarrow$S transition at the on-ramp 
in an initial free flow occurs
during the time interval $T_{0}$ at given flow rates $q_{\rm sum}$ and $q_{\rm on}$.

The flow rate $q_{\rm sum}$ was changed and the procedure with all
realizations was repeated
at the same flow rate to the on-ramp $q_{\rm on}$.
The flow rate $q_{\rm sum}$ at which the F$\rightarrow$S transition at the on-ramp
occurred in all realization is therefore related to the probability 
of the  F$\rightarrow$S transition $P_{\rm FS}=1$. We found that lower
flow rates $q_{\rm sum}$ correspond 
to $P_{\rm FS}<1$. 
As expected above we found indeed an exponential increase of the probability
$P_{\rm FS}$ as a function of the flow rate $q_{\rm sum}$ at a given flow rate
to the on-ramp $q_{\rm on}$. This confirms the above
assumptions~\cite{Kerner2000A} about the nature of the breakdown phenomenon at
the on-ramp.

Fig. \ref{Probability} shows that the nucleation rate is the higher
the larger $q_{\rm on}$. This can be inferred from the fact that the
range of flow rates downstream, $q_{\rm sum}$, over which the nucleation
probability changes by a given amount, is much narrower for higher
$q_{\rm on}$. The stronger permanent perturbation (higher $q_{\rm
  on}$) acts like a bias which makes it easier to overcome the
nucleation barrier.

The F$\rightarrow$S transition leads to the occurrence of different
synchronized flow patterns upstream 
of the on-ramp which have been considered above.

\subsection{The capacity drop \label{capacity} }

Empirical observations show that the speed  breakdown 
at a bottleneck is in general accompanied by a drop in highway
capacity. If there is free rather than congested flow upstream of the
bottleneck, the highway capacity is usually
higher. This phenomenon is called "the capacity drop"
(for a review see~\cite{Hall}). 

As explained in the previous section, the breakdown phenomenon in 
three-phase traffic theory is a F$\rightarrow$S transition at the
bottleneck, so that the capacity drop is the difference between 
highway capacity in free flow and in a situation, where there is
synchronized flow upstream and free flow downstream of the bottleneck
~\cite{Kerner1999B,Kerner2000A}. Thus the capacity drop is {\em not}
determined by the outflow $q_{\rm out}$ from a wide moving jam in
contrast to the fundamental diagram approach. 

Obviously, in order to study the capacity drop one has to
consider the outflow from a congested bottleneck
$q^{\rm (bottle)}_{\rm out}$, which is measured 
downstream of the bottleneck, where free flow conditions are reached.
In this paper we consider the special example of an on-ramp as
bottleneck. In~\cite{Kerner2000A} Kerner points out that $q^{\rm
  (bottle)}_{\rm out}$ is not just a characteristic property of the
type of bottleneck under consideration only. It also depends on the type of
congested pattern which 
actually is formed upstream of the bottleneck. Hence, in the case of
an on-ramp, $q^{\rm (bottle)}_{\rm out}$ is expected to vary with
$(q_{\rm on},q_{\rm in})$. Obviously, $q^{\rm (bottle)}_{\rm out}$
only limits the highway capacity, if it is smaller than the traffic
demand upstream of the on-ramp, $q_{\rm sum}=q_{\rm
  in}+q_{\rm on}$, i.e. if the condition
\begin{equation}
q^{\rm (bottle)}_{\rm out}(q_{\rm on},q_{\rm in}) 
< q_{\rm sum}(q_{\rm on},q_{\rm in})
\label{capacity_condition}
\end{equation}
is fulfilled. Then the 
congested pattern upstream from the on-ramp simply expands, while the
thoughput remains limited by $q^{\rm (bottle)}_{\rm out}$.
For example, if the general pattern (GP) is
formed at the bottleneck, an increase of $q_{\rm in}$ does not
influence the discharge flow rate $q^{\rm (bottle)}_{\rm out}$. Instead,
the width of the wide moving jam, which is most upstream in the
general pattern, simply grows.

Assuming that (\ref{capacity_condition}) is fulfilled, 
the capacity drop is given by 
\begin{equation}
\Delta q=q_{\rm max}-q^{\rm (bottle)}_{\rm out},
\label{drop2}
\end{equation}
where $q_{\rm max}$ denotes the highway capacity in free flow, i.e. the
maximum flow rate in free flow.
The minimum value, which $q^{\rm (bottle)}_{\rm
  out}$ can take, if one considers all kinds of
congested patterns upstream from a bottleneck, should be a
characteristic quantity for the type of bottleneck under
consideration. We denote this quantity by $q^{\rm (bottle)}_{\rm
  min}$.
The maximum of $q^{\rm (bottle)}_{\rm out}$ (denoted by  $q^{\rm
  (bottle)}_{\rm  max}$) is
predicted to be the maximum flow rate, which can be realized in
synchronized flow~\cite{Kerner2000A}, $q^{\rm (bottle)}_{\rm
  max} = q^{\rm (syn)}_{\rm max}$.
Hence, the capacity drop at a bottleneck cannot be smaller than   
\begin{equation}
\Delta q_{\rm min}=q_{\rm max}-q^{\rm (syn)}_{\rm max}.
\label{drop1}
\end{equation}

This general picture of the capacity drop proposed by Kerner for the
three-phase traffic theory has been confirmed by empirical
investigations~\cite{Kerner2000A,Kerner2002B} and in numerical studies
of congested patterns at an on-ramp~\cite{KKl}, as well as in this
paper.

In particular, we found that for the KKW-1 model (parameter-set I of
Table \ref{parameters}) $q^{\rm (syn)}_{\rm max} \approx 2250 \ {\rm
  vehicles}/h$. This flow rate corresponds to WSP at $q_{\rm
  in}=q_{\rm max}$ and low flow rate to the on-ramp ($q_{\rm
  on}\approx 10 \ {\rm vehicles}/h$ ).  The flow rate $q^{\rm
  (bottle)}_{\rm min} \approx 1600 \ {\rm vehicles}/h$ is related to
LSP.  Thus, the discharge flow rate $q^{\rm (bottle)}_{\rm out}$ in
(\ref{drop2}) can be changed in the range from $1600$ to $2250 \ {\rm
  vehicles}/h$.  As the capacity of the highway in free flow (for
parameter-set I) is $q_{\rm max} \approx 2400 \ {\rm vehicles}/h$, the
capacity drop (\ref{drop2}) can vary in the range from $150$ to $800 \ 
{\rm vehicles}/h$, depending on the type of congested pattern and the
pattern parameters.  In comparison with free flow conditions the
capacity drops by $6.25 \%$ for WSP at low flow rate to the on-ramp up
to $33.3 \%$ for LSP at a high flow rate to the on-ramp.

To show the dependence of the capacity drop on the congested pattern
type and the pattern parameters more clearly, let us consider a
specific example: Initially the flow to the on-ramp,  $q_{\rm on}=60 \ 
{\rm vehicles}/h$, is low, and the flow rate upstream of the on-ramp,
$q_{\rm in}=2160 \ {\rm vehicles}/h$, is relatively high. As a result, WSP
occurs (Fig.~\ref{Diagram1} (a)). In this case, the discharge flow
rate $q^{\rm (bottle)}_{\rm out}= 1950 \ {\rm vehicles}/h$, i.e., the
capacity drop (\ref{drop2}) is $\Delta q= 210 \ {\rm vehicles}/h$. Let us
now increase traffic demand upstream on the on-ramp, i.e., the flow
rate $q_{\rm on}$ increases up to $q_{\rm on}=120 \ {\rm vehicles}/h$ but
the flow rate $q_{\rm in}=2160 \ {\rm vehicles}/h$ remains the same. 
This causes the flow rate of synchronized flow in WSP to decrease {\it more}
than the traffic demand $(q_{\rm in}+q_{\rm on})$ increases. 
For this reason, although the flow rate $q_{on}$
increases, the discharge flow rate {\it decreases}: $q^{\rm
  (bottle)}_{\rm out}= 1800 \ {\rm vehicles}/h$.  This leads to a larger
capacity drop (\ref{drop2}): $\Delta q=360 \ {\rm vehicles}/h$.

If now the flow rate $q_{\rm on}$ is further increased up to $q_{\rm
  on}=240 \ {\rm vehicles}/h$ (the flow rate $q_{\rm in}=2160 \ {\rm
  vehicles}/h$ remaining unchanged), WSP transforms into GP
  (Fig.~\ref{Diagram1} (a)). 
This leads to a further decrease in the discharge flow rate: $q^{\rm
  (bottle)}_{\rm out}= 1700 \ {\rm vehicles}/h$, which is due to 
the pinch effect in synchronized flow of GP: The pinch effect strongly
reduces the flow rate through the pinch region of the GP: The
capacity drop (\ref{drop2}) increases to
$\Delta q=460 \ {\rm vehicles}/h$.

If the flow rate $q_{\rm on}$ is once more increased up to $q_{\rm on}
= 900 \ {\rm vehicles}/h$ and the flow rate $q_{\rm in}$ remains the
same, strong congestion occurs in the pinch region of the GP. This
decreases the discharge flow rate to $q^{\rm (bottle)}_{\rm out} =
1630 \ {\rm vehicles}/h$. Thus the capacity drop (\ref{drop2}) increases
to $\Delta q=530 \ {\rm vehicles}/h$. In all considered cases the
condition (\ref{capacity_condition}) is fulfilled.

\section{Discussion \label{Dis} }

The numerical simulations of the  CA-models within the
three-phase traffic theory, which we have developed in this paper,
show that the basic vehicle motion rules (\ref{next}),  (\ref{next1})
introduced by Kerner and Klenov in~\cite{KKl}
for a microscopic three-phase traffic theory
allow a {\it variety of simple specifications} for 
the synchronization distance, the fluctuations and 
for the vehicle acceleration and deceleration  which lead to
{\it qualitatively the same diagram of congested patterns and to the
  same pattern features} at on-ramps.
This is due to the introduction of the synchronization distance $D$ in
the basic vehicle motion rules (\ref{next}),  (\ref{next1}), which
leads to
a 2D-region of steady states in the flow density plane. The robustness of the
phenomena with respect to different specifications of the model details
leads to the expectation that these phenomena occur generically independent
of e.g. the different laws and driver behaviours in different countries.

In the KKW-model formulation of three-phase traffic theory
based on these basic vehicle motion 
rules (\ref{next}),  (\ref{next1})
specific functions for fluctuations and 
for the vehicle acceleration and deceleration 
can be very simple.
Nevertheless, the main features of  the diagrams 
of congested patterns, which these CA-models show, are
qualitatively the same as recently found 
within the three-phase traffic theory
and in empirical observations~\cite{Kerner2002A,Kerner2002B}.

In~\cite{KKl} it has been shown that these pattern features and the
related diagram of congested patterns at on-ramps in the three-phase
traffic theory are qualitatively different in comparison with the
diagram by Helbing {\it et al.}~\cite{Helbing,Helbing1999}, which has
been derived for a wide class of traffic flow models within the
fundamental diagram approach. In particular, in that diagram of
congested patterns (congested states) near a boundary which separates
trigger stop-and-go traffic (TSG) and oscillating congested traffic
(OCT) (in the terminology of~\cite{Helbing,Helbing1999}) a congested
pattern which is a ``mixture'' of TSG, OCT and HCT (homogeneous
congested traffic) should occur~\cite{Helbing}. This pattern, which at
first sight looks like GP, has been used in~\cite{Helbing} for an
explanation of the pinch effect in synchronized flow and for the jam
emergence observed in~\cite{Kerner1998}.  However, this mixture
pattern has no own region in the diagram of states
in~\cite{Helbing1999,Helbing2000,Helbing}: The pattern transforms into
TSG, if $q_{\rm on}$ decreases, or into OCT or else HCT, if $q_{\rm
  on}$ increases.  In our diagrams (Figs.~\ref{Diagram1}
and~\ref{Diagram2} and in~\cite{KKl,Kerner2002B}) there are no TSG, no
OCT and no HCT.  Instead, GP exists in a very large range of flow
rates $q_{\rm on}$ and $q_{\rm in}$.  At a given $q_{\rm in}$ GP in
the three-phase traffic theory does not transform into another
congested pattern, even if $q_{\rm on}$ increases up to the highest
possible values.  Thus GP in the CA-models under consideration and
in~\cite{KKl,Kerner2002B} has a qualitatively different nature in
comparison with the mixture of TSG, OCT and HCT
in~\cite{Helbing,Helbing1999}.  Note that empirical observations of
congested patterns at on-ramps~\cite{Kerner2002B} confirm the
theoretical features of GP found within the three-phase traffic
theory~\cite{KKl,Kerner2002B}, rather than the theoretical features of
either TSG, or OST, or HCT, or else of the mixture of TSG, OST, and
HCT within the fundamental diagram approach derived
in~\cite{Helbing1999,Helbing2000,Helbing}.

In 2000, Knospe {\it et al.}~\cite{Knospe2000} proposed a version
of the NaSch model where in addition to the previous versions
(e.g.,~\cite{Barlovic}) drivers react at intermediate distances to
speed changes of the next vehicle downstream, i.e., to ``brake
lights''. The steady states of this model with 
``comfortable driving'' 
belong to a fundamental diagram, i.e. the NaSch model
with ``comfortable driving'' is a CA-model in the fundamental diagram approach.
The NaSch model with ``comfortable driving'' 
has been applied in~\cite{Knospe2000,Knospe2001,Knospe2002} for a
description of three traffic phases: free flow, synchronized flow and
wide moving jams. 

In order to compare this model with the ones investigated here, 
the congested patterns, which spontaneously occur at an on-ramp, and
their evolution, when the flow rate to the on-ramp is changing
were calculated for the Nagel-Schreckenberg CA-model
with ``comfortable driving''~\cite{Knospe2000,Knospe2001,Knospe2002}. 
These new results will be discussed below.
It will be shown that both the diagrams of congested patterns and the pattern features
of the KKW-models in three-phase-traffic theory
are qualitatively different from those obtained for
the Nagel-Schreckenberg CA-model with ``comfortable driving'', with
one exception, concerning the wide moving jam propagation,
which will be discussed first.

\subsection{Wide moving jam propagation \label{Wide}}

\begin{figure}
\includegraphics[width=\textwidth]{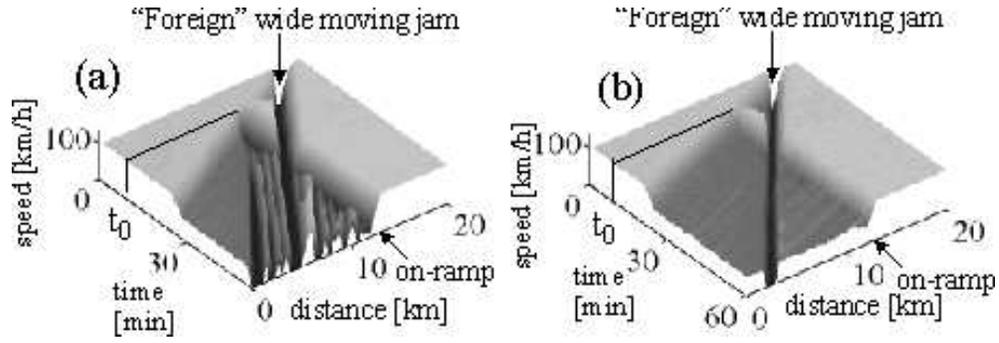}
\caption{Simulation of the propagation of a "foreign" wide moving jam:
(a) - Propagation through GP,
(b) - Propagation through WSP.
The on-ramp is located at $x=12 \ km$, and the inflow is switched on
at $t_{0}=7 \ min$. The flow rates 
$(q_{\rm on}, q_{\rm in})$ are:
(a) (480, 2300) and (b) (110, 2160). 
The initial location of the "foreign"
wide moving jam is $x=18 \ km$, the initial jam length is $0.7 \ km$.
Single  vehicle data are averaged over a space interval of
$40 \ m$ and a time interval of $1 \ min$.
KKW-1-model (Table \ref{KKW-1}): Parameter $k$ 
of the synchronization distance $D_{n}$ is $k=2.55$ for (a) and  $k=4$
for (b).  
Other model parameters are given in Table \ref{parameters}, parameter-set I.
\label{JamWide} }
\end{figure}

The characteristic parameters of wide moving jam propagation 
which were found in empirical 
observations~\cite{KR1996B,Kerner1998A,Kerner2001,Kerner2000A}
can be reproduced in many traffic flow models in the
fundamental diagram approach, where they have first been predicted
by Kerner and Konh\"auser in 1994~\cite{KK1994}
(see also the later papers by Bando, Sugiyama {\it et al.}~\cite{Bando}, 
by Krau{\ss} {\it et al.}~\cite{Kr}, by Barlovic {\it et al.}~\cite{Barlovic}, and the reviews
by Chowdhury {\it et al.}~\cite{Ch} and by Helbing~\cite{Helbing}). 
As already mentioned, 
the slow-to-start rules~\cite{TaTa,Barlovic,Knospe2000}  allow the wide moving jam propagation
through different traffic states and bottlenecks keeping the characteristic velocity of the downstream
jam front. This effect has recently
been simulated in  the NaSch model with
``comfortable driving''~\cite{Knospe2001}.
This is in accordance with 
empirical observations and the wide moving jam definition made 
above~\cite{Kerner1998A,Kerner2001,Kerner2000A}.
Because  the slow-to-start rules~\cite{Barlovic,Knospe2000}
are used in our CA-models as well, these CA-models within the
three-phase traffic theory also show 
the  effect of the wide moving jam
propagation through different congested patterns and bottlenecks 
(Fig.~\ref{JamWide}). In particular, if a wide moving jam 
which has been formed upstream of the on-ramp (the jam  is marked as ``foreign'' wide moving jam
in Fig.~\ref{JamWide}) then the jam propagates through the on-ramp  and through
GP (Fig.~\ref{JamWide} (a)) and also through 
WSP (Fig.~\ref{JamWide} (b)) keeping the velocity of the downstream
front of the jam.

However, the effect of the wide moving jam propagation~\cite{Kerner1998A} 
as well as the other characteristic 
parameters of wide moving jams are apparently the {\it only features}
which are the same in the fundamental diagram approach~\cite{KK1994,Kr,Barlovic,Knospe2001} 
and in the three-phase traffic theory~\cite{Kerner2001,Kerner2002C}.
All other known features of congested patterns which spontaneously
occur upstream of the on-ramp and their evolution for the Nagel-Schreckenberg 
CA-model with ``comfortable driving''~\cite{Knospe2000,Knospe2001,Knospe2002}
are qualitatively different from those, which follow from
the KKW-models within the three-phase-traffic theory, as will be shown
in the next section. This is due to  
the principal difference between the non-linear features of congested traffic
in the NaSch model with
``comfortable driving''~\cite{Knospe2000,Knospe2001} and in the KKW-models within
the three-phase traffic theory presented in our paper.

\subsection{Comparison with congested patterns in the Nagel-Schreckenberg 
cellular automata models with ``comfortable driving'' \label{Com} }

\begin{figure}
\includegraphics[width=0.8\textwidth]{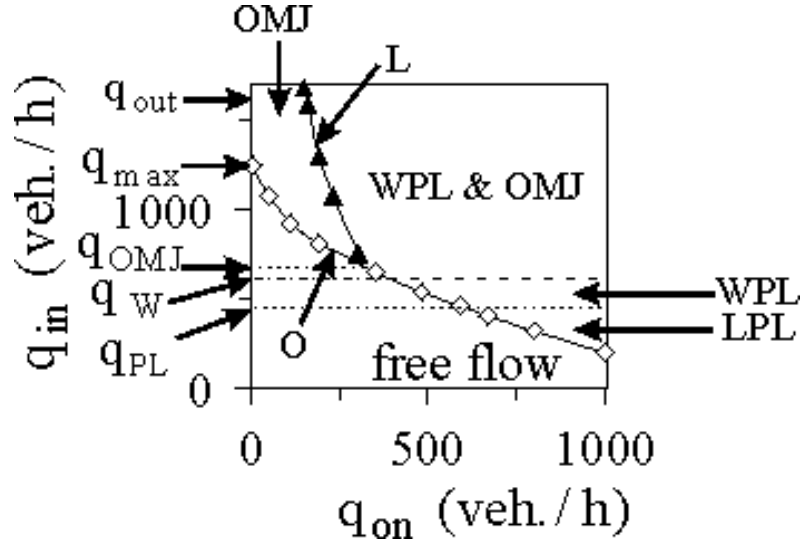}
\caption{Diagram of congested patterns
at the on-ramp for the NaSch CA-model with comfortable driving.
OMJ - oscillating moving jams, 
WPL - widening pinned layer, 
LPL - localized pinned layer.
The parameters of the model are 
taken from~\protect\cite{Knospe2000,Knospe2001}, in particular, 
the maximal speed $v_{\rm free}=27 \ m/s$ ($v_{\rm free}=97.2 \ km/h$), 
the minimal distance $d=7.5 \ m$, the time step $\tau=1 \ sec$, 
the maximum flow rate $q_{0}=2817 \ {\rm vehicles/h} $,
the flow rate out from the jam $q_{\rm out}=1580 \ {\rm vehicles/h} $. 
The other specific parameters of the CA-model with comfortable driving
(notations as in~\protect\cite{Knospe2000,Knospe2001})
are $p_{\rm d}=0.1$, $p_{\rm b}=0.95$,  $p_{0}=0.5$,  $h=8$, ${\rm
  gap}_{\rm security}=3$, the cell length is $1.5 \ m$. 
For the simulation of the on-ramp in the CA-model 
with comfortable driving the distance $dx^{\rm (min)}_{\rm on}$ 
is chosen as $dx^{\rm (min)}_{\rm on}=4d$.
The length of the road is $75 \ km$, the on-ramp is at $x=65.4 \ km$, the 
length of merging area $0.6 \ km$.
As in Figs.~\ref{Diagram1} and~\ref{Diagram2},
after the on-ramp has been switched on, there is  a delay time
for the congested pattern formation upstream of the on-ramp.
However, here the boundaries $O$ ("Oscillating")
and $L$ ("Layer") depend more strongly on the delay time
which is chosen for awaiting the congested pattern to occur
upstream of the on-ramp, after the on-ramp has been switched on. 
The boundary $O$
is determined as the points ($q_{\rm on}, q_{\rm in}$),  where either
OMJ or WPL or LPL occurred upstream of the on-ramp
in an initial flow with maximum speed $v=v_{\rm free}$
within 20 min after the on-ramp has been switched on. 
The time interval for the determination of the boundary $L$
was 60 min.
The qualitative features of the diagram 
do not depend on these time intervals.
The maximum flow rate in free flow $q_{\rm max}=1250 \ {\rm vehicles/h} $. 
\label{Diagram3_NaSch}  }
\end{figure}

\begin{figure}
\includegraphics[width=\textwidth]{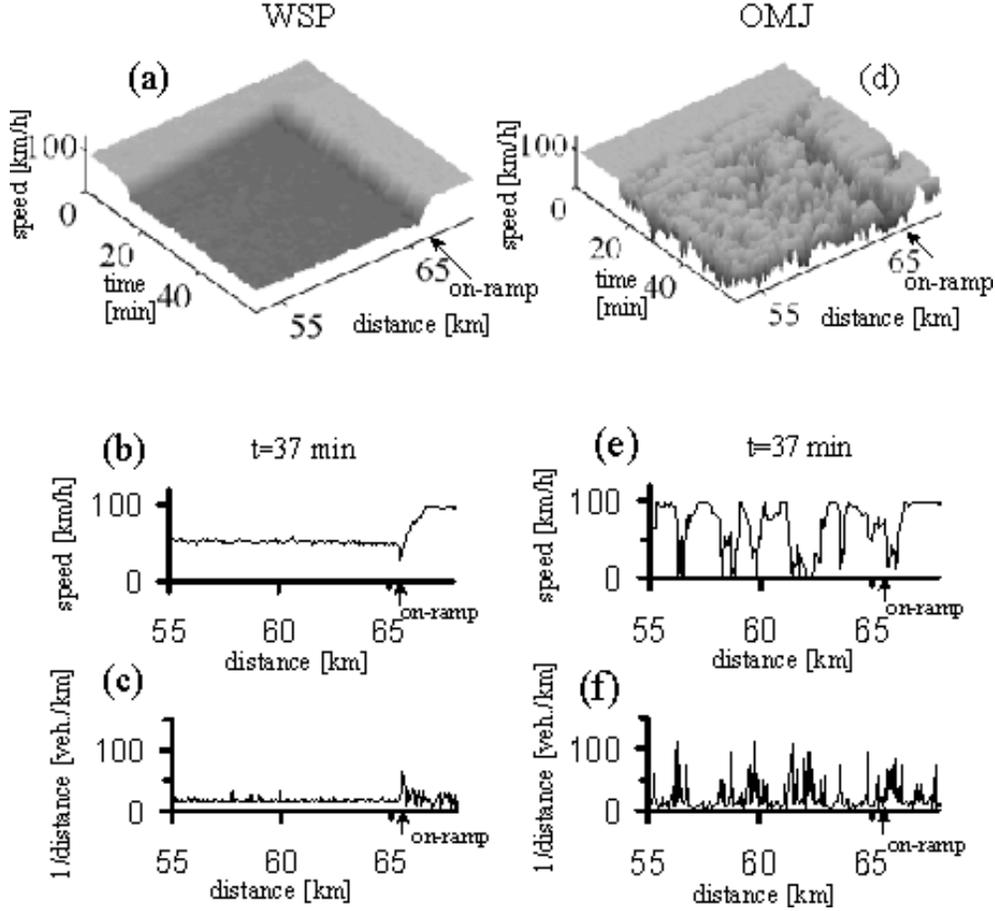}
\caption[]{Comparison of the widening synchronized flow pattern (WSP)
in our CA-model within the three-phase-traffic theory (a-c) with the
oscillating moving jam pattern (OMJ) at the on-ramp
in the  NaSch CA-model with comfortable driving (d-f).
(a, d) - the vehicle speed as function of distance and time;
at $t_{0}=8 \ min$ the on-ramp inflow is switched on.
(b, e) - the vehicle speed as function of distance at a given time.
(c, f) - the inverse distance between vehicles as function of distance
along the road.
In (a-f) single vehicle data are used.
For (a-c), the KKW-1-model (Table \ref{KKW-1}) with the following parameters
is used: 
$k=4.5$,
$p=0.04$, $p_{0}=0.5$, $v_{p}=14 \ m/s$   
$p_{a1}=0.5$ and $p_{a2}=0.05$. The parameter of the merging condition
(\ref{lambda}) at the on-ramp is $\lambda=0.4$.
For (d-f) the CA-model with comfortable driving given
in~\protect\cite{Knospe2000,Knospe2001,Knospe2002} was used.  
In both models,
$v_{\rm free}=27 \ m/s$, 
$d=7.5 \ m$, 
$\tau=1 \ sec$, 
$q_{0}=2817 \ {\rm vehicles/h}$,
$q_{\rm out}=1580 \ {\rm vehicles/h}$ (Table \ref{symbols}), 
the length of the road is $75 \ km$, the on-ramp is at $x=65.4 \ km$, the 
length of merging area $0.6 \ km$.
The other specific parameters of the CA-model with comfortable driving 
are the same as in Fig.~\ref{Diagram3_NaSch}.
With these parameters one obtains the results in (a-c) for 
$(q_{\rm in},q_{\rm on}) = (1246,600) \ {\rm vehicles/h} $,
in (d-f) for
$(q_{\rm in},q_{\rm on}) = (1246,190) \ {\rm vehicles/h} $.
\label{Mesh_OMJ_WSP}  }
\end{figure}

\begin{figure}
\includegraphics[width=\textwidth]{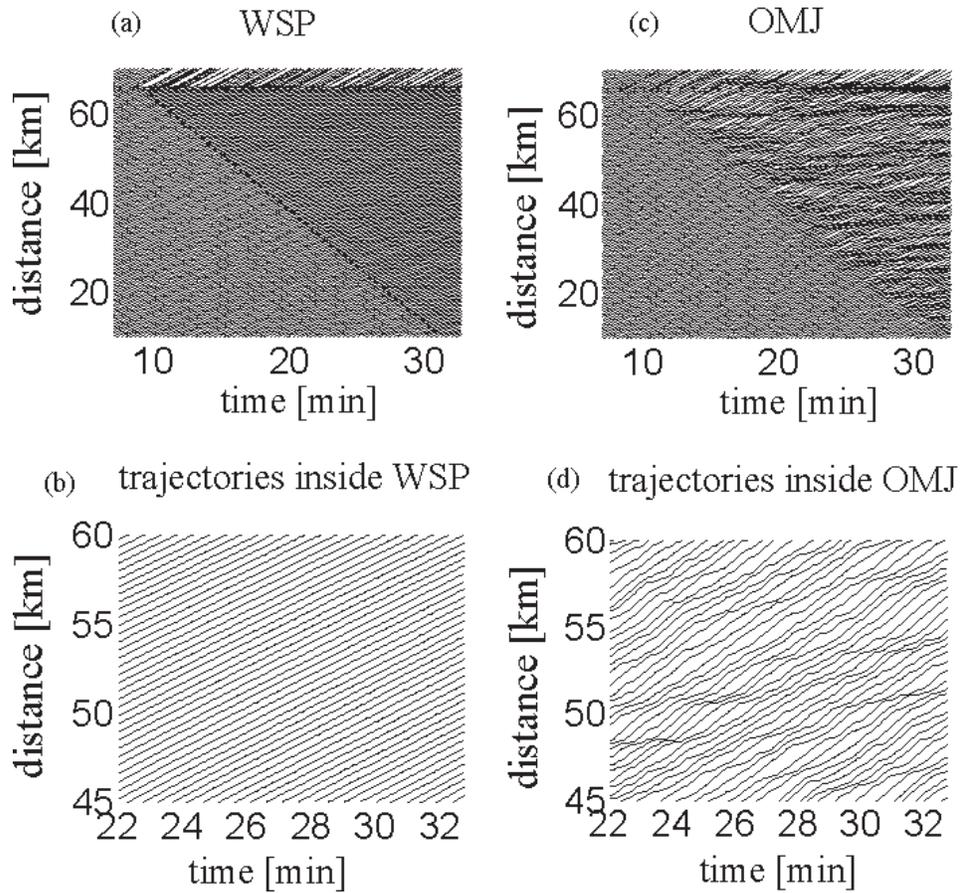}
\caption[]{Comparison of the vehicle trajectories related to the patterns in
  Fig.~\ref{Mesh_OMJ_WSP}: Widening synchronized flow pattern (WSP) in
  the KKW-1-model (Table \ref{KKW-1}) within the
  three-phase-traffic theory (left), oscillating moving jams (OMJ) 
  in the NaSch CA-model with comfortable driving 
  within the fundamental diagram approach (right).
  (a) - vehicle trajectories (overview) of WSP shown in
  Fig.~\ref{Mesh_OMJ_WSP} (a).  (b) - vehicle trajectories inside of
  WSP.  (c) - vehicle trajectories (overview) of OMJ upstream of the
  on-ramp shown in Fig.~\ref{Mesh_OMJ_WSP} (d).  (d) - vehicle
  trajectories inside OMJ.  Only trajectories of every 7th vehicle are
  shown.
\label{XT_OMJ_WSP}  }
\end{figure}

\begin{figure}
\includegraphics[width=\textwidth]{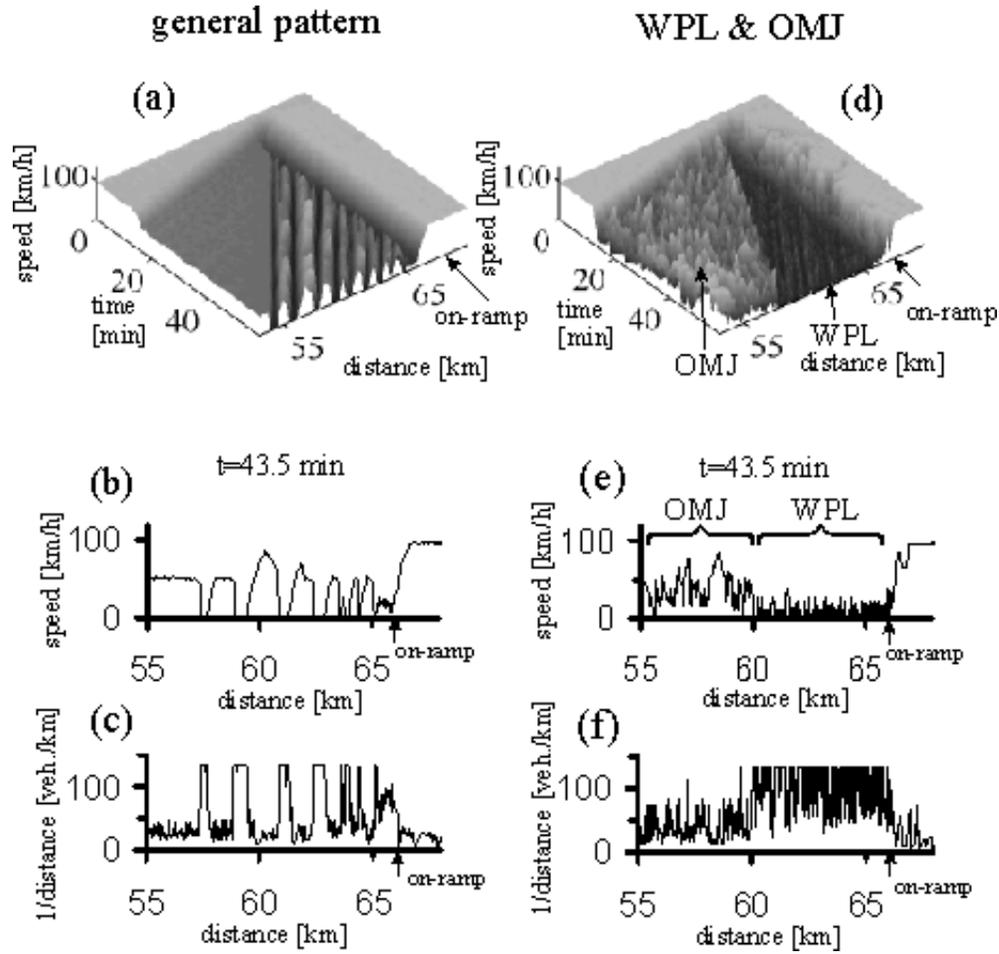}
\caption[]{Comparison of the general pattern (GP)
in  the KKW-1-model (Table \ref{KKW-1})
within the three-phase-traffic theory (left) 
with the congested pattern at the on-ramp
 in the  NaSch CA-model with comfortable driving 
  within the fundamental diagram approach (right).
The latter consists of a widening pinned layer (WPL) going over into OMJ
further upstream. 
(a, d) - vehicle speed as function of distance and time;
at $t_{0}=1 \ min$ the on-ramp inflow is switched on.
(b, e) - vehicle speed as function of distance at a given time.
(c, f) - inverse distance between vehicles as function of distance
along the road.
In (a-f) single vehicle data are used.
The simulation parameters for both models are  given in
Fig.~\ref{Diagram3_NaSch} and Fig.~\ref{Mesh_OMJ_WSP}. In both models
$(q_{\rm in},q_{\rm on}) = (1964,600) \ {\rm vehicles/h} $.
\label{Mesh_WPL_GP}  }
\end{figure}

\begin{figure}
\includegraphics[width=\textwidth]{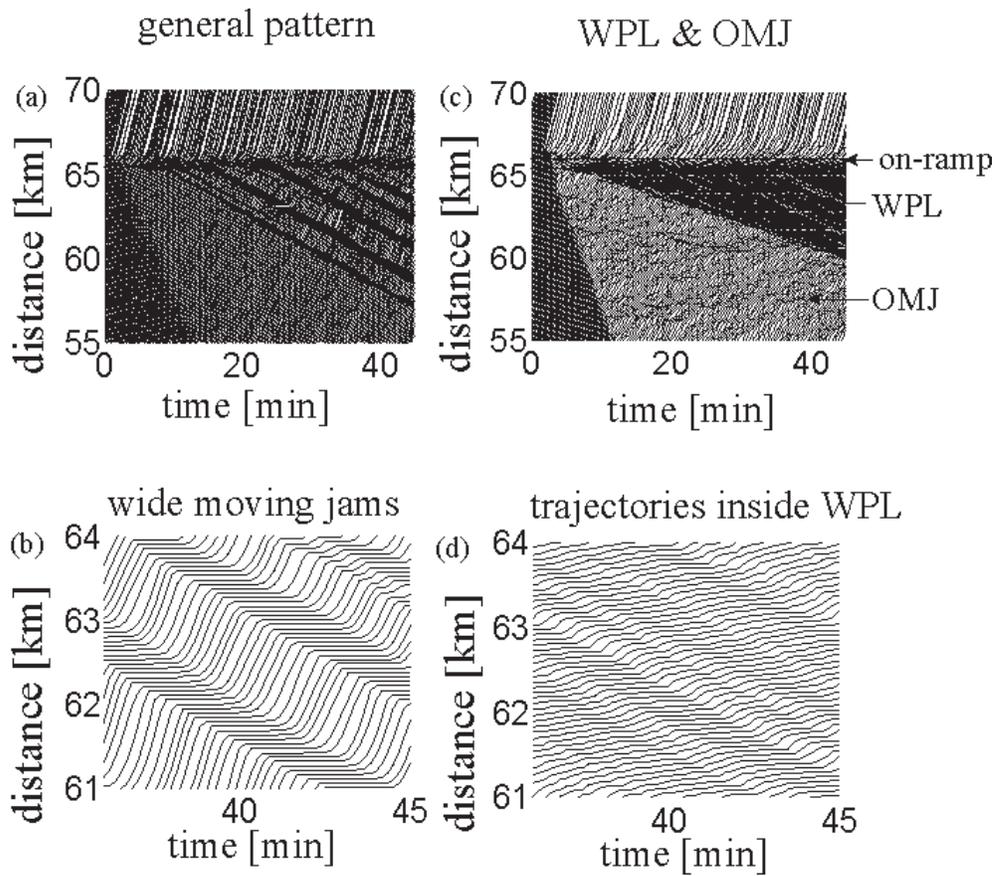}
\caption[]{Comparison of the vehicle trajectories related to the
  patterns in Fig.~\ref{Mesh_WPL_GP}: GP in the KKW-1-model (Table \ref{KKW-1})
 (figures left), 
 congested pattern at the on-ramp
 in the  NaSch CA-model with comfortable driving 
(figures right). Model specifications as in
Fig.~\ref{Diagram3_NaSch}
and Fig.~\ref{Mesh_OMJ_WSP}.
(a) - vehicle trajectories (overview) of GP
shown in Fig.~\ref{Mesh_WPL_GP} (a).
(b) - vehicle trajectories inside the region of wide moving jams in GP.
(c) - vehicle trajectories (overview) of WPL and further upstream OMJ
corresponding to Fig.~\ref{Mesh_WPL_GP} (d).
(d) - vehicle trajectories inside WPL.
Only trajectories of every 7th vehicle are shown.
\label{XT_WPL_GP}  }
\end{figure}

In Figs.~\ref{Diagram3_NaSch} - \ref{XT_WPL_GP}
we compare the congested patterns
and their evolution obtained in the KKW-models of three phase traffic
theory with those for a Nagel-Schreckenberg type CA-model with
``comfortable driving'' with one lane, for which we use the  rules and parameters  
presented in~\cite{Knospe2000,Knospe2001,Knospe2002}.
All  calculations for the Nagel-Schreckenberg CA-model
with comfortable driving  are made for two different models of the on-ramp:
(1) The lane changing rules described in~\cite{Knospe2001,Knospe2002} are applied,
or (2) the model of the on-ramp described in Sect.~\ref{On-ramp} is used.
In the latter case the distance between two consecutive vehicles 
on the main road, which permits  a vehicle to
enter from the on-ramp, is chosen as $dx^{\rm (min)}_{\rm on}=4d$ .
This corresponds 
to the condition applied in~\cite{Knospe2001,Knospe2002} that ``an effective gap to 
the predecessor and a gap to the successor on the destination lane'' is 
larger than or equal to the vehicle length $d$.
It has been found that all  features of the congested patterns
and their evolution, when the flow rate to the on-ramp is increasing,
remain qualitatively the same for both models of the on-ramp.
Therefore, only one set of results for the model of the on-ramp
described in Sect.~\ref{On-ramp} is shown in Figs.~\ref{Diagram3_NaSch} -
\ref{XT_WPL_GP}. This comparison shows the following results.

(i) In the Nagel-Schreckenberg CA-model
with comfortable driving ~\cite{Knospe2001} without on-ramps
and other bottlenecks, i.e.  on a homogeneous one-lane road,
when the flow rate is gradually increasing, the free flow motion
spontaneously transforms into a very complex dynamical behaviour
at some critical flow rate on the road, $q_{\rm max}$:
If a narrow moving jam emerges spontaneously, this jam dissolves within
a very short time interval (about one-two time steps, i.e. 1-2 s), 
then a new narrow moving  jam emerges which again dissolves quickly and so on 
at different locations and at different times.
This behaviour resembles the
oscillating congested traffic  which was reported for other
traffic flow models within the fundamental diagram approach by Lee
{\it et al.}~\cite{Lee1998}, Tomer {\it et al.}~\cite{Tomer}
and in the diagram of congested patterns 
at on-ramps by Helbing  {\it et al.}~\cite{Helbing,Helbing1999}.

(ii) In the Nagel-Schreckenberg CA-model with comfortable driving
on a one-lane road with an on-ramp, a complex oscillation pattern
occurs upstream of the on-ramp  spontaneously for $q_{\rm in}\geq
q_{\rm max}$, which is qualitatively the same as in (i): 
Narrow moving  jams first emerge and then dissolve within a very
short time interval at different highway locations and at different
times. This pattern exists already for very small values of the
flow rate to the on-ramp $q_{\rm on}$ (Fig.~\ref{Diagram3_NaSch}).
We will call this pattern  in the
Nagel-Schreckenberg  CA-model
with comfortable driving ``oscillating moving jams'' (OMJ)
(Figs.~\ref{Mesh_OMJ_WSP} (d-f) and~\ref{XT_OMJ_WSP} (c, d)).

The OMJ upstream of the on-ramp shows the same features as on a
homogeneous road without on-ramp as mentioned above.  The random
emergence and dissolution of narrow moving jams on a short time scale,
which is characteristic for OMJ (Figs.~\ref{Mesh_OMJ_WSP} (figures
right) and~\ref{XT_OMJ_WSP} (figures right)), is qualitatively
different from the behaviour inside the widening synchronized flow
pattern (WSP), which occurs spontaneously at the same parameters in
the KKW-models within the three-phase traffic theory
(Figs.~\ref{Mesh_OMJ_WSP} (figures left) and~\ref{XT_OMJ_WSP} (figures
left)).

Indeed, whereas inside WSP
in the KKW-models vehicles can move with nearly constant vehicle speed
(Figs.~\ref{Mesh_OMJ_WSP} (b) and~\ref{XT_OMJ_WSP} (b)),
in the Nagel-Schreckenberg 
CA-model
with comfortable driving inside OMJ the vehicles must randomly slow down sharply 
sometimes up to a stop and then accelerate within a short time scale, and so on
(Figs.~\ref{Mesh_OMJ_WSP} (e) and~\ref{XT_OMJ_WSP} (d)). 
The inverse distance between vehicles shows the same high
amplitude oscillating behaviour inside OMJ in the Nagel-Schreckenberg 
CA-model with comfortable driving (Fig.~\ref{Mesh_OMJ_WSP} (f)).
Its amplitude is much higher than would be expected due to bare model
fluctuations. 
This is the result of nonlinear amplification of fluctuations
in the  Nagel-Schreckenberg CA-model with comfortable driving.
In contrast, in the KKW-models the inverse distance between vehicles  
shows only small changes in WSP, the amplitude of which is comparable to the
bare model fluctuations. 
Thus, WSP in our CA-models within the three-phase traffic theory has a
qualitatively different nature in comparison with OMJ in the Nagel-Schreckenberg 
CA-model with comfortable driving.

(iii) If the flow rate $q_{\rm on}$ is further gradually increasing  first no 
transition to another congested traffic pattern occurs in the Nagel-Schreckenberg 
CA-model with comfortable driving:
OMJ persists in some range of the flow rate $q_{\rm on}$
(the region between the boundaries
$O$ and $L$ which is marked ``OMJ'' in
the diagram of congested patterns in Fig.~\ref{Diagram3_NaSch}).

(iv) However, above some flow rate $q_{\rm on}$ 
a widening region of very low mean vehicle speed 
(about $v=10 \ km/h$) and very low mean flow rate 
($q_{\rm LP} \approx 480$ {\rm vehicles/h})
occurs spontaneously upstream of the on-ramp in the Nagel-Schreckenberg 
CA-model with comfortable driving (Fig.~\ref{Mesh_WPL_GP} (d-f) 
and~\ref{XT_WPL_GP} (c, d)). 

The downstream front of this congested pattern is pinned at the
on-ramp and the upstream front is slowly moving upstream.  Therefore,
this patterns may be called ``the widening pinned layer'' (WPL for
short).  Inside WPL as well as in OMJ a very complex non-stationary
behaviour of vehicles occurs.  This random behaviour resembles the one
in OMJ, however with two differences: (1) the maximal vehicle
speed in WPL is much lower than in OMJ, and (2) the
vehicles come much more frequently to a stop in WPL. Because of
the extremely low mean vehicle speed and large fluctuations in
WPL, we cannot discern a regular pattern in WPL. States in which
vehicles stop can emerge and dissolve stochastically but in a
correlated way at different locations in WPL with time scale of
about 5-10 min.  These correlations in WPL seem to propagate with
a velocity faster than the propagation of the front between WPL and
OMJ.

(v) If the whole range of the flow rates $q_{\rm in}$ on the road
upstream of the on-ramp and the flow rate $q_{\rm on}$ to the on-ramp is studied,
then the conclusion can be drawn that there are only  two different congested patterns
in the Nagel-Schreckenberg CA-model
with comfortable driving at the on-ramp: The pinned layer (PL)
which can be either widening (WPL)
or localized (LPL) and OMJ. In addition a  combination
of WPL and OMJ is possible 
(the latter pattern is marked as ``WPL $\&$ OMJ'' in
Fig.~\ref{Mesh_WPL_GP} (d-f) 
and~\ref{XT_WPL_GP} (c)).

In particular, if the flow rate $q_{\rm in}\geq q_{\rm OMJ}$
(at $q_{\rm in}= q_{\rm OMJ}$ the boundaries $O$ and $L$ intersect one
another), then to the right of the  
boundary $L$ in the diagram of patterns for  the Nagel-Schreckenberg 
CA-model with comfortable driving 
WPL occurs at the on-ramp, and upstream of this widening pinned layer OMJ is 
realized (in the region marked as ``WPL $\&$ OMJ'' in
Fig.~\ref{Diagram3_NaSch}). The case of such a spatial combination of
WPL and OMJ is shown in  Fig.~\ref{Mesh_WPL_GP} (d-f) 
and~\ref{XT_WPL_GP} (c)). 

The same combination of WPL and OMJ occurs
spontaneously from an initial state of free flow, if
the flow rate $q_{\rm in}$ is within the range $q_{\rm W} <q_{\rm
  in}\leq q_{\rm OMJ}$.
This occurs, when the flow rate to the on-ramp $q_{\rm on}$ 
is increasing and becomes larger than the
one which is related to the boundary $O$.

However if the flow rate  $q_{\rm in}$ is within the range $q_{PL}
<q_{\rm in}\leq q_{W}$, 
then no OMJ occurs upstream of the WPL right of the boundary $O$
(the region marked ``WPL'' in the diagram of congested patterns in
Fig.~\ref{Diagram3_NaSch}).
In the latter case free flow is realized upstream of the WPL.

If the flow rate  $q_{\rm in}< q_{\rm PL}$, then  right of the boundary
$O$ the pinned layer occurs 
whose upstream front does not move continuously upstream.
Indeed, in this case the flow rate $q_{\rm in}$ upstream of the pinned layer
is lower than the mean flow rate inside the pinned layer $q_{\rm PL}$.
This means that the localized pinned layer (LPL) rather than WPL occurs in the region
marked ``LPL'' in the diagram in Fig.~\ref{Diagram3_NaSch}.
The vehicle behaviour inside LPL is qualitatively the same as inside WPL.
Note that  OMJ and WPL, and also LPL, which occur in the Nagel-Schreckenberg CA-model
with comfortable driving at the on-ramp,
have {\it not} been observed in empirical observations~\cite{Kerner2002B}.

These congested patterns in the Nagel-Schreckenberg CA-model
with comfortable driving are qualitatively different from the ones
which occur spontaneously
for the same conditions in the KKW-models within three-phase
traffic theory 
(Table \ref{KKW-1}). 
These differences are summarized below:

(1) At the same given flow rate on the one-lane road upstream 
of the on-ramp, $q_{\rm in}$, and at some flow rate $q_{\rm on}$,
where OMJ is formed in the Nagel-Schreckenberg 
CA-model with comfortable driving, 
WSP occurs spontaneously upstream of the on-ramp in the KKW-models.
It can be seen from Figs.~\ref{Mesh_OMJ_WSP} (a-c)
and~\ref{XT_OMJ_WSP} (a, b), where  the spatial
vehicle speed distribution in the WSP is shown that these pattern
characteristics are different from OMJ shown in 
Figs.~\ref{Mesh_OMJ_WSP} (d-f) and~\ref{XT_OMJ_WSP} (c, d). In particular,
whereas OMJ in the Nagel-Schreckenberg CA-model
with comfortable driving 
is characterized by a complex birth and decay of narrow moving jams,
no moving jams were seen in WSP.

(2) If the flow rate $q_{\rm on}$ is further gradually increasing, WSP
spontaneously transforms either into DGP or into GP  in our CA-models
within three-phase traffic theory. At high flow rate to the
on-ramp $q_{\rm on}$ the GP does not transform into any other kind
of congested pattern: the GP remains GP no matter how high the 
flow rate $q_{\rm on}$ upstream of the on-ramp  is.
In contrast, if the flow rate to the on-ramp $q_{\rm on}$  increases
for a given value $q_{\rm in}>q_{\rm OMJ}$  in the Nagel-Schreckenberg 
CA-model with comfortable driving, then  to the right of the
boundary $L$ in Fig.~\ref{Diagram3_NaSch} a widening pinned layer WPL
develops upstream of the on-ramp.  

(3) In the Nagel-Schreckenberg CA-model with comfortable driving
upstream of WPL the oscillating moving jams (OMJ) occur spontaneously,
if $q_{\rm in}>q_{\rm OMJ}$ (Figs.~\ref{Diagram3_NaSch} and
Fig.~\ref{Mesh_WPL_GP} (d)).  In the KKW-models within three-phase
traffic theory, however, upstream of the most upstream wide moving jam
in the GP a region of synchronized flow is realized, where no
oscillations and no moving jams occur (Fig.~\ref{Mesh_WPL_GP} (a)).

(4) In the pinch region of the GP  in our CA-models
within three-phase traffic theory narrow moving jams emerge.
Some of these narrow moving jams grow and transform into wide moving jams 
spontaneously at the upstream front of the pinch region
(Figs.~\ref{Patterns1_GP} (a, b) and \ref{Mesh_Patterns1} (a)).
These wide moving jams propagate further upstream without any limitation.
Thus, in the KKW-models wide moving jams
which possess the characteristic parameters mentioned above (Sect.~\ref{Wide})
 {\it spontaneously} occur
in GP. 
In contrast, in the Nagel-Schreckenberg 
CA-model
with comfortable driving narrow moving jams 
do  not transform into wide moving jams. Instead, OMJ or WPL patterns are formed.
In other words, in contrast to our CA-models
within three-phase traffic theory,
in the Nagel-Schreckenberg CA-model
with comfortable driving {\it no spontaneous} emergence of wide moving jams
which possess the characteristic parameters mentioned above (Sect.~\ref{Wide})
occurs: In their model, such a wide moving jam 
can only be excited by an additional external perturbation of a very large
amplitude (which was done in~\cite{Knospe2001} in order to create the
wide moving jam), i.e. by a perturbation 
 which forces one of the vehicles to stop for several time steps.

(5)  The fact that no wide moving jams
can {\it spontaneously} occur in the Nagel-Schreckenberg CA-model
with comfortable driving  can be seen  from a comparison of the left
pictures for GP with 
the right pictures for WPL in Fig.~\ref{Mesh_WPL_GP} and~\ref{XT_WPL_GP}.
In particular, there is a clear regular spatio-temporal structure of 
wide moving jams which alternate with the regions where vehicles move sometimes 
with nearly the maximal vehicle speed inside GP (left figures),
whereas there is no such region inside WPL  (right figures).

Moreover, wide moving jams which have spontaneously occurred in GP 
(Fig.~\ref{Mesh_WPL_GP} (a) and~\ref{XT_WPL_GP} (a)) propagate further without any limitation
upstream. In contrast, in the Nagel-Schreckenberg CA-model
with comfortable driving regions inside WPL, 
where vehicles come to a stop, can emerge  and dissolve randomly
during several minutes. These regions
do  {\it not} propagate through the upstream boundary of WPL.

\subsection{Conclusions about features of the KKW-models in
the three-phase traffic theory}

Several CA-models belonging to three-phase traffic theory were
proposed in this paper. Their simulation gives results that allow to
draw  the following conclusions:

(i) The conditions (\ref{next}),  (\ref{next1}) from~\cite{KKl}, 
where due to the introduction of the synchronization distance $D$ 
a 2D-region of
the steady states in the flow density plane appears,
allow the formulation of {\it several different sets of specific functions} for
fluctuations and  
for the vehicle acceleration and deceleration  which lead to
{\it qualitatively the same diagram of congested patterns} at on-ramps.

(ii) In the KKW-model formulation, specific functions for fluctuations and 
for the vehicle acceleration and deceleration in the basic model
(\ref{next}),  (\ref{next1})  can be much simpler than in~\cite{KKl}.
Nevertheless, the main features of  the diagrams of congested patterns are
qualitatively the same as within the three-phase traffic theory
in~\cite{Kerner2002B,KKl}.

(iii) The diagrams of congested patterns in the KKW-models within  
three-phase traffic theory have the following features which differ
from the case considered in~\cite{KKl}.
First, there is no region in the diagrams of patterns where the moving
synchronized flow pattern (MSP) occurs exclusively. However,
MSP can spontaneously randomly emerge in the region ``WSP'', where 
the widening synchronized flow pattern (WSP)  occurs.
Second, when the initial flow rate on the road upstream 
of the on-ramp $q_{\rm in}$ is higher than 
the mean flow rate in the wide moving jam outflow
then the general pattern (GP) occurs at lower flow rates
to the on-ramp in comparison with the case of the model in~\cite{KKl}.

(iv) The congested patterns and the diagram of these patterns of the
KKW-models explain empirical pattern features~\cite{Kerner2002B} and
their evolution, when the flow rate to the on-ramp is changing.

(v) Both the diagrams of congested patterns and the pattern features
of the KKW-models, which belong to three-phase traffic theory,
are  qualitatively different from those derived by Helbing {\it et al.} 
for a wide class of traffic flow models in the fundamental diagram approach
(for more details see~\cite{KKl}).

(vi) The features of congested patterns, which occur upstream of the
on-ramp in the KKW-models differ from those in Nagel-Schreckenberg
CA-models, which belong to the fundamental diagram approach, including
the one with comfortable driving (see Sect.~\ref{Com}). This is due to
qualitatively different rules of vehicle motion of the basic model
(\ref{next}),  (\ref{next1})~\cite{KKl} in the three-phase traffic
theory in comparison with Nagel-Schreckenberg
CA-models~\cite{Helbing,Ch,Knospe2000,Knospe2001,NaSch,Nagel1995,
  Schreckenberg,Barlovic,Wolf,Nagel,Knospe2002}.\\

{\bf Acknowledgements}
We like to thank Kai Nagel for useful comments. BSK acknowledges
funding by BMBF within project DAISY.\\

\Figures

\clearpage


\clearpage

\end{document}